\documentclass[
aps,prd,
,nofootinbib,
floats]{revtex4}
\usepackage{graphicx}
\usepackage{amssymb,latexsym}
\usepackage[draft=false]{hyperref}

\def\lsim{\hbox{ \raise.35ex\rlap{$<$}\lower.6ex\hbox{$\sim$}\ }}
\def\gsim{\hbox{ \raise.35ex\rlap{$>$}\lower.6ex\hbox{$\sim$}\ }}
\def\eg{{\sl e.g.}}  
\def\ie{{\sl i.e.}}     
\def\be{\begin{equation}}\def\bea{\begin{eqnarray}}\def\beaa{\begin{eqnarray*}}
\def\ee{\end{equation}}  \def\eea{\end{eqnarray}} \def\eeaa{\end{eqnarray*}}        
\def\la{\bigl\langle} \def\ra{\bigr\rangle}    \def\etal{{\sl et al.}}   
\def\D{ {\Delta T \over T} }     
\def\bbox#1{\hbox{\boldmath{$#1$}}}
\def\edth{\;\raise1.5pt\hbox{$'$}\hskip-6pt\partial\;}
\def\baredth{\;\overline{\raise1.5pt\hbox{$'$}\hskip-6pt \partial}\;}   

\begin{document}
\title{Cosmology from Topological Defects
\footnote{Write-up of a set of two Lecture Notes delivered at the Xth Brazilian School on Cosmology and
          Gravitation, Mangaratiba, Rio de Janeiro, Brazil, July 29 -- August 9, 2002. $~~$ 
         {\tt http://www.cbpf.br/\~{}cosmogra/Xescola/Xschool.html} \\ 
A complete and updated version of these notes with colour figures can be found at
{\tt www.iafe.uba.ar/relatividad/gangui/xescola/} }}
\author{Alejandro Gangui}
\email{gangui@iafe.uba.ar}
\affiliation{Instituto de Astronom\'{\i}a y F\'{\i}sica del Espacio,
Ciudad Universitaria, 1428 Buenos Aires, Argentina, and 
\\  
Dept. de F\'{\i}sica, Universidad de Buenos Aires, Ciudad Universitaria --
Pab. 1, 1428 Buenos Aires, Argentina.}
\begin{abstract} 
\noindent 
The potential role of cosmic topological defects has raised interest in the astrophysical community
for many years now.  In this set of notes, we give an introduction to the subject of cosmic
topological defects and some of their possible observable signatures. We begin with a review of the
basics of general defect formation and evolution, we briefly comment on some general features of
conducting cosmic strings and vorton formation, as well as on the possible role of defects as dark
energy, to end up with cosmic structure formation from defects and some specific imprints in the
cosmic microwave background radiation from simulated cosmic strings. A detailed, pedagogical
explanation of the mechanism underlying the tiny level of polarization discovered in the cosmic
microwave background by the DASI collaboration (and recently confirmed by WMAP) is also given, and a
first rough comparison with some predictions from defects is provided.
\end{abstract}
\maketitle


\section{Introduction}
\label{sec-intro}      

\noindent
On a cold day, ice forms quickly on the surface of a pond. But it does not grow as a smooth,
featureless covering.  Instead, the water begins to freeze in many places independently, and the
growing plates of ice join up in random fashion, leaving zig--zag boundaries between them. These
irregular margins are an example of what physicists call ``topological defects'' -- {\sl defects}
because they are places where the crystal structure of the ice is disrupted, and {\sl topological}
because an accurate description of them involves ideas of symmetry embodied in topology, the branch
of mathematics that focuses on the study of continuous surfaces.

Current theories of particle physics likewise predict that a variety of topological defects would
almost certainly have formed during the early evolution of the universe. Just as water turns to ice
(a phase transition) when the temperature drops, so the interactions between elementary particles
run through distinct phases as the typical energy of those particles falls with the expansion of the
universe. When conditions favor the appearance of a new phase, it generally crops up in many places
at the same time, and when separate regions of the new phase run into each other, topological
defects are the result. The detection of such structures in the modern universe would provide
precious information on events in the earliest instants after the Big Bang. Their absence, on the
other hand, would force a major revision of current physical theories.

The aim of this set of Lectures is to introduce the reader to the subject of cosmology from
topological defects. We begin with a review of the basics of defect formation and evolution, to get
a grasp of the overall picture.  We will see that defects are generically predicted to exist in most
interesting models of high energy physics trying to describe the early universe.  The basic elements
of the standard cosmology, with its successes, shortcomings, and new developments, are covered
elsewhere in this volume. See for example the lecture notes by Rocky Kolb on Astroparticle Physics,
Ed Copeland's material on String / M-Theory Cosmology, and Jim Bartlett's Observational
Cosmology. So we will not devote much space to these topics here. Rather, we will focus on some
specific subjects. We will first briefly comment on conducting cosmic strings and one of their most
important predictions for cosmology, namely, the existence of equilibrium configurations of string
loops, dubbed vortons. We will then pass on to study some key signatures that a network of defects
would produce on the cosmic microwave background (CMB) radiation, \eg, the CMB bispectrum of the
temperature anisotropies from a simulated model of cosmic strings.  Miscellaneous topics also
reviewed below are, for example, the way in which these cosmic entities lead to large--scale
structure formation and some astrophysical footprints left by the various defects, and we will
discuss the possibility of isolating their effects by astrophysical observations. Also, we include a
short, detailed discussion of CMB polarization and some brief comparison with the predictions from
cosmic defects. 

\newpage

Many areas of modern research directly related to cosmic defects are unfortunately not covered in
these notes. The subject is now so vast -and beyond the possibilities of a single review- that we
suggest the reader to consult some of the excellent recent literature already available. So, have a
look, for example, to the report by Ach\'ucarro \& Vachaspati [2000] for a treatment of semilocal and
electroweak strings~\footnote{Animations of semilocal and electroweak string formation and evolution
can be found at {\tt http://www.nersc.gov/\~{}borrill/}}, and to [Vachaspati, 2001] for a review of
certain topological defects, like monopoles, domain walls and, again, electroweak strings, virtually
not covered here.  For conducting defects, cosmic strings in particular, see for example [Gangui \&
Peter, 1998] for a brief overview of many different astrophysical and cosmological phenomena, [Gangui,
2001b], from which this review borrows (and updates) a great deal of material, for a treatment of
conducting cosmic strings and one of their most important predictions for cosmology, namely, the
existence of equilibrium configurations of string loops, dubbed vortons. Finally, refer to the
comprehensive colorful lecture notes by Carter [1997] on the dynamics of branes with applications to
conducting cosmic strings and vortons.  If your are in cosmological structure formation, Durrer [2000]
presents a good review of modern developments on global topological defects and their relation to CMB
anisotropies, while Magueijo \& Brandenberger [2000] give a set of imaginative lectures with an update
on local string models of large-scale structure formation and also baryogenesis with cosmic
defects. Finally, Durrer, Kunz \& Melchiorri [2002] give a complete update of cosmic structure
formation with global defects, including detailed analyses of correlators, mixed models, and the
resulting matter and CMB power spectra.

The interdisciplinary subject of topological defects in the cosmos and the lab is nicely covered in
the proceedings of the school held {\sl aux} Houches on topological defects and non-equilibrium
dynamics, edited by Bunkov \& Godfrin [2000]; the ensemble of lectures in this volume, together with
the recent review by Kibble [2002], give an exhaustive illustration of this fast developing area of
research, which includes various fields of physics, like low--temperature condensed--matter, liquid
crystals, astrophysics and high--energy physics.
Finally, all of the above can also be found in the concise review by Hindmarsh \& Kibble [1995],
particularly concerned with the physics and cosmology of cosmic strings, and in the monograph by
Vilenkin \& Shellard [2000] on cosmic strings and other topological defects.

\subsection{How defects form}
\label{subsec-howdefects}   

A central concept of particle physics theories attempting to unify all the fundamental interactions is
the concept of symmetry breaking.  As the universe expanded and cooled, first the gravitational
interaction, and subsequently all other known forces would have begun adopting their own identities.
In the context of the standard hot Big Bang theory the spontaneous breaking of fundamental symmetries
is realized as a phase transition in the early universe.  Such phase transitions have several exciting
cosmological consequences and thus provide an important link between particle physics and cosmology.

There are several symmetries which are expected to break down in the course of time.  In each of these
transitions the space--time gets `oriented' by the presence of a hypothetical force field called the
`Higgs field', named for Peter Higgs, pervading all the space. This field orientation signals the
transition from a state of higher symmetry to a final state where the system under consideration obeys
a smaller group of symmetry rules.  As an every--day analogy we may consider the transition from
liquid water to ice; the formation of the crystal structure ice (where water molecules are arranged in
a well defined lattice), breaks the symmetry possessed when the system was in the higher temperature
liquid phase, when every direction in the system was equivalent.  In the same way, it is precisely the
orientation in the Higgs field which breaks the highly symmetric state between particles and forces.
                                                                          
Having built a model of elementary particles and forces, particle physicists and cosmologists are
today embarked on a difficult search for a theory that unifies all the fundamental interactions. As we
mentioned, an essential ingredient in all major candidate theories is the concept of symmetry
breaking. Experiments have determined that there are four physical forces in nature; in addition to
gravity these are called the strong, weak and electromagnetic forces. Close to the singularity of the
hot Big Bang, when energies were at their highest, it is believed that these forces were unified in a
single, all--encompassing interaction. As the universe expanded and cooled, first the gravitational
interaction, then the strong interaction, and lastly the weak and the electromagnetic forces would
have broken out of the unified scheme and adopted their present distinct identities in a series of
symmetry breakings.

Theoretical physicists are still struggling to understand how gravity can be united with the other
interactions, but for the unification of the strong, weak and electromagnetic forces plausible
theories exist. Indeed, force--carrying particles whose existence demonstrated the fundamental
unification of the weak and electromagnetic forces into a primordial ``electroweak'' force -- the W
and Z bosons -- were discovered at CERN, the European accelerator laboratory, in 1983.  In the context
of the standard Big Bang theory, cosmological phase transitions are produced by the spontaneous
breaking of a fundamental symmetry, such as the electroweak force, as the universe cools. For example,
the electroweak interaction broke into the separate weak and electromagnetic forces when the
observable universe was $10^{-12}$ seconds old, had a temperature of $10^{15}$ degrees Kelvin, and was
only one part in $10^{15}$ of its present size. There are also other phase transitions besides those
associated with the emergence of the distinct forces. The quark-hadron confinement transition, for
example, took place when the universe was about a microsecond old. Before this transition, quarks --
the particles that would become the constituents of the atomic nucleus -- moved as free particles;
afterward, they became forever bound up in protons, neutrons, mesons and other composite particles.

\begin{figure}[htbp]
\includegraphics[width=8cm]{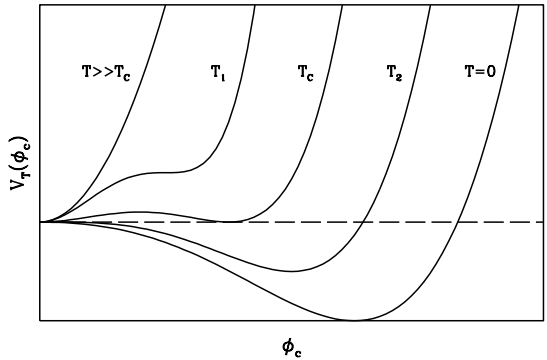}\\
\caption{ {\sl Temperature--dependent effective potential for a first--order phase transition for the
Higgs field. For very high temperatures, well above the critical one $T_c$, the potential possesses
just one minimum for the vanishing value of the Higgs field. Then, when the temperature decreases, a
whole set of minima develops (it may be two or more, discrete or continuous, depending of the type of
symmetry under consideration). Below $T_c$, the value $\phi = 0$ stops being the global minimum and
the system will spontaneously choose a new (lower) one, say $\phi = \eta \exp(i\theta)$ (for complex
$\phi$) for some angle $\theta$ and nonvanishing $\eta$, amongst the available ones.  This choice
signals the breakdown of the symmetry in a cosmic phase transition and the generation of random
regions of conflicting field orientations $\theta$. In a cosmological setting, the merging of these
domains gives rise to cosmic defects.}  }
\label{fig-pot_phtrans}
\end{figure}

As we said, the standard mechanism for breaking a symmetry involves the hypothetical Higgs field that
pervades all space. As the universe cools, the Higgs field can adopt different ground states, also
referred to as different vacuum states of the theory. In a symmetric ground state, the Higgs field is
zero everywhere. Symmetry breaks when the Higgs field takes on a finite value (see Figure
\ref{fig-pot_phtrans}).

Kibble [1976] first saw the possibility of defect formation when he realized that in a cooling
universe phase transitions proceed by the formation of uncorrelated domains that subsequently
coalesce, leaving behind relics in the form of defects.  In the expanding universe, widely separated
regions in space have not had enough time to `communicate' amongst themselves and are therefore not
correlated, due to a lack of causal contact.  It is therefore natural to suppose that different
regions ended up having arbitrary orientations of the Higgs field and that, when they merged together,
it was hard for domains with very different preferred directions to adjust themselves and fit
smoothly. In the interfaces of these domains, defects form.  Such relic `flaws' are unique examples of
incredible amounts of energy and this feature attracted the minds of many cosmologists.

\subsection{Phase transitions and finite temperature field theory}
\label{sec-PhaseTrans}      

Phase transitions are known to occur in the early universe.  Examples we mentioned are the quark to
hadron (confinement) transition, which QCD predicts at an energy around 1 GeV, and the electroweak
phase transition at about 250 GeV.  Within grand unified theories (GUT), aiming to describe the
physics beyond the standard model, other phase transitions are predicted to occur at energies of order
$10^{15}$ GeV; during these, the Higgs field tends to fall towards the minima of its potential while
the overall temperature of the universe decreases as a consequence of the expansion.

A familiar theory to make a bit more quantitative the above considerations is the $\lambda |\phi|^4$
theory,
\be
\label{lambda4}
{\cal L} = {1\over 2} |\partial_\mu\phi|^2 + {1\over 2} m_0^2 |\phi|^2
- {\lambda \over 4!} |\phi|^4 ~, \ee with $m^2_0 > 0$.  
The second and third terms on the right hand side yield the usual
`Mexican hat' potential for the complex scalar field.  For energies
much larger than the critical temperature, $T_c$, the fields are in
the so--called `false' vacuum: a highly symmetric state characterized
by a vacuum expectation value $\la | \phi | \ra = 0$.  But when
energies decrease the symmetry is spontaneously broken: a new `true'
vacuum develops and the scalar field rolls down the potential and sits
onto one of the degenerate new minima. In this situation the vacuum
expectation value becomes $\la | \phi | \ra^2 = 6 m_0^2 / \lambda$.

Research done in the 1970's in finite--temperature field theory
[Weinberg, 1974; Dolan \& Jackiw, 1974; Kirzhnits \& Linde, 1974] has
led to the result that the temperature--dependent effective potential
can be written down as \be
\label{VfiniT}
V_T( | \phi | ) = -{1\over 2} m^2(T) |\phi|^2 + {\lambda \over 4!}
|\phi|^4 \ee 
with $T_c^2 = 24 m_0^2 / \lambda $, $m^2(T) = m_0^2 (1 -
T^2 / T_c^2)$, and $\la | \phi | \ra^2 = 6 m^2(T) / \lambda$.  We
easily see that when $T$ approaches $T_c$ from below the symmetry is
restored, and again we have $\la | \phi | \ra = 0$.  In
condensed--matter jargon, the transition described above is
second--order
[Mermin, 1979].\footnote{In a first--order phase transition the order parameter
(\eg, $<|\phi|>$ in our case) is not continuous.  It may
proceed by bubble nucleation [Callan \& Coleman, 1977; Linde, 1983b]
or by spinoidal decomposition [Langer, 1992].  Phase transitions can
also be continuous second--order processes. The `order' depends
sensitively on the ratio of the coupling constants appearing in the
Lagrangian.}

\subsection{The Kibble mechanism}
\label{sec-Kibbbb}                                 

The model described in the last subsection is an example in which the
transition may be second--order.  As we saw, for temperatures much
larger than the critical one the vacuum expectation value of the
scalar field vanishes at all points of space, whereas for $T < T_c$ it
evolves smoothly in time towards a non vanishing $\la | \phi | \ra$.
Both thermal and quantum fluctuations influence the new value taken by
$\la | \phi | \ra$ and therefore it has no reasons to be uniform in
space.  This leads to the existence of domains wherein the $\la | \phi
(\vec x) | \ra$ is coherent and regions where it is not.  The
consequences of this fact are the subject of this subsection.
                                 
Phase transitions can also be first--order proceeding via bubble
nucleation.  At very high energies the symmetry breaking potential has
$\la | \phi | \ra = 0$ as the only vacuum state. When the temperature
goes down to $T_c$ a set of vacua, degenerate to the previous one,
develops. However this time the transition is not smooth as before,
for a potential barrier separates the old (false) and the new (true)
vacua (see, \eg\ Figure \ref{fig-pot_phtrans}).  
Provided the barrier at this small temperature is high enough,
compared to the thermal energy present in the system, the field $\phi$
will remain trapped in the false vacuum state even for small ($< T_c$)
temperatures. Classically, this is the complete picture.  However,
quantum tunneling effects can liberate the field from the old vacuum
state, at least in some regions of space: there is a probability per
unit time and volume in space that at a point $\vec x$ a bubble of
true vacuum will nucleate.  The result is thus the formation of
bubbles of true vacuum with the value of the field in each bubble
being independent of the value of the field in all other bubbles.
This leads again to the formation of domains where the fields are
correlated, whereas no correlation exits between fields belonging to
different domains.  Then, after creation the bubble will expand at the
speed of light surrounded by a `sea' of false vacuum domains.  As
opposed to second--order phase transitions, here the nucleation
process is extremely inhomogeneous and $\la | \phi (\vec x) | \ra$ is
not a continuous function of time.

Let us turn now to the study of correlation lengths and their r\^ole
in the formation of topological defects.  One important feature in
determining the size of the domains where $\la | \phi (\vec x) | \ra$
is coherent is given by the spatial correlation of the field $\phi$.
Simple field theoretic considerations [see, \eg, Copeland, 1993]
for long wavelength fluctuations of $\phi$ lead to different functional
behaviors for the correlation function $G(r) \equiv \la
\phi(r_1)\phi(r_2) \ra$, where we noted $r = |r_1 - r_2|$.  What is
found depends radically on whether the wanted correlation is
computed between points in space separated by a distance $r$ much
smaller or much larger than a characteristic length $\xi^{-1} = m(T)
\simeq \sqrt{\lambda} ~ |\la\phi\ra |$, known as the {\sl correlation
length}.
Then, we have 
$G(r) \simeq  {T_c \over 4 \pi r} \exp (- {r\over \xi})$ for $r >> \xi$ , 
while 
$G(r) \simeq    {T^2 \over 2 \pi^2}$ for $r << \xi$ .

This tells us that domains of size $ \xi \sim m^{-1}$ arise where the
field $\phi$ is correlated.  On the other hand, well beyond $\xi$ no
correlations exist and thus points separated apart by $r >> \xi$ will
belong to domains with in principle arbitrarily different orientations
of the Higgs field.  This in turn leads, after the merging of these
domains in a cosmological setting, to the existence of defects, where
field configurations fail to match smoothly.

However, when $T \to T_c$ we have $m\to 0$ and so $\xi\to\infty$,
suggesting perhaps that for all points of space the field $\phi$
becomes correlated. This fact clearly violates causality.  The
existence of particle horizons in cosmological models (proportional to
the inverse of the Hubble parameter $H^{-1}$) constrains microphysical
interactions over distances beyond this causal domain.  Therefore we
get an upper bound to the correlation length as $\xi < H^{-1} \sim t$.

The general feature of the existence of uncorrelated domains has
become known as the Kibble mechanism [Kibble, 1976] and it seems to be
generic to most types of phase transitions.

\subsection{A survey of topological defects}
\label{sec-ASurv}

Different models for the Higgs field lead to the formation of a whole
variety of topological defects, with very different characteristics
and dimensions.  Some of the proposed theories have symmetry breaking
patterns leading to the formation of `domain walls' (mirror reflection
discrete symmetry): incredibly thin planar surfaces trapping enormous
concentrations of mass--energy which separate domains of conflicting
field orientations, similar to two--dimensional sheet--like structures
found in ferromagnets.  Within other theories, cosmological fields get
distributed in such a way that the old (symmetric) phase gets confined
into a finite region of space surrounded completely by the new
(non--symmetric) phase. This situation leads to the generation of
defects with linear geometry called `cosmic strings'.  Theoretical
reasons suggest these strings (vortex lines) do not have any loose
ends in order that the two phases not get mixed up.  This leaves
infinite strings and closed loops as the only possible alternatives
for these defects to manifest themselves in the early
universe\footnote{`Monopole' is another possible topological defect;
we defer its discussion to the next subsection.  Cosmic strings
bounded by monopoles is yet another possibility in GUT phase
transitions of the kind, \eg, ${\bf G}\to {\bf K}\times U(1)\to {\bf
K}$.  The first transition yields monopoles carrying a magnetic charge
of the $U(1)$ gauge field, while in the second transition the magnetic
field in squeezed into flux tubes connecting monopoles and
antimonopoles [Langacker \& Pi, 1980].}.

\begin{figure}[htbp]
\includegraphics[width=14cm]{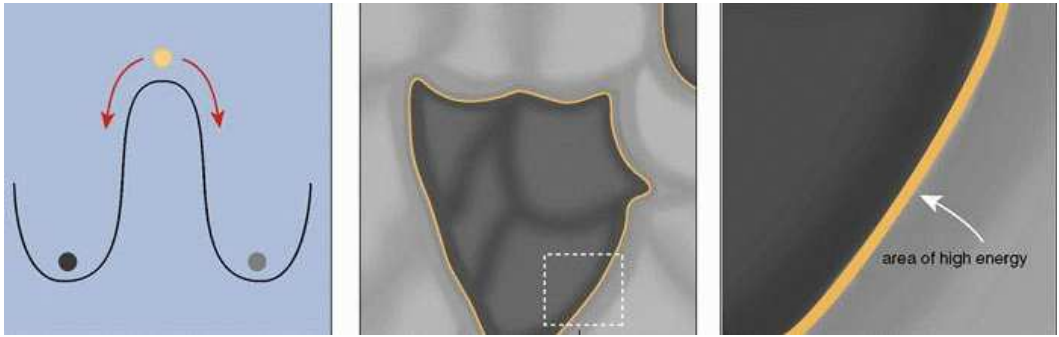}\\
\caption{ 
{\sl In a simple model of symmetry breaking, the initial symmetric ground state of the Higgs field
(yellow central 
dot) can fall into the left- or right-hand valley of a double-well energy potential (light and
dark dots). In a cosmic phase transition, regions of the new phase appear randomly and begin to grow
and eventually merge as the transition proceeds toward completion (middle). Regions in which the
symmetry has broken the same way can coalesce, but where regions that have made opposite choices
encounter each other, a topological defect known as a domain wall forms (right). Across the wall, the
Higgs field has to go from one of the valleys to the other (in the left panel), and must therefore
traverse the energy peak. This creates a narrow planar region of very high energy, in which the
symmetry is locally unbroken.}
}
\label{fig-dwcolor}
\end{figure}            

With a bit more abstraction scientists have even conceived other (semi) topological defects, called
`textures'. These are conceptually simple objects, yet, it is not so easy to imagine them for they are
just global field configurations living on a three--sphere vacuum manifold (the minima of the
effective potential energy), whose non linear evolution perturbs spacetime.  Turok [1989] was the
first to realize that many unified theories predicted the existence of peculiar Higgs field
configurations known as (texture) knots, and that these could be of potential interest for cosmology.
Several features make these defects interesting.  In contrast to domain walls and cosmic strings,
textures have no core and thus the energy is more evenly distributed over space.  Secondly, they are
unstable to collapse and it is precisely this last feature which makes these objects cosmologically
relevant, for this instability makes texture knots shrink to a microscopic size, unwind and radiate
away all their energy.  In so doing, they generate a gravitational field that perturbs the surrounding
matter in a way which can seed structure formation.

\subsection{Conditions for their existence: topological criteria}
\label{subsec-condexist}      

Let us now explore the conditions for the existence of topological defects. It is widely accepted that
the final goal of particle physics is to provide a unified gauge theory comprising strong, weak and
electromagnetic interactions (and some day also gravitation).  This unified theory is to describe the
physics at very high temperatures, when the age of the universe was slightly bigger than the Planck
time.  At this stage, the universe was in a state with the highest possible symmetry, described by a
symmetry group {\bf G}, and the Lagrangian modeling the system of all possible particles and
interactions present should be invariant under the action of the elements of {\bf G}.

As we explained before, the form of the finite temperature effective potential of the system is
subject to variations during the cooling down evolution of the universe.  This leads to a chain of
phase transitions whereby some of the symmetries present in the beginning are not present anymore at
lower temperatures.  The first of these transitions may be described as {\bf G}$\to${\bf H}, where now
{\bf H} stands for the new (smaller) unbroken symmetry group ruling the system.  This chain of
symmetry breakdowns eventually ends up with SU(3)$\times$SU(2)$\times$U(1), the symmetry group
underlying the `standard model' of particle physics.

A broken symmetry system (with a Mexican-hat potential for the Higgs field) may have many different
minima (with the same energy), all related by the underlying symmetry.  Passing from one minimum to
another is included as one of the symmetries of the original group {\bf G}, and the system will not
change due to one such transformation.  If a certain field configuration yields the lowest energy
state of the system, transformations of this configuration by the elements of the symmetry group will
also give the lowest energy state.  For example, if a spherically symmetric system has a certain
lowest energy value, this value will not change if the system is rotated.

The system will try to minimize its energy and will spontaneously choose one amongst the available
minima.  Once this is done and the phase transition achieved, the system is no longer ruled by {\bf G}
but by the symmetries of the smaller group {\bf H}.  So, if {\bf G}$\to${\bf H} and the system is in
one of the lowest energy states (call it $S_1$), transformations of $S_1$ to $S_2$ by elements of {\bf
G} will leave the energy unchanged. However, transformations of $S_1$ by elements of {\bf H} will
leave $S_1$ {\it itself} (and not just the energy) unchanged.  The many distinct ground states of the
system $S_1 , S_2 , \ldots $ are given by all transformations of {\bf G} that are {\it not} related by
elements in {\bf H}. This space of distinct ground states is called the {\sl vacuum manifold} and
denoted $\cal M$.
%
So, $\cal M$ is the space of all elements of {\bf G} in which elements related by transformations in
{\bf H} have been identified.  Mathematicians call it the {\sl coset space} and denote it {\bf
G}$/${\bf H}. We then have ${\cal M}=$ {\bf G}$/${\bf H}.

The importance of the study of the vacuum manifold lies in the fact that it is precisely the {\sl
topology} of ${\cal M}$ what determines the type of defect that will arise.  Homotopy theory tells us
how to map ${\cal M}$ into physical space in a non--trivial way, and what ensuing defect will be
produced.  For instance, the existence of non contractible loops in ${\cal M}$ is the requisite for
the formation of cosmic strings.  In formal language this comes about whenever we have the first
homotopy group $\pi_1 ({\cal M}) \neq$ {\bf 1}, where {\bf 1} corresponds to the trivial group.  If
the vacuum manifold is disconnected we then have $\pi_0 ({\cal M}) \neq$ {\bf 1}, and domain walls are
predicted to form in the boundary of these regions where the field $\phi$ is away from the minimum of
the potential.  Analogously, if $\pi_2 ({\cal M}) \neq$ {\bf 1} it follows that the vacuum manifold
contains non contractible two--spheres, and the ensuing defect is a monopole.  Textures arise when
${\cal M}$ contains non contractible three--spheres and in this case it is the third homotopy group,
$\pi_3 ({\cal M})$, the one that is non trivial.  We summarize this in Table \ref{table-topo} .

\begin{table}[t]\begin{center}
\begin{tabular}{|c l l|}
\hline
$\pi_0 ({\cal M}) \neq${\bf 1} & ${\cal M}$ {\it disconnected}
     & {\sc Domain Walls}  \\
$\pi_1 ({\cal M}) \neq${\bf 1} & {\it non contractible loops} in
${\cal M}$
     & {\sc Cosmic Strings}  \\
$\pi_2 ({\cal M}) \neq${\bf 1} & {\it non contractible 2--spheres} in
${\cal M}$
     & {\sc Monopoles}   \\
$\pi_3 ({\cal M}) \neq${\bf 1} & {\it non contractible 3--spheres} in
${\cal M}$
     & {\sc Textures}  \\
\hline
\end{tabular}\end{center}
\caption{The topology of ${\cal M}$ determines the type of defect
that will arise.}
\label{table-topo}
\end{table}        

\section{Defects in the universe}
\label{sec-definuni}     

Generically topological defects will be produced if the conditions for their existence are met. Then
for example if the unbroken group {\bf H} contains a disconnected part, like an explicit U(1) factor
(something that is quite common in many phase transition schemes discussed in the literature),
monopoles will be left as relics of the transition. This is due to the fundamental theorem on the
second homotopy group of coset spaces [Mermin, 1979], which states that for a simply--connected
covering group {\bf G} we have\footnote{The isomorfism between two groups is noted as $\cong$.  Note
that by using the theorem we therefore can reduce the computation of $\pi_2$ for a coset space to the
computation of $\pi_1$ for a group.  A word of warning: the focus here is on the physics and the
mathematically--oriented reader should bear this in mind, especially when we will become a bit sloppy
with the notation.  In case this happens, consult the book [Steenrod, 1951] for a clear exposition of
these matters.}  \be \pi_2({\bf G} / {\bf H}) \cong \pi_1({\bf H}_0) ~, \ee with ${\bf H}_0$ being the
component of the unbroken group connected to the identity.  Then we see that since monopoles are
associated with unshrinkable surfaces in {\bf G}$/${\bf H}, the previous equation implies their
existence if {\bf H} is multiply--connected.  The reader may guess what the consequences are for GUT
phase transitions: in grand unified theories a semi--simple gauge group {\bf G} is broken in several
stages down to {\bf H} = SU(3)$\times$U(1).  Since in this case $\pi_1({\bf H}) \cong {\cal Z}$, the
integers, we have $\pi_2 ({\bf G} / {\bf H}) \neq$ {\bf 1} and therefore gauge monopole solutions
exist [Preskill, 1979].

\subsection{Local and global monopoles and domain walls}
\label{sec-monoANDdo}
                        
Monopoles are yet another example of stable topological defects.  Their formation stems from the fact
that the vacuum expectation value of the symmetry breaking Higgs field has random orientations
($\la\phi^a\ra$ pointing in different directions in group space) on scales greater than the horizon.
One expects therefore to have a probability of order unity that a monopole configuration will result
after the phase transition (cf. the Kibble mechanism).  Thus, about one monopole per Hubble volume
should arise and we have for the number density $n_{monop} \sim 1 / H^{-3} \sim T_c^6 / m_P^3$, where
$T_c$ is the critical temperature and $m_P$ is Planck mass, when the transition occurs.  We also know
the entropy density at this temperature, $s \sim T_c^3$, and so the monopole to entropy ratio is
$n_{monop} / s \simeq 100 (T_c / m_P)^3$.  In the absence of non--adiabatic processes after monopole
creation this constant ratio determines their present abundance.  For the typical value $T_c\sim
10^{14}$ GeV we have $n_{monop} / s \sim 10^{-13}$. This estimate leads to a present $\Omega_{monop}
h^2 \simeq 10^{11}$, for the superheavy monopoles $m_{monop}\simeq 10^{16}$ GeV that are
created\footnote{These are the actual figures for a gauge SU(5) GUT second--order phase
transition. Preskill [1979] has shown that in this case monopole antimonopole annihilation is not
effective to reduce their abundance. Guth \& Weinberg [1983] did the case for a first--order phase
transition and drew qualitatively similar conclusions regarding the excess of monopoles.}.  This value
contradicts standard cosmology and the presently most attractive way out seems to be to allow for an
early period of inflation: the massive entropy production will hence lead to an exponential decrease
of the initial $n_{monop} / s$ ratio, yielding $\Omega_{monop}$ consistent with
observations.\footnote{The inflationary expansion reaches an end in the so--called reheating process,
when the enormous vacuum energy driving inflation is transferred to coherent oscillations of the
inflaton field. These oscillations will in turn be damped by the creation of light particles (\eg, via
preheating) whose final fate is to thermalise and reheat the universe.} In summary, the broad--brush
picture one has in mind is that of a mechanism that could solve the monopole problem by `weeping'
these unwanted relics out of our sight, to scales much bigger than the one that will eventually become
our present horizon today.

Note that these arguments do not apply for global monopoles as these (in the absence of gauge fields)
possess long--range forces that lead to a decrease of their number in comoving coordinates. The large
attractive force between global monopoles and antimonopoles leads to a high annihilation probability
and hence monopole over--production does not take place.  Simulations performed by Bennett \& Rhie
[1990] showed that global monopole evolution rapidly settles into a scale invariant regime with only a
few monopoles per horizon volume at all times.

Given that global monopoles do not represent a danger for cosmology one may proceed in studying their
observable consequences. The gravitational fields of global monopoles may lead to matter clustering
and CMB anisotropies. Given an average number of monopoles per horizon of $\sim 4$, Bennett \& Rhie
[1990] estimate a scale invariant spectrum of fluctuations $( \delta\rho / \rho )_H \sim 30 G \eta^2$
at horizon crossing\footnote{The spectrum of density fluctuations on smaller scales has also been
computed.  They normalize the spectrum at $8 h^{-1}$ Mpc and agreement with observations lead them to
assume that galaxies are clustered more strongly than the overall mass density, this implying a
`biasing' of a few [see Bennett, Rhie \& Weinberg, 1993 for details].}.  In a subsequent paper they
simulate the large--scale CMB anisotropies and, upon normalization with {\sl COBE}--DMR, they get
roughly $G \eta^2 \sim 6 \times 10^{-7}$ in agreement with a GUT energy scale $\eta$ [Bennett \& Rhie,
1993]. However, as we will see in the CMB sections below, current estimates for the angular power
spectrum of global defects do not match the most recent observations, their main problem being the
lack of power on the degree angular scale once the spectrum is normalized to {\sl COBE} on large
scales [Durrer \etal, 1996; Durrer \etal, 2002].

Let us concentrate now on domain walls, and briefly try to show why they are not welcome in any
cosmological context [at least in the simple version we here consider -- there is always room for more
complicated (and contrived) models]. If the symmetry breaking pattern is appropriate at least one
domain wall per horizon volume will be formed.  The mass per unit surface of these two-dimensional
objects is given by $\sim \lambda^{1/2} \eta^3$, where $\lambda$ as usual is the coupling constant in
the symmetry breaking potential for the Higgs field.  Domain walls are generally horizon--sized and
therefore their mass is given by $\sim \lambda^{1/2} \eta^3 H^{-2}$. This implies a mass energy
density roughly given by $\rho_{DW}\sim \eta^3 t^{-1}$ and we may readily see now how the problem
arises: the critical density goes as $\rho_{crit} \sim t^{-2}$ which implies $\Omega_{DW}(t) \sim
(\eta / m_P)^2 \eta t$.  Taking a typical GUT value for $\eta$ we get $\Omega_{DW}(t\sim 10^{-35}{\rm
sec}) \sim 1$ {\sl already} at the time of the phase transition. It is not hard to imagine that today
this will be at variance with observations; in fact we get $\Omega_{DW}(t \sim 10^{18}{\rm sec}) \sim
10^{52}$. This indicates that models where domain walls are produced are tightly constrained, and the
general feeling is that it is best to avoid them altogether [see Kolb \& Turner, 1990 for further
details; see also Dvali \etal, 1998, Pogosian \& Vachaspati, 2000
\footnote{Animations of monopoles colliding with domain walls can be found in 
`LEP' page at {\tt http://theory.ic.ac.uk/\~{}LEP/figures.html}}
and Alexander \etal, 1999 for an alternative solution].

\subsection{Are defects inflated away?}
\label{sec-topoandinfla}
                              
It is important to realize the relevance that the Kibble's mechanism has for cosmology; nearly every
sensible grand unified theory (with its own symmetry breaking pattern) predicts the existence of
defects.  We know that an early era of inflation helps in getting rid of the unwanted relics.  One
could well wonder if the very same Higgs field responsible for breaking the symmetry would not be the
same one responsible for driving an era of inflation, thereby diluting the density of the relic
defects.  This would get rid not only of (the unwanted) monopoles and domain walls but also of any
other (cosmologically appealing) defect.  Let us follow [Brandenberger, 1993] and sketch why this
actually does not occur.  Take first the symmetry breaking potential of Eq. (\ref{VfiniT}) at zero
temperature and add to it a harmless $\phi$--independent term $3 m^4 / (2\lambda)$. This will not
affect the dynamics at all. Then we are led to \be
\label{VfiniT2}
V(  \phi  ) =
{\lambda \over 4!} \left(
\phi^2 - {\eta }^2
\right)^2 ~,
\ee
with $\eta = ( 6 m^2 / \lambda )^{1/2}$
the symmetry breaking energy scale,
and where for the present heuristic digression we just took a real
Higgs field. Consider now the equation of motion for $\phi$,
\be                                             
\label{Higgapprox}
\ddot \phi \simeq - {\partial V\over\partial\phi }
= - {\lambda\over 3!} \phi^3 + m^2 \phi
\approx  m^2 \phi ~,
\ee
for $\phi << \eta$ very near the false vacuum of the effective Mexican hat potential and where, for
simplicity, the expansion of the universe and possible interactions of $\phi$ with other fields were
neglected.  The typical time scale of the solution is $\tau\simeq m^{-1}$.  For an inflationary epoch
to be effective we need $\tau >> H^{-1}$, \ie, a sufficiently large number of e--folds of
slow--rolling solution. Note, however, that after some e--folds of exponential expansion the curvature
term in the Friedmann equation becomes subdominant and we have $H^2 \simeq 8\pi G ~V(0) / 3 \simeq
(2\pi m^2 / 3 )(\eta / m_P)^2$.  So, unless $\eta > m_P$, which seems unlikely for a GUT phase
transition, we are led to $\tau << H^{-1}$ and therefore the amount of inflation is not enough for
getting rid of the defects generated during the transition by hiding them well beyond our present
horizon.

Recently, there has been a large amount of work in getting defects, particularly cosmic strings, after
post-inflationary preheating.  Reaching the latest stages of the inflationary phase, the inflaton
field oscillates about the minimum of its potential. In doing so, parametric resonance may transfer a
huge amount of energy to other fields leading to cosmologically interesting nonthermal phase
transitions.  Just like thermal fluctuations can restore broken symmetries, here also, these large
fluctuations may lead to the whole process of defect formation again. Numerical simulations employing
potentials similar to that of Eq. (\ref{VfiniT2}) have shown that strings indeed arise for values
$\eta\sim 10^{16}$ GeV [Tkachev \etal , 1998, Kasuya \& Kawasaki, 1998]. Hence, preheating after
inflation helps in generating cosmic defects.

\subsection{Cosmic strings}
\label{sec-cosmi}         

Cosmic strings are without any doubt the topological defect most thoroughly studied, both in cosmology
and solid--state physics (vortices).  The canonical example, also describing flux tubes in
superconductors, is given by the Lagrangian 
\be
\label{lagraCS}
{\cal L} = -{1\over 4} F_{\mu\nu} F^{\mu\nu} + {1\over 2} |D_\mu\phi|^2 - {\lambda \over 4!} \left(
|\phi |^2 - {\eta }^2 \right)^2 ~, \ee 
with $F_{\mu\nu} = \partial_{[\mu}A_{\nu ]}$, where $A_{\nu}$ is the gauge field and the covariant
derivative is $D_\mu = \partial_\mu + i e A_{\mu}$, with $e$ the gauge coupling constant.  This
Lagrangian is invariant under the action of the Abelian group {\bf G} = U(1), and the spontaneous
breakdown of the symmetry leads to a vacuum manifold ${\cal M}$ that is a circle, $S^1$, \ie, the
potential is minimized for $\phi = \eta\exp (i\theta)$, with arbitrary $0\leq\theta\leq 2\pi$.  Each
possible value of $\theta$ corresponds to a particular `direction' in the field space.

\begin{figure}[htbp]
\includegraphics[width=16cm]{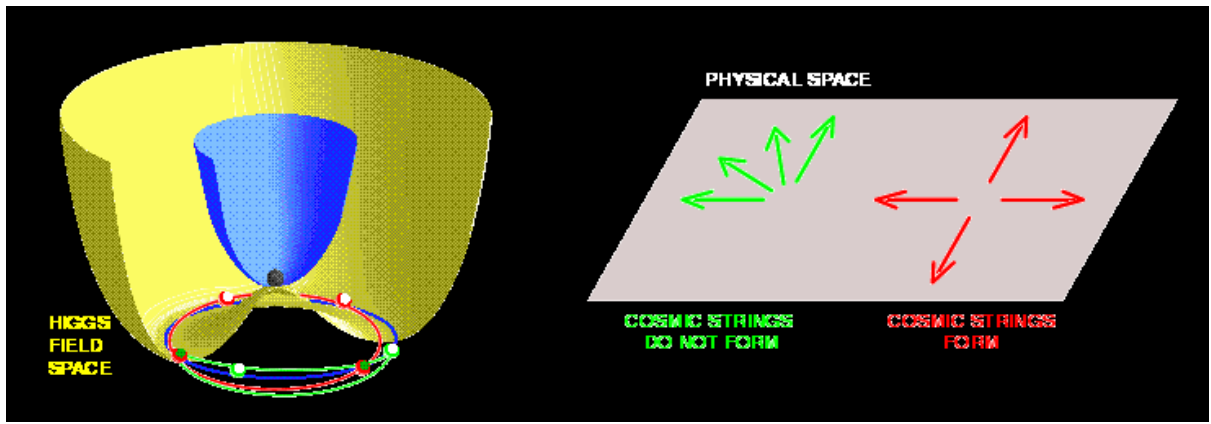}\\  
\caption{{\sl The complex scalar Higgs field evolves in a
temperature-dependent potential $V(\phi)$. At high temperatures (violet narrow 
surface) the vacuum expectation value of the field lies at the
bottom of $V$. For lower temperatures, the potential adopts the ``Mexican hat'' form (yellow surface)
and the field spontaneously chooses one amongst the new available (degenerate) lowest energy states
(the violet circle along the valley of the hat). This isolates a single value/direction for the phase
of the field, spontaneously breaking the symmetry possessed by the system at high energies. Different
regions of the universe, with no causal connection, will end up having arbitrarily different
directions for the field (arrows on the right). As separate regions of broken symmetry merge, it is
not always possible for the field orientations to match. It may happen that a closed loop in physical
space intersects regions where the Higgs phase varies from 0 to 2$\pi$ (red arrows, corresponding to
the red dashed-line on the left panel). In that situation, a cosmic string will pass somewhere inside
the loop. On the contrary, green arrows (and green dashed-line on the left panel) show a situation
where no string is formed after the phase transition.}}
\label{fig-recuadro-a}
\end{figure}               

Now, as we have seen earlier, due to the overall cooling down of the universe, there will be regions
where the scalar field rolls down to different vacuum states.  The choice of the vacuum is totally
independent for regions separated apart by one correlation length or more, thus leading to the
formation of domains of size $\xi\sim \eta^{-1}$.  When these domains coalesce they give rise to edges
in the interface.  If we now draw a imaginary circle around one of these edges and the angle $\theta$
varies by $2\pi$ then by contracting this loop we reach a point where we cannot go any further without
leaving the manifold ${\cal M}$. This is a small region where the variable $\theta$ is not defined
and, by continuity, the field should be $\phi = 0$.  In order to minimize the spatial gradient energy
these small regions line up and form a line--like defect called cosmic string.

The width of the string is roughly $m_\phi^{-1} \sim (\sqrt{\lambda} \eta)^{-1}$, $m_\phi$ being the
Higgs mass. The string mass per unit length, or tension, is $\mu \sim \eta^2$. This means that for GUT
cosmic strings, where $\eta\sim 10^{16}$ GeV, we have $G\mu \sim 10^{-6}$.  We will see below that the
dimensionless combination $G\mu$, present in all signatures due to strings, is of the right order of
magnitude for rendering these defects cosmologically interesting.

There is an important difference between global and gauge (or local) cosmic strings: local strings
have their energy confined mainly in a thin core, due to the presence of gauge fields $A_\mu$ that
cancel the gradients of the field outside of it. Also these gauge fields make it possible for the
string to have a quantized magnetic flux along the core.  On the other hand, if the string was
generated from the breakdown of a {\sl global} symmetry there are no gauge fields, just Goldstone
bosons, which, being massless, give rise to long--range forces. No gauge fields can compensate the
gradients of $\phi$ this time and therefore there is an infinite string mass per unit length.

Just to get a rough idea of the kind of models studied in the
literature, consider the case ${\bf G} = SO(10)$ that is broken to
${\bf H} = SU(5)\times {\cal Z}_2$.  For this pattern we have
$\pi_1({\cal M}) = {\cal Z}_2$, which is clearly non trivial and
therefore
cosmic strings are formed [Kibble \etal, 1982].\footnote{In the
        analysis one uses the
        fundamental theorem stating that, for a simply--connected Lie
        group {\bf G} breaking down to {\bf H}, we have
        $\pi_1({\bf G} / {\bf H}) \cong \pi_0({\bf H})$;
        see [Hilton, 1953].}                          

\subsection{String loops and scaling}
\label{sec-loopheurist}      

We saw before the reasons why gauge monopoles and domain walls were a bit of a problem for
cosmology. Essentially, the problem was that their energy density decreases more slowly than the
critical density with the expansion of the universe. This fact resulted in their contribution to
$\Omega_{\rm def}$ (the density in defects normalized by the critical density) being largely in excess
compared to 1, hence in blatant conflict with modern observations. The question now arises as to
whether the same might happened with cosmic strings. Are strings dominating the energy density of the
universe? Fortunately, the answer to this question is {\sl no}; strings evolve in such a way to make
their density $\rho_{\rm strings}\propto \eta^2 t^{-2}$. Hence, one gets the same temporal behavior as
for the critical density. The result is that $\Omega_{\rm strings} \sim G\mu \sim (\eta/m_P)^2 \sim
10^{-6}$ for GUT strings, \ie, we get an interestingly small enough, constant fraction of the critical
density of the universe and strings never upset standard observational cosmology.

\begin{figure}[htbp]
\includegraphics[width=12cm]{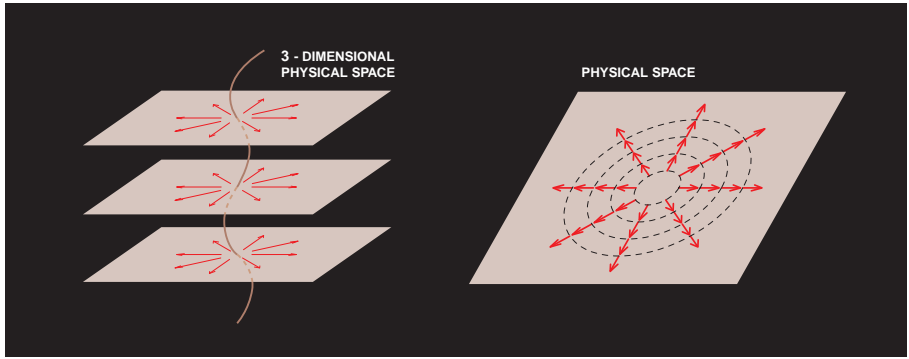}\\  
\caption{{\sl We can now extend the mechanism shown in the previous figure to the full
three-dimensional space. Regions of the various planes that were traversed by strings can be
superposed to show the actual location of the cosmic string (left panel).  The figure on the right
panel shows why we are sure a string crosses the plane inside the loop in physical space (the case
with red arrows in the previous figure). Continuity of the field imposes that if we gradually contract
this loop the direction of the field will be forced to wind ``faster''. In the limit in which the loop
reduces to a point, the phase is no longer defined and the vacuum expectation value of the Higgs field
has to vanish. This corresponds to the central tip of the Mexican hat potential in the previous figure
and is precisely the locus of the false vacuum. Cosmic strings are just that, narrow, extremely
massive line-like regions in physical space where the Higgs field adopts its high-energy false vacuum
state.}}
\label{fig-recuadro-b}
\end{figure}               

Now, why this is so? The answer is simply the efficient way in which a network of strings looses
energy. The evolution of the string network is highly nontrivial and loops are continuously chopped
off from the main infinite strings as the result of (self) intersections within the infinite--string
network. Once they are produced, loops oscillate due to their huge tension and slowly decay by
emitting gravitational radiation. Thus, energy is transferred from the cosmic string network to
radiation.\footnote{High--resolution cosmic string simulations can be found in the Cambridge cosmology
page at {\tt http://www.damtp.cam.ac.uk/user/gr/public/cs\_{}evol.html}}

It turns out from simulations that most of the energy in the string network (roughly a 80\%) is in the
form of infinite strings. Soon after formation one would expect long strings to have the form of
random-walk with characteristic step given by the correlation length $\xi$.  Also, the typical
distance between long string segments should also be of order $\xi$. Monte Carlo simulations show that
these strings are Brownian on sufficiently large scales, which means that the length $\ell$ of a
string is related to the end-to-end distance ${\rm d}$ of two given points along the string (with
${\rm d} \gg \xi$) in the form 
\be \ell = {\rm d}^2/\xi .  \ee 
What remains of the energy is given in the form of closed loops with no preferred length scale (a
scale invariant distribution) which implies that the number density of loops having sizes between $R$
and $R+ dR$ follows just from dimensional analysis \be d n_{\rm loops} \propto {dR\over R^4} \ee which
is just another way of saying that $n_{\rm loops}\propto 1/R^3$, loops behave like normal
nonrelativistic matter.  The actual coefficient, as usual, comes from string simulations.

There are both analytical and numerical indications in favor of the existence of a stable ``scaling
solution'' for the cosmic string network. After generation, the network quickly evolves in a self
similar manner with just a few infinite string segments per Hubble volume and Hubble time.  A
heuristic argument for the scaling solution due to Vilenkin [1985] is as follows.

If we take $\nu(t)$ to be the mean number of infinite string segments per Hubble volume, then the
energy density in infinite strings $\rho_{\rm strings} = \rho_{\rm s}$ is \be
\label{vilen1}
\rho_{\rm s}(t) = \nu(t) \eta^2 t^{-2} = \nu(t) \mu t^{-2} .  
\ee
Now, $\nu$ strings will typically have $\nu$ intersections, and so 
the number of loops $n_{\rm loops}(t) =  n_{\rm l}(t)$ produced per 
unit volume will be proportional to $\nu^2$. We find
\be
d n_{\rm l} \sim \nu^2 R^{-4} dR .
\ee
Hence, recalling now that the loop sizes grow with the expansion like
$R\propto t$ we have 
\be
\label{vilen2}
{d n_{\rm l}(t)\over dt} \sim p \nu^2 t^{-4} 
\ee
where $p$ is the probability of loop formation per intersection, a
quantity related to the intercommuting probability, both roughly of
order 1. 
We are now in a position to write an energy conservation equation for
strings plus loops in the expanding universe. Here it is
\be
\label{vilen3}
{d \rho_{\rm s} \over dt} + {3 \over 2 t} \, \rho_{\rm s} \sim 
 - m_{\rm l} {d n_{\rm l} \over dt } \sim 
 - \mu t     {d n_{\rm l} \over dt }
\ee
where $m_{\rm l} = \mu t$ is just the loop mass and where 
the second on the left hand side is the dilution term 
$3 H \rho_{\rm s}$ for an expanding radiation--dominated universe. 
The term on the right hand side amounts to the loss of energy from
the long string network by the generation of small closed loops. 
Plugging Eqs. (\ref{vilen1}) and (\ref{vilen2}) into (\ref{vilen3})
Vilenkin finds the following kinetic equation for $\nu(t)$
\be
{d\nu \over dt} - \, {\nu\over{2 t}} \sim - p {\nu^2 \over t}
\ee
with $p\sim 1$. Thus if $\nu \gg 1$ then ${d\nu / dt} < 0$ and $\nu$
tends to decrease in time, while if $\nu \ll 1$ then ${d\nu / dt} > 0$ and
$\nu$ increases. Hence, there will be a stable solution with 
$\nu \sim {\rm a ~few}$. 

\begin{figure}[htbp]
\includegraphics[width=16cm]{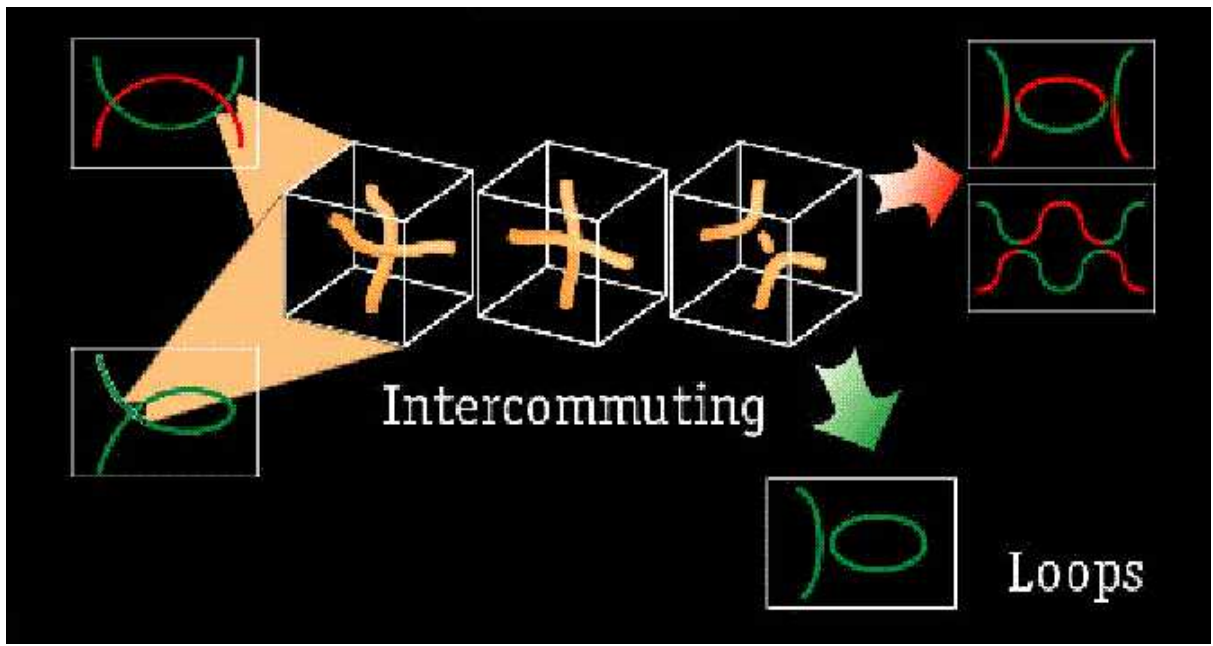}\\ 
\caption{{\sl Global string interactions leading to loop formation. Whenever two string segments
intersect, they reconnect or intercommute (green and red strings -- upper part of the
figure). Analogously, if a string intersects itself, it can break off a closed loop (green string --
bottom part of the figure). In both cases, the interacting string segments first suffer a slight
deformation (due to the long--range forces present for global strings), they subsequently fuse and
finally exchange partners. A ephemeral unstable amount of energy in the form of a small loop remains
in the middle where the energy is high enough to place the Higgs field in the false vacuum. It then
quickly collapses, radiating away its energy. The situation is roughly the same for local strings, as
simulations have shown.}}
\label{fig-loopsform}
\end{figure}               

\subsection{Global textures}
\label{sec-texuuu}      

Whenever a global non--Abelian symmetry is spontaneously and completely broken (\eg\ at a grand
unification scale), global defects called textures are generated.  Theories where this global symmetry
is only partially broken do not lead to global textures, but instead to global monopoles and
non--topological textures.  As we already mentioned global monopoles do not suffer the same
constraints as their gauge counterparts: essentially, having no associated gauge fields, the
long--range forces between pairs of monopoles lead to the annihilation of their eventual excess and as
a result monopoles scale with the expansion.  On the other hand, non--topological textures are a
generalization that allows the broken subgroup {\bf H} to contain non--Abelian factors. It is then
possible to have $\pi_3$ trivial as in, \eg, SO(5)$\to$SO(4) broken by a vector, for which case we
have ${\cal M} = S^4$, the four--sphere [Turok, 1989].  Having explained this, let us concentrate in
global topological textures from now on.

Textures, unlike monopoles or cosmic strings, are not well localized in space. This is due to the fact
that the field remains in the vacuum everywhere, in contrast to what happens for other defects, where
the field leaves the vacuum manifold precisely where the defect core is.  Since textures do not
possess a core, all the energy of the field configuration is in the form of field gradients.  This
fact is what makes them interesting objects {\sl only} when coming from global theories: the presence
of gauge fields $A_\mu$ could (by a suitable reorientation) compensate the gradients of $\phi$ and
yield $D_\mu\phi = 0$, hence canceling out (gauging away) the energy of the
configuration\footnote{This does not imply, however, that the classical dynamics of a gauge texture is
trivial. The evolution of the $\phi$--$A_\mu$ system will be determined by the competing tendencies of
the global field to unwind and of the gauge field to compensate the $\phi$ gradients. The result
depends on the characteristic size $L$ of the texture: in the range $m_\phi^{-1} << L << m_A^{-1} \sim
(e\eta)^{-1}$ the behavior of the gauge texture resembles that of the global texture, as it should,
since in the limit $m_A$ very small ($e\to 0$) the gauge texture turns into a global one [Turok \&
Zadrozny, 1990].}.

One feature endowed by textures that really makes these defects peculiar is their being unstable to
collapse.  The initial field configuration is set at the phase transition, when $\phi$ develops a
nonzero vacuum expectation value.  $\phi$ lives in the vacuum manifold ${\cal M}$ and winds around
${\cal M}$ in a non--trivial way on scales greater than the correlation length, $\xi \lsim t$.  The
evolution is determined by the nonlinear dynamics of $\phi$.  When the typical size of the defect
becomes of the order of the horizon, it collapses on itself.  The collapse continues until eventually
the size of the defect becomes of the order of $\eta^{-1}$, and at that point the energy in gradients
is large enough to raise the field from its vacuum state.  This makes the defect unwind, leaving
behind a trivial field configuration.  As a result $\xi$ grows to about the horizon scale, and then
keeps growing with it.  As still larger scales come across the horizon, knots are constantly formed,
since the field $\phi$ points in different directions on ${\cal M}$ in different Hubble volumes.  This
is the scaling regime for textures, and when it holds simulations show that one should expect to find
of order 0.04 unwinding collapses per horizon volume per Hubble time [Turok, 1989].  However,
unwinding events are not the most frequent feature [Borrill \etal, 1994], and when one considers
random field configurations without an unwinding event the number raises to about 1 collapse per
horizon volume per Hubble time.

\subsection{Evolution of global textures}
\label{sec-texevol} 

We mentioned earlier that the breakdown of any non--Abelian global symmetry led to the formation of
textures. The simplest possible example involves the breakdown of a global SU(2) by a complex doublet
$\phi^a$, where the latter may be expressed as a four--component scalar field, \ie, $a=1\ldots 4$.  We
may write the Lagrangian of the theory much in the same way as it was done in Eq. (\ref{lagraCS}), but
now we drop the gauge fields (thus the covariant derivatives become partial derivatives).  Let us take
the symmetry breaking potential as follows, $V( \phi ) = {\lambda \over 4} \left( |\phi|^2 - {\eta }^2
\right)^2$. The situation in which a global SU(2) in broken by a complex doublet with this potential
$V$ is equivalent to the theory where SO(4) is broken by a four--component vector to SO(3), by making
$\phi^a$ take on a vacuum expectation value.  We then have the vacuum manifold ${\cal M}$ given by
SO(4)/SO(3) = $S^3$, namely, a three--sphere with $\phi^a\phi_a = \eta^2$.  As $\pi_3 (S^3) \not= {\bf
1}$ (in fact, $\pi_3 (S^3) = {\cal Z}$) we see we will have non--trivial solutions of the field
$\phi^a$ and global textures will arise.

As usual, variation of the action with respect to the
field  $\phi^a$ yields the equation of motion
\be
\label{phieqn}
{\phi^b}'' + 2 {a' \over a} {\phi^b}' - \nabla^2 \phi^b = - a^2
{\partial V \over\partial\phi^b } ~, \ee where primes denote
derivatives with respect to conformal time and $\nabla$ is computed in
comoving coordinates.  When the symmetry in broken three of the
initially four degrees of freedom go into massless Goldstone bosons
associated with the three directions tangential to the vacuum
three--sphere. The `radial' massive mode that remains ($m_\phi \sim
\sqrt{\lambda}\eta$) will not be excited, provided we
concentrate on length scales much larger than $m_\phi^{-1}$.

To solve for the dynamics of the field $\phi^b$, two different
approaches have been implemented in the literature.  The first one
faces directly the full equation (\ref{phieqn}), trying to solve it
numerically.  The alternative to this exploits the fact that, at
temperatures smaller than $T_c$, the field is constrained to live in
the true vacuum.  By implementing this fact via a Lagrange
multiplier\footnote{In fact, in the action the coupling constant
$\lambda$ of the `Mexican hat' potential is interpreted as the
Lagrange multiplier.}  we get
\be
\label{phieqnsigma}
\nabla^\mu\nabla_\mu\phi^b =
- {\nabla^\mu\phi^c\nabla_\mu\phi_c \over \eta^2} \phi^b ~~ ; ~~
\phi^2 = \eta^2  ~,
\ee                               
with $\nabla^\mu$ the covariant derivative operator.
Eq. (\ref{phieqnsigma}) represents a non--linear sigma model
for the  interaction of the three massless modes
[Rajaraman, 1982].
This last approach  is only valid when probing length scales
larger than the inverse of the mass $m_\phi^{-1}$.
As we mentioned before, when this condition is not met the
gradients of the field are strong enough to make it leave
the vacuum manifold and unwind.                   

The approach (cf. Eqs. (\ref{phieqnsigma})) is suitable for analytic inspection. In fact, an exact
flat space solution was found assuming a spherically symmetric ansatz. This solution represents the
collapse and subsequent conversion of a texture knot into massless Goldstone bosons, and is known as
the spherically symmetric self--similar (SSSS) exact unwinding solution.  We will say no more here
with regard to the this solution, but just refer the interested reader to the original articles [see,
\eg, Turok \& Spergel, 1990; Notzold, 1991].  Simulations taking full account of the energy stored in
gradients of the field, and not just in the unwinding events, like in Eq. (\ref{phieqn}), were
performed, for example, in [Durrer \& Zhou, 1995]. \footnote{Simulations of the collapse of `exotic'
textures can be found at {\tt http://camelot.mssm.edu/\~{}ats/texture.html}}

\section{Currents along strings}
\label{sec-currstrings}      

In the past few years it has become clear that topological defects, and in particular strings, will be
endowed with a considerably richer structure than previously envisaged.  In generic grand unified
models the Higgs field, responsible for the existence of cosmic strings, will have interactions with
other fundamental fields. This should not surprise us, for well understood low energy particle
theories include field interactions in order to account for the well measured masses of light
fermions, like the familiar electron, and for the masses of gauge bosons $W$ and $Z$ discovered at
CERN in the eighties.  Thus, when one of these fundamental (electromagnetically charged) fields
present in the model condenses in the interior space of the string, there will appear electric
currents flowing along the string core.

Even though these strings are the most attractive ones, the fact of them having electromagnetic
properties is not actually fundamental for understanding the dynamics of circular string loops. In
fact, while in the uncharged and non current-carrying case symmetry arguments do not allow us to
distinguish the existence of rigid rotations around the loop axis, the very existence of a small
current breaks this symmetry, marking a definite direction, which allows the whole loop configuration
to rotate.  This can also be viewed as the existence of spinning particle--like solutions trapped
inside the core.  The stationary loop solutions where the string tension gets balanced by the angular
momentum of the charges is what Davis and Shellard [1988] dubbed {\it vortons}.

Vorton configurations do not radiate classically.  Because they have loop shapes, implying periodic
boundary conditions on the charged fields, it is not surprising that these configurations are
quantized. At large distances these vortons look like point masses with quantized electric charge
(actually they can have more than a hundred times the electron charge) and angular momentum.  They are
very much like particles, hence their name.  They are however very peculiar, for their characteristic
size is of order of their charge number (around a hundred) times their thickness, which is essentially
some fourteen orders of magnitude smaller than the classical electron radius.  Also, their mass is
often of the order of the energies of grand unification, and hence vortons would be some twenty orders
of magnitude heavier than the electron.

But why should strings become conducting in the first place?  The physics inside the core of the
string differs somewhat from outside of it.  In particular the existence of interactions among the
Higgs field forming the string and other fundamental fields, like that of charged fermions, would make
the latter loose their masses inside the core. Then, only small energies would be required to produce
pairs of trapped fermions and, being effectively massless inside the string core, they would propagate
at the speed of light.  These zero energy fermionic states, also called zero modes, endow the string
with currents and in the case of closed loops they provide the mechanical angular momentum support
necessary for stabilizing the contracting loop against collapse.

\subsection{Goto--Nambu Strings}
\label{subsec-gnstrings}   

Our aim now would be to introduce extra fields into the problem and see what new features arise. We
would expect to find --among other novelties-- currents flowing along the defect cores, as
advertised before. However, doing this in detail would unfortunately take us too much away from the
main topic of these notes, and we just refer the reader to some recent work [Lemperi\`ere \&
Shellard 2002] (see also ~\cite{CPG,AgPpCb}) and to the recent review in \cite{boli01}, as well as
to the other references given in the introduction, for a detailed treatment. Here below, we just
give some few additional features of cosmic string field theory.

The simple Lagrangian we saw in previous sections was a good approximation for ideal structureless
strings, known under the name of Goto--Nambu strings [Goto, 1971; Nambu, 1970].  Additional fields
coupled with the string--forming Higgs field often lead to interesting effects in the form of
generalized currents flowing along the string core. But before taking into full consideration the
internal structure of strings (given in \cite{boli01}) it is appropriate to start by setting the
scene with the simple Abelian Higgs model (which describes scalar electrodynamics). This is a
prototype of gauge field theory with spontaneous symmetry breaking G = U(1) $\to$ \{1\}.  The
Lagrangian reads [Higgs, 1964] \be
\label{higgs964}   
{\cal L}_{_{\rm H}}
           = -{1\over 2}[{D}^\mu \Phi][{D}_\mu\Phi]^*
           - {1\over 4} (F^{(\phi)}_{\mu \nu})^2
           - {\lambda_{\phi}\over 8}(\vert\Phi\vert^2 - \eta^2)^2 , 
\ee
with gauge covariant derivative 
$D_\mu = \partial_\mu + i q A^{(\phi)}_\mu$,
antisymmetric tensor 
$F^{(\phi)}_{\mu \nu}= \nabla_{\mu}A^{(\phi)}_{\nu}-
\nabla_{\nu}A^{(\phi)}_{\mu}$ for the gauge vector field
$A^{(\phi)}_{\nu}$, and complex scalar field 
$\Phi=\vert\Phi\vert e^{i\alpha}$ with gauge coupling $q$. 

The first solutions for this theory were found by 
Nielsen \& Olesen [1973]. A couple of relevant properties are
noteworthy: 
\begin{itemize}
\item{}  
the mass per unit length for the string is $\mu = U \sim \eta^2$. 
For GUT local strings this gives $\mu \sim 10^{22}{\rm g}/{\rm cm}$, 
while one finds  
$\mu \sim \eta^2\ln(r/m_{\rm s}^{-1}) \to \infty$ if strings are
global, due to the absence of compensating gauge fields. This
divergence is in general not an issue, because global strings only
in few instances are isolated; in a string network, a natural cutoff is the
distance to the neighboring string.

\item{} 
There are essentially two characteristic mass scales (or inverse
length scales) in the problem: $m_{\rm s}\sim
\lambda_{\phi}^{1/2}\eta$ and $m_{\rm v}\sim q\eta$, corresponding to
the inverse of the Compton wavelengths of the scalar (Higgs) and vector
($A^{(\phi)}_{\nu}$) particles, respectively.

\item{}    
There exists a sort of screening of the energy, called `Higgs
screening', implying a finite energy configuration, thanks to the way in
which the vector field behaves far from the string core: 
$A_\theta\to (1 / qr) d\alpha / d\theta \,\, ,  {\rm ~for~} r\to\infty$.

After a closed path around the vortex one has $\Phi(2\pi)=\Phi(0)$,
which implies that the winding phase $\alpha$ should be an integer
times the cylindrical angle $\theta$, namely $\alpha=n\theta$. 
This integer $n$ is dubbed the `winding number'. In turn, from this fact
it follows that there exists a tube of quantized  `magnetic' flux,
given by   
\be
\Phi_{_{\rm B}}= \oint \vec A . \vec {d\ell}
= {1\over q}\int_0^{2\pi}{d\alpha\over d\theta} d\theta
= {2\pi n\over q}
\ee

\end{itemize}

In the string there is a sort of competing effect between the fields: 
the gauge field acts in a repulsive manner; the flux doesn't like to
be confined to the core and $B$ lines repel each other. On the other
hand, the scalar field behaves in an attractive way; it tries to
minimize the area where $V(\Phi)\not=0$, that is, where the field
departs from the true vacuum. 

\begin{figure}[htbp]
\includegraphics[width=5cm]{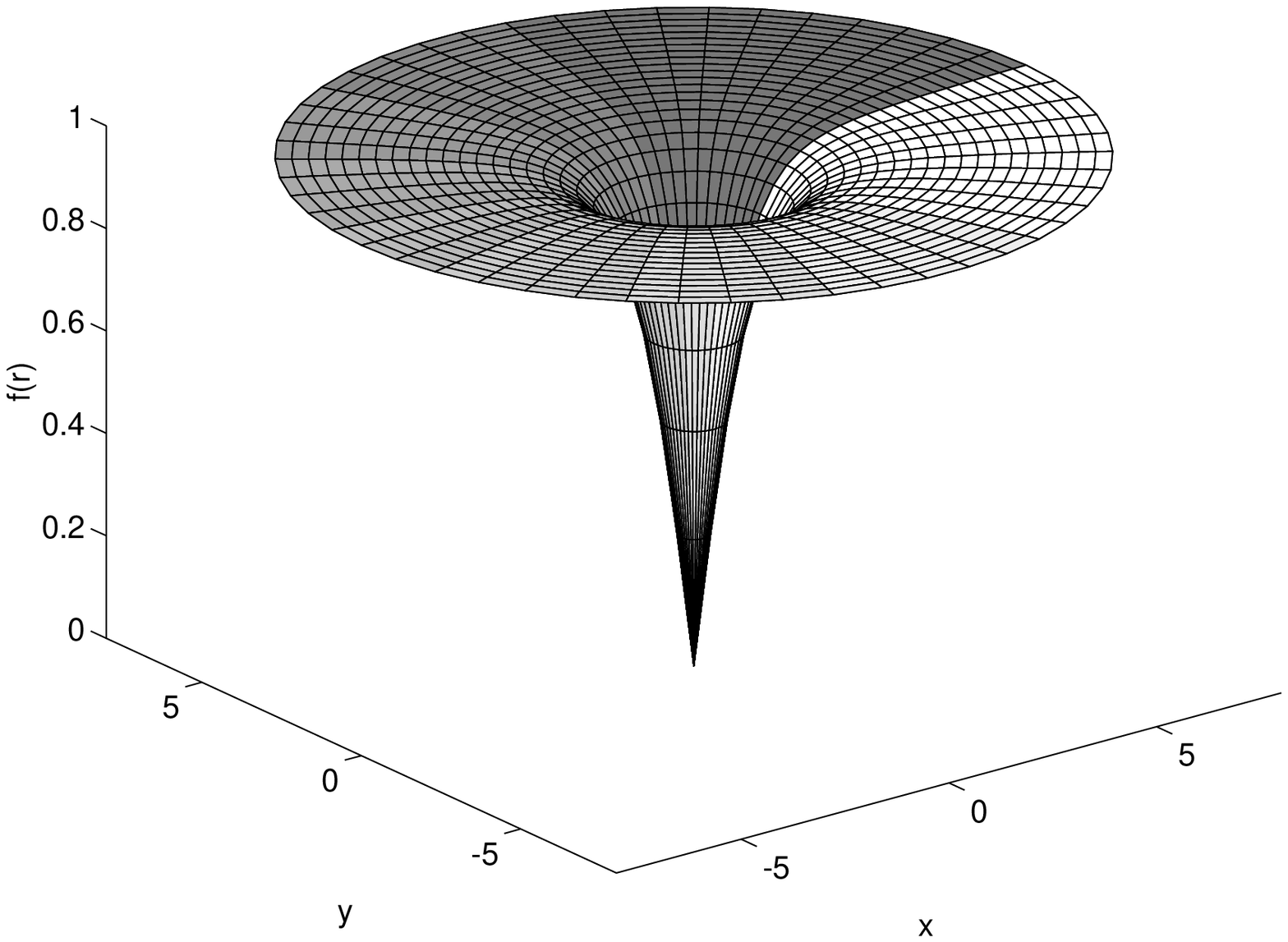}
%
\includegraphics[width=5cm]{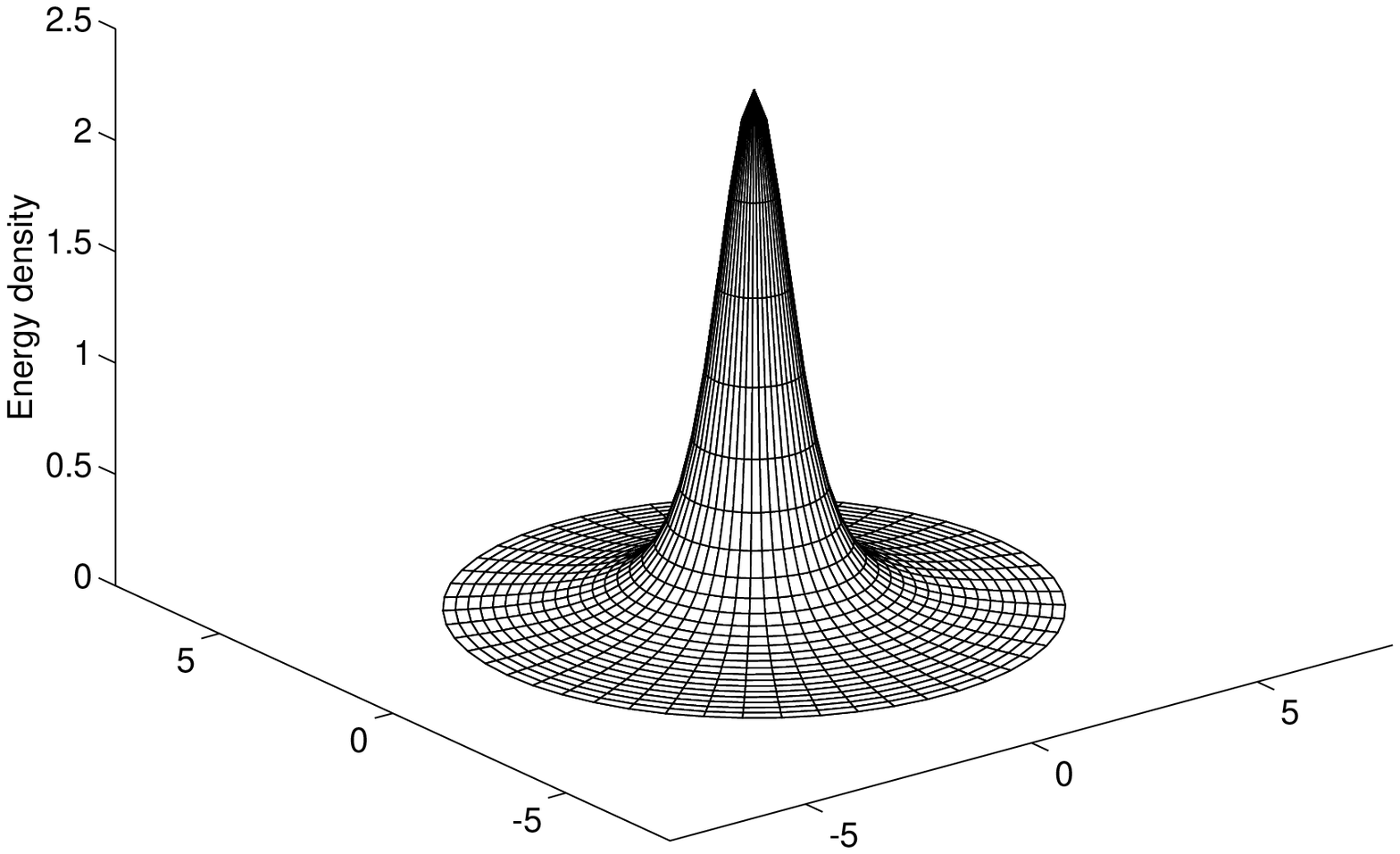}\\ 
\caption{{\sl 
Higgs field and energy profiles for Goto--Nambu cosmic strings.
The left panel shows the amplitude of the Higgs field around the
string. The field vanishes at the origin (the false vacuum) and
attains its asymptotic value (normalized to unity in the figure) far
away from the origin. The phase of the scalar field (changing from 0
to $2\pi$) is shown by the shading of the surface. In the right panel
we show the energy density of the configuration. The maximum value is
reached at the origin, exactly where the Higgs is placed in the false
vacuum. [Hindmarsh \& Kibble, 1995].}}
\label{fig-HinKib}\end{figure}   

Finally, we can mention a few condensed--matter `cousins' of Goto--Nambu strings: flux tubes in
superconductors [Abrikosov, 1957] for the nonrelativistic version of gauge strings ($\Phi$ corresponds
to the Cooper pair wave function). Also, vortices in superfluids, for the nonrelativistic version of
global strings ($\Phi$ corresponds to the Bose condensate wave function).  Moreover, the only two
relevant scales of the problem we mentioned above are the Higgs mass $m_{\rm s}$ and the gauge vector
mass $m_{\rm v}$. Their inverse give an idea of the characteristic scales on which the fields acquire
their asymptotic solutions far away from the string `location'. In fact, the relevant core widths of
the string are given by $m_{\rm s}^{-1}$ and $m_{\rm v}^{-1}$. It is the comparison of these scales
that draws the dividing line between two qualitatively different types of solutions. If we define the
parameter $\beta = ({m_{\rm s} / m_{\rm v}})^2$, superconductivity theory says that $\beta <1$
corresponds to Type I behavior while $\beta >1$ corresponds to Type II. For us, $\beta <1$ implies
that the characteristic scale for the vector field is smaller than that for the Higgs field and so
magnetic field $B$ flux lines are well confined in the core; eventually, an $n$--vortex string with
high winding number $n$ stays stable. On the contrary, $\beta >1$ says that the characteristic scale
for the vector field exceeds that for the scalar field and thus $B$ flux lines are not confined; the
$n$--vortex string will eventually split into $n$ vortices of flux $2\pi/q$. In summary: \be \beta =
({m_{\rm s}\over m_{\rm v}})^2 \cases {<1 {\it ~n\!\!-\!\!vortex ~stable~ (B ~flux ~lines ~confined
~in ~core) {\rm ~~- Type ~I}} \cr >1 {\it ~Unstable: ~splitting ~into ~n ~vortices ~of ~flux~} 2\pi/q
{\rm ~- Type ~II}} \ee
                          
\section{Structure formation from defects}
\label{sec-lls}      

\subsection{Cosmic strings}
\label{sec-llsstrings}       

In this section we will provide just a quick description of the remarkable cosmological features of
cosmic strings.  Many of the proposed observational tests for the existence of cosmic strings are
based on their gravitational interactions.  In fact, the gravitational field around a straight static
string is very unusual [Vilenkin, 1981].  As is well known, the Newtonian limit of Einstein field
equations with source term given by $T^\mu_\nu = {\rm diag}(\rho, -p_1, -p_2, -p_3)$ in terms of the
Newtonian potential $\Phi$ is given by $\nabla^2\Phi = 4\pi G (\rho + p_1 + p_2 + p_3)$, just a
statement of the well known fact that pressure terms also contribute to the `gravitational mass'. For
an infinite string in the $z$--direction one has $p_3 = -\rho$, \ie, strings possess a large
relativistic tension (negative pressure). Moreover, averaging on the string core results in vanishing
pressures for the $x$ and $y$ directions yielding $\nabla^2\Phi = 0$ for the Poisson equation. This
indicates that space is flat outside of an infinite straight cosmic string and therefore test
particles in its vicinity should not feel any gravitational attraction.

In fact, a full general relativistic analysis confirms this and test particles in the space around the
string feel no Newtonian attraction; however there exists something unusual, a sort of wedge missing
from the space surrounding the string and called the `deficit angle', usually noted $\Delta$, that
makes the topology of space around the string that of a cone.  To see this, consider the metric of a
source with energy--momentum tensor [Vilenkin 1981, Gott 1985] \be T_\mu ^\nu = \delta (x) \delta (y)
{\rm diag}(\mu ,0,0,T) \ . \ee In the case with $T= \mu$ (a rather simple equation of state) this is
the effective energy--momentum tensor of an unperturbed string with string tension $\mu$ as seen from
distances much larger than the thickness of the string (a Goto--Nambu string).  However, real strings
develop small--scale structure and are therefore not well described by the Goto--Nambu action. When
perturbations are taken into account $T$ and $\mu$ are no longer equal and can only be interpreted as
effective quantities for an observer who cannot resolve the perturbations along its length. And in
this case we are left without an effective equation of state. Carter [1990] has proposed that these
`noisy' strings should be such that both its speeds of propagation of perturbations coincide. Namely,
the transverse (wiggle) speed $c_{\rm T}=(T/\mu)^{1/2}$ for extrinsic perturbations should be equal to
the longitudinal (woggle) speed $c_{\rm L}=(-dT/d\mu)^{1/2}$ for sound--type perturbations.  This
requirement yields the new equation of state \be \mu T = \mu_0^2 \ee and, when this is satisfied, it
describes the energy-momentum tensor of a wiggly string as seen by an observer who cannot resolve the
wiggles or other irregularities along the string [Carter 1990, Vilenkin 1990].

\begin{figure}[htbp]
\includegraphics[width=5cm]{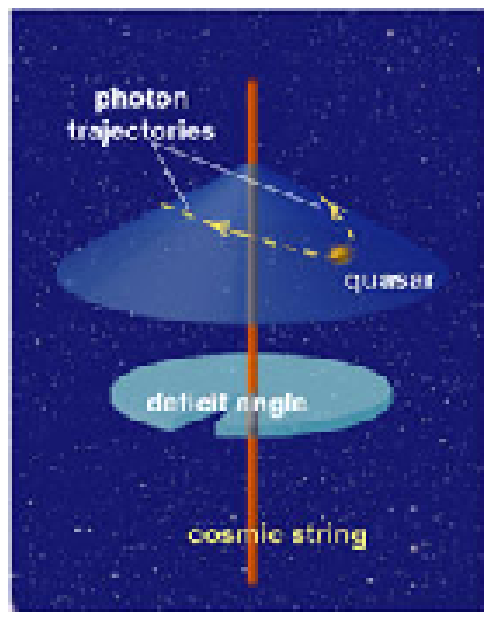}
%
\includegraphics[width=4cm]{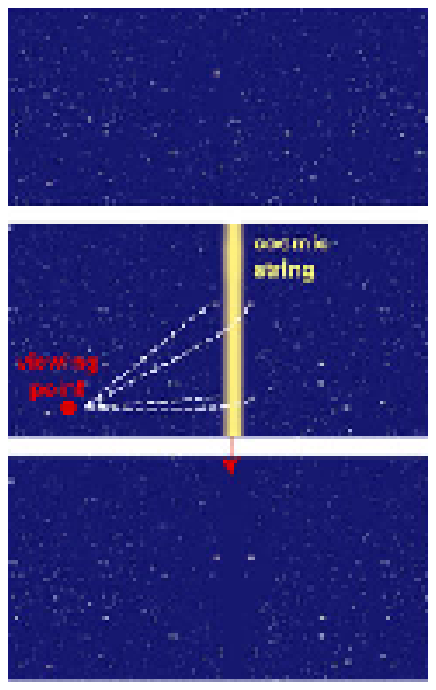}\\    
\caption{{\sl Cosmic strings affect surrounding spacetime by removing a small angular wedge, creating
a conelike geometry (left). Space remains flat everywhere, but a circular path around the string
encompasses slightly less than 360 degrees. The deficit angle is tiny, about $10^{-5}$ radian. To an
observer, the presence of a cosmic string would be betrayed by its effect on the trajectory of passing
light rays, which are deflected by an amount equal to the deficit angle. The resultant gravitational
lensing reveals itself in the doubling of images of objects behind the string (right panel).}}
\label{fig-fig5ab}\end{figure}   

The gravitational field around the cosmic string [neglecting terms of
order $(G\mu)^2$] is found by solving the linearized Einstein
equations with the above $T_\mu ^\nu$. One gets
\be
h_{00} = h_{33} = 4G(\mu -T) \ln(r/r_0 ) ,
\label{h00}
\ee
\be
h_{11} = h_{22} = 4G(\mu +T) \ln(r/r_0 ) ,
\ee
where $h_{\mu \nu} = g_{\mu \nu} - \eta _{\mu \nu}$ is the metric
perturbation, the radial distance from the string is 
$r = (x^2 + y^2 )^{1/2}$, and $r_0$ is a constant of
integration.                                                    

For an ideal, straight, unperturbed string, the tension and mass per
unit length are $T = \mu = \mu_0$ and one gets
\be
h_{00} = h_{33} = 0, \ \ \
h_{11} = h_{22} = 8G\mu_0 \ln(r/r_0 ) .
\ee
By a coordinate transformation one can bring this metric to a locally flat form
\be
ds^2 = dt^2 - dz^2 - dr^2 - (1-8 G\mu_0 ) r^2 d\phi ^2 ,
\ee
which describes a conical and flat (Euclidean) space with a wedge of
angular size $\Delta = 8\pi G \mu_0$ (the deficit angle) removed from
the plane and with the two faces of the wedge identified.

\subsubsection{Wakes and gravitational lensing}

We saw above that test particles\footnote{If one takes into account the own gravitational field of the
particle living in the spacetime around a cosmic string, then the situation changes. In fact, the
presence of the conical `singularity' introduced by the string distorts the particle's own
gravitational field and results in the existence of a weak attractive force proportional to $G^2\mu
m^2/r^2$, where $m$ is the particle's mass [Linet, 1986].} at rest in the spacetime of the straight
string experience no gravitational force, but if the string moves the situation radically changes. Two
particles initially at rest while the string is far away, will suddenly begin moving towards each
other after the string has passed between them. Their head--on velocities will be proportional to
$\Delta$ or, more precisely, the particles will get a boost $v = 4\pi G\mu_0 v_s \gamma$ in the
direction of the surface swept out by the string.  Here, $\gamma = (1-v_s^2)^{-1/2}$ is the Lorentz
factor and $v_s$ the velocity of the moving string.  Hence, the moving string will built up a {\sl
wake} of particles behind it that may eventually form the `seed' for accreting more matter into
sheet--like structures [Silk \& Vilenkin 1984].

\begin{figure}[htbp]
\includegraphics[width=5cm]{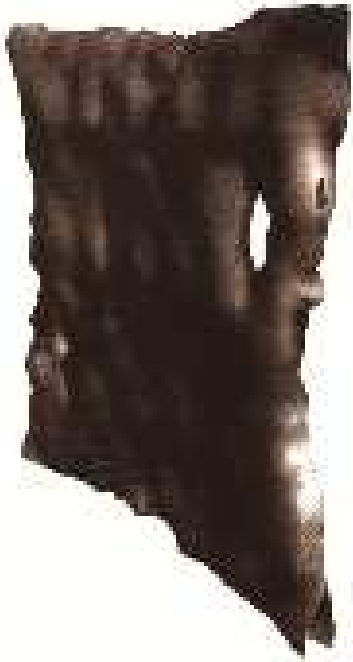}\\    
\caption{{\sl By deflecting the trajectory of ordinary matter, strings offer an interesting means of
forming large-scale structure. A string sweeping through a distribution of interstellar dust will draw
particles together in its wake, giving them lateral velocities of a few kilometers per second. The
trail of the moving string will become a planar region of high-density matter, which, after
gravitational collapse, could turn into thin, sheetlike distributions of galaxies [Image courtesy of
Pedro Avelino and Paul Shellard].}}
\label{fig-wake}\end{figure} 

Also, the peculiar topology around the string makes it act as a cylindric gravitational lens that may
produce double images of distant light sources, \eg, quasars.  The angle between the two images
produced by a typical GUT string would be $\propto G\mu$ and of order of a few arcseconds, independent
of the impact parameter and with no relative magnification between the images [see Cowie \& Hu, 1987,
for a first observational attempt]. Surprisingly enough, a recent detection of an extragalactic double
source with the appropriate characteristics (few arcseconds angle, lack of excess light in between the
images, etc), lead Sazhin \etal ~[2003] to propose the serendipitous discovery of a cosmic string lens
event. Their data suggest that both images belong to early-type giant elliptical galaxies with
redshift 0.46 (or some 1900 Mpc away, for a reduced Hubble constant $h=0.65$) meaning that these
galaxies (if indeed there is such chance projection) are some 20 Kpc away from each other. On the
contrary, if this pair is caused by the splitting due to an intervening cosmic string, the
energy-scale of the symmetry-breaking transition giving bith to such string can be computed, and it
turns out to be of a typical GUT scale. Doubtless, more independent observations are needed to confirm
this interesting case.

Turning back now to the peculiar matter structures generated by moving strings, the situation
described above gets even more interesting when we allow the string to have small--scale structure,
which we called wiggles, as in fact simulations indicate. Wiggles not only modify the string's
effective mass per unit length, $\mu$, but also built up a Newtonian attractive term in the velocity
boost inflicted on nearby test particles. To see this, let us consider the formation of a wake behind
a moving wiggly string. Assuming the string moves along the $x$--axis, we can describe the situation
in the rest frame of the string. In this frame, it is the particles that move, and these flow past the
string with a velocity $v_s$ in the opposite direction. Using conformally Minkowskian coordinates we
can express the relevant components of the metric as \be ds^2 = (1+h_{00})[dt^2 - (dx^2 + dy^2)] , \ee
where the missing wedge is reproduced by identifying the half-lines $y=\pm 4 \pi G \mu x$, $x \ge 0$.
The linearized geodesic equations in this metric can be written as \be 2 \ddot x = - ( 1- { \dot x }^2
- {\dot y} ^2 ) \partial _x h_{00} , \ee \be 2 \ddot y = - ( 1- { \dot x }^2 - {\dot y} ^2 ) \partial
_y h_{00} , \ee where over--dots denote derivatives with respect to $t$.  Working to first order in
$G\mu$, the second of these equations can be integrated over the unperturbed trajectory $x = v_s t$,
$y = y_0$.  Transforming back to the frame in which the string has a velocity $v_s$ yields the result
for the velocity impulse in the $y$--direction after the string has passed [Vachaspati \& Vilenkin,
1991; Vollick, 1992] \be v = - {{2\pi G (\mu -T)} \over {v_s \gamma}} - 4 \pi G \mu v_s \gamma \ee

The second term is the velocity impulse due to the conical deficit angle we saw above. This term will
dominate for large string velocities, case in which big planar wakes are predicted.  In this case, the
string wiggles will produce inhomogeneities in the wake and may easy the fragmentation of the
structure. The `top--down' scenario of structure formation thus follows naturally in a universe with
fast-moving strings.  On the contrary, for small velocities, it is the first term that dominates over
the deflection of particles. The origin of this term can be easily understood [Vilenkin \& Shellard,
2000]. From Eqn. (\ref{h00}), the gravitational force on a non--relativistic particle of mass $m$ is
$F \sim m G(\mu - T) /r$. A particle with an impact parameter $r$ is exposed to this force for a time
$\Delta t \sim r/v_s$ and the resulting velocity is $v \sim (F/m) \Delta t \sim G(\mu - T) / v_s$.

\subsection{Textures}
\label{sec-llstextures}       

During the radiation era, and when the correlation length is already growing with the Hubble radius,
the texture field has energy density $\rho_{texture}\sim (\nabla\phi)^2 \sim \eta^2 / H^{-2}$, and
remains a fixed fraction of the total density $\rho_{c} \sim t^{-2}$ yielding $\Omega_{texture} \sim G
\eta^2$. This is the scaling behavior for textures and thus we do not need to worry about textures
dominating the universe.

But as we already mentioned, textures are unstable to collapse, and this collapse generates
perturbations in the metric of spacetime that eventually lead to large scale structure
formation. These perturbations in turn will affect the photon geodesics leading to CMB anisotropies,
the clearest possible signature to probe the existence of these exotic objects being the appearance of
hot and cold {\sl spots} in the microwave maps.  Due to their scaling behavior, the density
fluctuations induced by textures on any scale at horizon crossing are given by $(\delta\rho / \rho )_H
\sim G \eta^2$.  CMB temperature anisotropies will be of the same amplitude.  Numerically--simulated
maps, with patterns smoothed over $10^\circ$ angular scales, by Bennett \& Rhie [1993] yield, upon
normalization to the {\sl COBE}--DMR data, a dimensionless value $G \eta^2 \sim 10^{-6}$, in good
agreement with a GUT phase transition energy scale. It is fair to say, however, that the texture
scenario is having problems in matching current data on smaller scales [see, \eg, Durrer, 2000].

\subsection{Defects as dark energy}
\label{sec-lldenergy}       

There is recent mounting evidence that our current universe is being dominated by a unexpectedly large
amount of dark energy [\eg, Riess \etal, 1998; Perlmutter \etal, 1999]. Recent observations with type
Ia supernovae, together with other astrophysical tests, suggest that more than 65 percent of the
critical energy density is made up by some yet unknown energy component. 

Cosmic defects can also be seen as a novel form of dark energy. For example, a tangled web of cosmic
strings with fixed mass per unit length, which self--intersects without having reconnection. Non
intercommuting strings means no production of loops, and therefore the main channel for loosing energy
is not active. The model proposed in [Vilenkin, 1984] has the mean mass density in strings scaling as
$\rho_{\rm strings}\propto (ta(t))^{-1}$ instead of $\rho_{\rm strings}\propto \eta^2 t^{-2}$ as
we saw above.  
From this, one has $\rho_{\rm strings} t^2 \propto t^{1/2}$ in the radiation--dominated era and 
$\rho_{\rm strings} t^2 \propto t^{1/3}$ during matter domination, which means that the energy in
strings $\Omega_{\rm strings}$ grows with time and, after a certain $t_s$, strings would 
dominate the universe.
With $t_s$ falling in the matter--domination era, we have 
$\rho_{\rm strings}/\rho (t_s)\sim (t_{\rm eq}/t_*)^{1/2} (t_s/t_{\rm eq})^{1/3} G\mu\sim 1$, with the
background $\rho\propto 1/Gt^2$. In the case $0 < z \lsim 3$, roughly $t_s\sim 10^{17}$ sec., with 
$t_{\rm eq}\sim 10^{11}$ sec. and $t_*\sim m_P/\eta^2$, we get $G\mu\sim 10^{-20}$ for the
characteristic energy scale of these non--intercommuting strings. 

After $t_s$ the Friedmann's equation can be cast as $(\dot a / a)^2\propto G\rho_{\rm strings}\propto
G/t a(t)$, which implies that the scale factor goes as $a(t)\propto t$ and then $\rho_{\rm
strings}\propto 1/t^2$. 
Now, recalling the local energy conservation law 
$\dot\rho = -3 (\rho+p)\dot a / a$, and applying it for a dark ``$x$'' component, $w_x=p_x/\rho_x$, we
get $\rho_x\propto a^{-3(1+w_x)}$. If this dark component is made up by strings, one then deduces that
it should be $w_x=-1/3$. Of course, this gives $\ddot a = 0$ for the scale factor, so it cannot
explain the recent acceleration phase. It nevertheless goes in the right direction.  

Similar arguments have been studied for other defects, like textures [Davis, 1987] and can also be
devised for domain walls [Zel'dovich \etal, 1974; Battye \etal, 1999], in this latter case yielding
$w_x=-2/3$ which points closer to the observational ``equation of state'' currently selected by the
analysis of the different astrophysical surveys. For these and other reasons, with the words of the
recent authoritative review by Peebles \& Ratra [2002], the class of cosmic defect models is worth
bearing in mind.


\section{CMB signatures from defects}
\label{sec-cmbdefects}      

\begin{figure}[htbp]
\includegraphics[width=5cm]{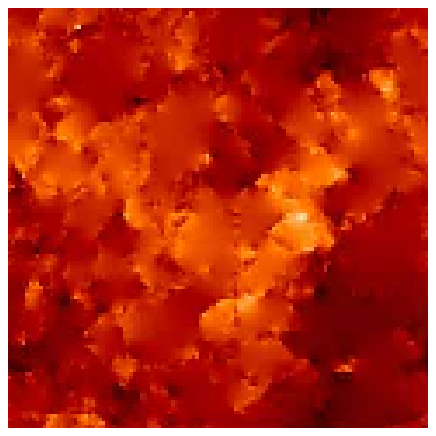}
%
\includegraphics[width=5cm]{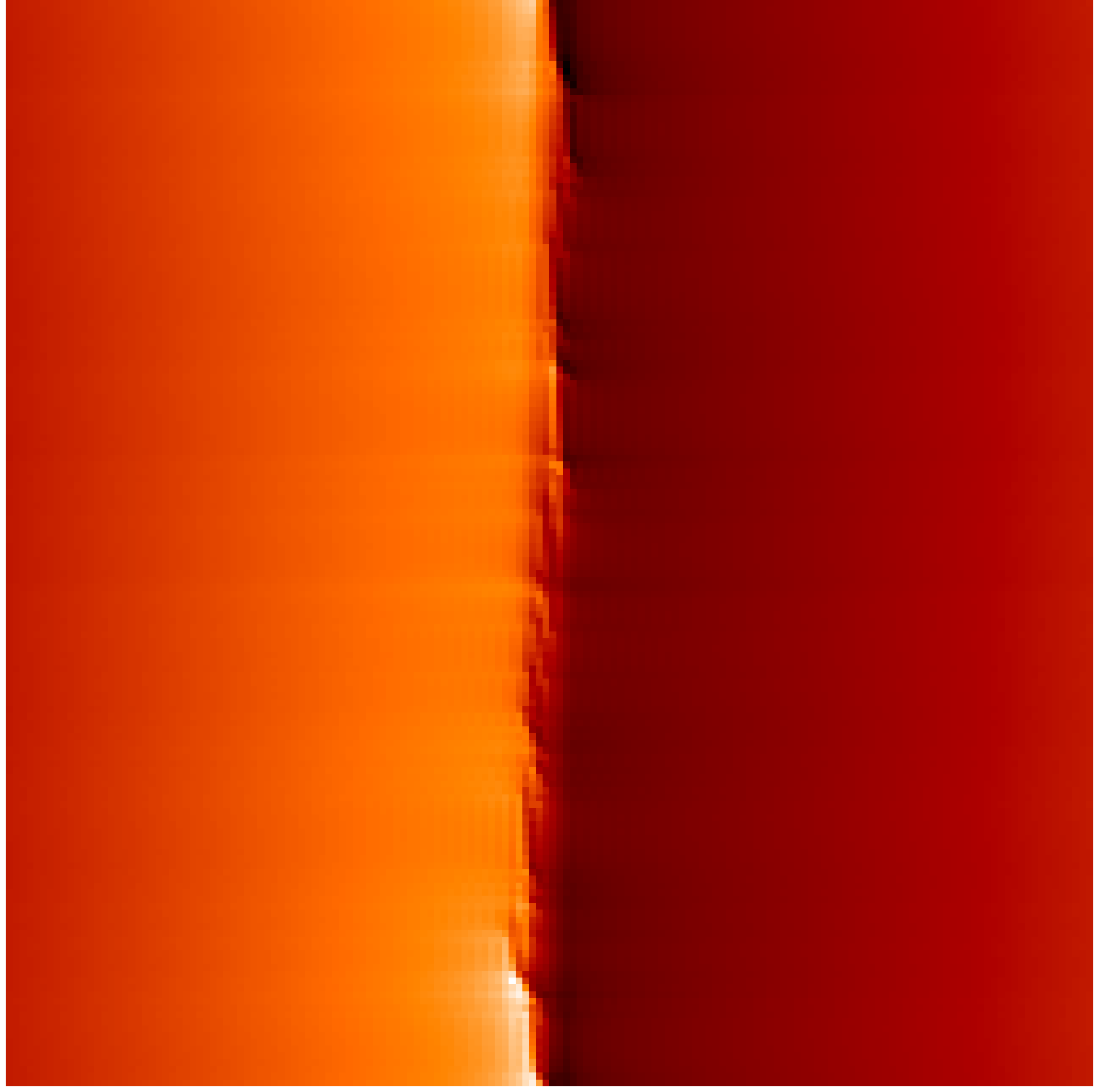}\\    
\caption{{\sl The Kaiser-Stebbins effect for cosmic strings.  A string network evolves into a
self-similar scaling regime, perturbing matter and radiation during its evolution. The effect on the
CMB after recombination leads to distinct steplike discontinuities on small angular scales that were
first studied by Kaiser \& Stebbins [1984]. The left panel shows a simulated patch of the sky that
fits in one of the pixels of the COBE experiment.  Hence, higher resolution observatories are needed
in order to detect strings. The right panel shows a patch on the CMB sky of order 20' across. However,
recent studies indicate that this clean tell-tale signal gets obscured at subdegree angular scales due
to the temperature fluctuations generated before recombination.  [Magueijo \& Ferreira 1997].}}
\label{fig-KSs}\end{figure}

If cosmic defects have really formed in the early universe and some of them are still within our
present horizon today, the anisotropies in the CMB they produce would have a characteristic signature.
Strings, for example, would imprint the background radiation in a very particular way due to the
Doppler shift that the background radiation suffers when a string intersects the line of sight.  The
conical topology of space around the string will produce a differential redshift of photons passing on
different sides of it, resulting in step--like discontinuities in the effective CMB temperature, given
by $\D \approx 8 \pi G \mu v_s \gamma$ with, as before, $\gamma = (1-v_s^2)^{-1/2}$ the Lorentz factor
and $v_s$ the velocity of the moving string.  This `stringy' signature was first studied by Kaiser \&
Stebbins [1984] and Gott [1985] (see Figure \ref{fig-KSs}).

Anisotropies of the CMB are directly related to the origin of structure in the universe. Galaxies and
clusters of galaxies eventually formed by gravitational instability from primordial density
fluctuations, and these same fluctuations left their imprint on the CMB. Recent balloon [de Bernardis,
\etal, 2000; Hanany, \etal, 2000], ground-based interferometer [Halverson, \etal, 2001] and satellite
[Bennett, \etal ~2003] experiments have produced reliable estimates of the power spectrum of the CMB
temperature anisotropies.  While they helped eliminate certain candidate theories for the primary
source of cosmic perturbations, the power spectrum data is still compatible with the theoretical
estimates of a relatively large variety of models, such as $\Lambda$CDM, quintessence models or some
hybrid models including cosmic defects.

There are two main classes of models of structure formation --\textit{passive} and \textit{active}
models. In passive models, density inhomogeneities are set as initial conditions at some early time,
and while they subsequently evolve as described by Einstein--Boltzmann equations, no additional
perturbations are seeded. On the other hand, in active models the sources of density perturbations are
time--dependent.

All specific realizations of passive models are based on the idea of inflation. In simplest
inflationary models it is assumed that there exists a weakly coupled scalar field $\phi$, called the
inflaton, which ``drives'' the (quasi) exponential expansion of the universe. The quantum fluctuations
of $\phi$ are stretched by the expansion to scales beyond the horizon, thus ``freezing'' their
amplitude.  Inflation is followed by a period of thermalization, during which standard forms of matter
and energy are formed. Because of the spatial variations of $\phi$ introduced by quantum fluctuations,
thermalization occurs at slightly different times in different parts of the universe. Such
fluctuations in the thermalization time give rise to density fluctuations. Because of their quantum
nature and because of the fact that initial perturbations are assumed to be in the vacuum state and
hence well described by a Gaussian distribution, perturbations produced during inflation are expected
to follow Gaussian statistics to a high degree [Gangui, Lucchin, Matarrese \& Mollerach, 1994], or
either be products of Gaussian random variables. This is a fairly general prediction that is being
tested currently with data from WMAP and will be tested more thoroughly in the future with
Planck.\footnote{Useful CMB resources can be found at {\tt
http://www.mpa-garching.mpg.de/\~{}banday/CMB.html}} 

Active models of structure formation are motivated by cosmic topological defects with the most
promising candidates being cosmic strings. As we saw in previous sections, it is widely believed that
the universe underwent a series of phase transitions as it cooled down due to the expansion. If our
ideas about grand unification are correct, then some cosmic defects should have formed during phase
transitions in the early universe. Once formed, cosmic strings could survive long enough to seed
density perturbations.  Defect models possess the attractive feature that they have no parameter
freedom, as all the necessary information is in principle contained in the underlying particle physics
model. Generically, perturbations produced by active models are not expected to be Gaussian
distributed [Gangui, Pogosian \& Winitzki, 2001a].

\subsection{CMB power spectrum from strings}
\label{sec-powerspectrum}      

The narrow main peak and the presence of the second and the third peaks in the CMB angular power
spectrum, as measured by BOOMERANG, MAXIMA, DASI and WMAP [de Bernardis, \etal, 2000; Hanany, \etal,
2000; Halverson, \etal, 2001; Page, \etal, 2003], is an evidence for coherent oscillations of the
photon--baryon fluid at the beginning of the decoupling epoch [see, \eg, Gangui, 2001]. While such
coherence is a property of all passive model, realistic cosmic string models produce highly incoherent
perturbations that result in a much broader main peak. This excludes cosmic strings as the primary
source of density fluctuations unless new physics is postulated, \textit{e.g.}~models with a varying
speed of light [Avelino \& Martins, 2000]. In addition to purely active or passive models, it has been
recently suggested that perturbations could be seeded by some combination of the two mechanisms. For
example, cosmic strings could have formed just before the end of inflation and partially contributed
to seeding density fluctuations. It has been shown [Contaldi, \etal, 1999; Battye \& Weller, 2000;
Bouchet, \etal, 2001] that, although not compelling, such hybrid models can be rather successful in
fitting the CMB power spectrum data.

Calculating CMB anisotropies sourced by topological defects is a rather difficult task. In
inflationary scenario the entire information about the seeds is contained in the initial conditions
for the perturbations in the metric.  In the case of cosmic defects, perturbations are continuously
seeded over the period of time from the phase transition that had produced them until today. The exact
determination of the resulting anisotropy requires, in principle, the knowledge of the
energy--momentum tensor [or, if only two point functions are being calculated, the unequal time
correlators, Pen, Seljak, \& Turok, 1997] of the defect network and the products of its decay at all
times. This information is simply not available. Instead, a number of clever simplifications, based on
the expected properties of the defect networks ({\it e.g.} scaling), are used to calculate the
source. The latest data from BOOMERANG, MAXIMA and WMAP experiments clearly disagree with the
predictions of these simple models of defects [Durrer, Gangui \& Sakellariadou, 1996].

The shape of the CMB angular power spectrum is determined by three main factors: the geometry of the
universe, coherence and causality.  The curvature of the universe directly affects the paths of light
rays coming to us from the surface of last scattering. In a closed universe, because of the lensing
effect induced by the positive curvature, the same physical distances between points on the sky would
correspond to larger angular scales. As a result, the peak structure in the CMB angular power spectrum
would shift to larger angular scales or, equivalently, to smaller values of the multipoles $\ell$'s.

\begin{figure}[htbp]
\includegraphics[width=8cm]{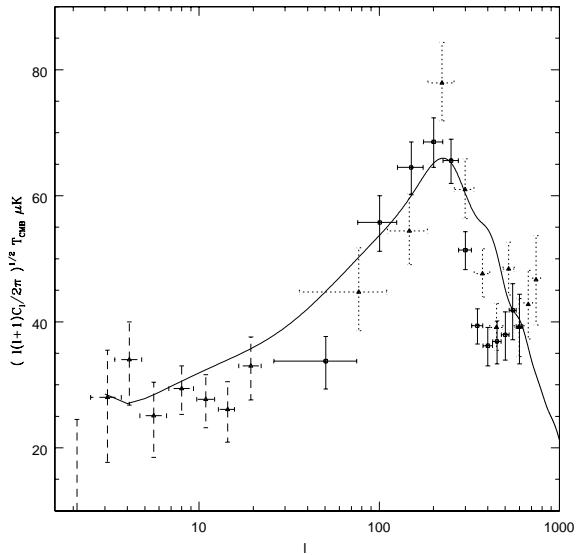}\\ 
\caption{{\sl The CMB power spectrum produced by the wiggly string model of [Pogosian \& Vachaspati,
1999] in a closed universe with $\Omega_{\rm total}=1.3$, $\Omega_{\rm baryon}=0.05$, $\Omega_{\rm
CDM}=0.35$, $\Omega_{\Lambda}=0.9$, and $H_0 = 65 {\rm ~km ~s}^{-1} {\rm Mpc}^{-1}$ [Pogosian, 2001].}}
\label{omega}
\end{figure}

The prediction of the cosmic string model of [Pogosian \& Vachaspati, 1999] for $\Omega_{\rm
total}=1.3$ is shown in Figure \ref{omega}.  As can be seen, the main peak in the angular power
spectrum can be matched by choosing a reasonable value for $\Omega_{\rm total}$. However, even with
the main peak in the right place the agreement with the data is far from satisfactory. The peak is
significantly wider than that in the data and there is no sign of a rise in power at multipole roughly
$l\approx 400$ as the actual data from WMAP suggests. The sharpness and the height of the main peak
in the angular spectrum can be enhanced by including the effects of gravitational radiation [Contaldi,
Hindmarsh \& Magueijo, 1999] and wiggles [Pogosian \& Vachaspati, 1999].  More precise
high--resolution numerical simulations of string networks in realistic cosmologies with a large
contribution from $\Omega_{\Lambda}$ are needed to determine the exact amount of small--scale
structure on the strings and the nature of the products of their decay [Landriau \& Shellard, 2002].
It is, however, unlikely that including these effects alone would result in a sufficiently narrow main
peak and some presence of a second peak.

This brings us to the issues of causality and coherence and how the random nature of the string
networks comes into the calculation of the anisotropy spectrum.  Both experimental and theoretical
results for the CMB power spectra involve calculations of averages. When estimating the correlations
of the observed temperature anisotropies, it is usual to compute the average over all available
patches on the sky. When calculating the predictions of their models, theorists find the average over
the {\em ensemble} of possible outcomes provided by the model.

In inflationary models, as in all passive models, only the initial conditions for the perturbations
are random.  The subsequent evolution is the same for all members of the ensemble. For wavelengths
higher than the Hubble radius, the linear evolution equations for the Fourier components of such
perturbations have a growing and a decaying solution. The modes corresponding to smaller wavelengths
have only oscillating solutions. As a consequence, prior to entering the horizon, each mode undergoes
a period of phase ``squeezing'' which leaves it in a highly coherent state by the time it starts to
oscillate. Coherence here means that all members of the ensemble, corresponding to the same Fourier
mode, have the same temporal phase. So even though there is randomness involved, as one has to draw
random amplitudes for the oscillations of a given mode, the time behavior of different members of the
ensemble is highly correlated. The total spectrum is the ensemble--averaged superposition of all
Fourier modes, and the predicted coherence results in an interference pattern seen in the angular
power spectrum as the well--known acoustic peaks.

\begin{figure}[htbp]
\includegraphics[width=8cm]{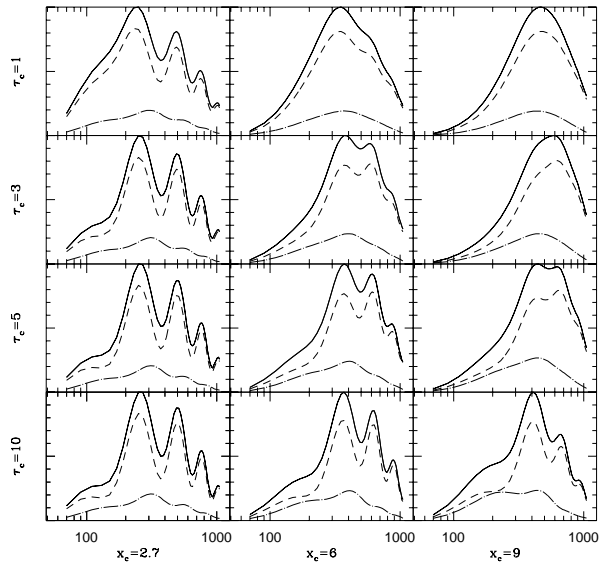}\\    
\caption{{\sl The predictions of the toy model of Magueijo, \etal\ [1996] for different values of
parameters $x_c$, the coherence length, and $\tau_c$, the coherence time. $x_c\propto \eta /
\lambda_c(\eta)$, where $\eta$ is the conformal time and $\lambda_c(\eta)$ is the correlation length
of the network at time $\eta$. One can obtain oscillations in the CMB power spectrum by fixing either
one of the parameters and varying the other.}}
\label{magetal}\end{figure}

In contrast, the evolution of the string network is highly non-linear. Cosmic strings are expected to
move at relativistic speeds, self--intersect and reconnect in a chaotic fashion. The consequence of
this behavior is that the unequal time correlators of the string energy--momentum vanish for time
differences larger than a certain coherence time ($\tau_c$ in Figure \ref{magetal}). Members of the
ensemble corresponding to a given mode of perturbations will have random temporal phases with the
``dice'' thrown on average once in each coherence time. The coherence time of a realistic string
network is rather short. As a result, the interference pattern in the angular power spectrum is
completely washed out.

Causality manifests itself, first of all, through the initial conditions for the string sources, the
perturbations in the metric and the densities of different particle species. If one assumes that the
defects are formed by a causal mechanism in an otherwise smooth universe then the correct initial
condition are obtained by setting the components of the stress--energy pseudo--tensor $\tau_{\mu \nu}$
to zero [Veeraraghavan \& Stebbins, 1990; Pen, Spergel \& Turok, 1994].  These are the same as the
isocurvature initial conditions [Hu, Spergel \& White, 1997]. A generic prediction of isocurvature
models (assuming perfect coherence) is that the first acoustic peak is almost completely hidden. The
main peak is then the second acoustic peak and in flat geometries it appears at $\ell\approx 300 -
400$. This is due to the fact that after entering the horizon a given Fourier mode of the source
perturbation requires time to induce perturbations in the photon density.  Causality also implies that
no superhorizon correlations in the string energy density are allowed. The correlation length of a
``realistic'' string network is normally between 0.1 and 0.4 of the horizon size.

An interesting study was performed by Magueijo, Albrecht, Ferreira \& Coulson [1996], where they
constructed a toy model of defects with two parameters: the coherence length and the coherence
time. The coherence length was taken to be the scale at which the energy density power spectrum of the
strings turns from a power law decay for large values of $k$ into a white noise at low $k$. This is
essentially the scale corresponding to the correlation length of the string network. The coherence
time was defined in the sense described in the beginning of this section, in particular, as the time
difference needed for the unequal time correlators to vanish.
Their study showed (see Figure \ref{magetal}) that by accepting any value for one of the parameters
and varying the other (within the constraints imposed by causality) one could reproduce the
oscillations in the CMB power spectrum. Unfortunately for cosmic strings, at least as we know them
today, they fall into the parameter range corresponding to the upper right corner in Figure
\ref{magetal}.

In order to get a better fit to present--day observations, cosmic strings must either be more coherent
or they have to be stretched over larger distances, which is another way of making them more
coherent. To understand this imagine that there was just one long straight string stretching across
the universe and moving with some given velocity. The evolution of this string would be linear and the
induced perturbations in the photon density would be coherent. By increasing the correlation length of
the string network we would move closer to this limiting case of just one long straight string and so
the coherence would be enhanced.

The question of whether or not defects can produce a pattern of the CMB power spectrum similar to, and
including the acoustic peaks of, that produced by the adiabatic inflationary models was repeatedly
addressed in the literature [Contaldi, Hindmarsh \& Magueijo 1999; Magueijo, \etal\ 1996; Liddle,
1995; Turok, 1996; Avelino \& Martins, 2000]. In particular, it was shown [Magueijo, \etal\ 1996;
Turok, 1996] that one can construct a causal model of active seeds which for certain values of
parameters can reproduce the oscillations in the CMB spectrum. The main problem today is that current
realistic models of cosmic strings fall out of the parameter range that is needed to fit the
observations.  At the moment, only the (non-minimal) models with either a varying speed of light or
hybrid contribution of strings+inflation are the only ones involving topological defects that to some
extent can match the observations\footnote{With the recent first-year WMAP data, specially the
cross-correlation between the temperature and polarization maps (TE correlation), the situation gets
worse for the whole class of causal models. In fact, there exists a clear nontrivial cross-correlation
signature for angular scales above the size of the acoustic horizon [as shown for example in Kogut,
\etal\ 2003] which is virtually impossible to reproduce with any class of non contrived causal models,
cosmic defects included.}. One possible way to distinguish their predictions from those of
inflationary models would be by computing key non--Gaussian statistical quantities, such as the CMB
bispectrum.

\subsection{CMB bispectrum from active models}
\label{sec-}      

Different cosmological models differ in their predictions for the statistical distribution of the
anisotropies beyond the power spectrum. Current WMAP data and the future Planck satellite mission will
provide high-precision data allowing definite estimates of non-Gaussian signals in the CMB. It is
therefore important to know precisely which are the predictions of all candidate models for the
statistical quantities that will be extracted from the new data and identify their specific
signatures.

Of the available non-Gaussian statistics, the CMB bispectrum, or the three-point function of Fourier
components of the temperature anisotropy, has been perhaps the one best studied in the literature
[Gangui \& Martin, 2000a]. There are a few cases where the bispectrum may be deduced analytically from
the underlying model.  The bispectrum can be estimated from simulated CMB sky maps; however, computing
a large number of full-sky maps resulting from defects is a much more demanding task.  Recently, a
precise numerical code to compute it, not using CMB maps and similar to the \textsc{CMBFAST}
code\footnote{{\tt http://physics.nyu.edu/matiasz/CMBFAST/cmbfast.html}} for the power spectrum, was
developed in [Gangui, Pogosian \& Winitzki 2001b]. What follows below is an account of this work.

In a few words, given a suitable model, one can generate a statistical \emph{ensemble} of realizations
of defect matter perturbations. We used a modified Boltzmann code based on \textsc{CMBFAST} to compute
the effect of these perturbations on the CMB and found the bispectrum estimator for a given
realization of sources. We then performed statistical averaging over the ensemble of realizations to
compute the expected CMB bispectrum. (The CMB power spectrum was also obtained as a byproduct.)  As a
first application, we then computed the expected CMB bispectrum from a model of simulated string
networks first introduced by Albrecht \etal\ [1997] and further developed in [Pogosian \& Vachaspati,
1999] and in [Gangui, Pogosian \& Winitzki 2001].

We assume that, given a model of active perturbations, such as a string simulation, we can calculate
the energy-momentum tensor $T_{\mu \nu }({\mathbf x},\tau )$ for a particular realization of the
sources in a finite spatial volume $V_{0}$.  Here, ${\bf x}$ is a 3-dimensional coordinate and $\tau $
is the cosmic time.  Many simulations are run to obtain an ensemble of random realizations of sources
with statistical properties appropriate for the given model. The spatial Fourier decomposition of
$T_{\mu \nu }$ can be written as
\begin{equation} \label{fouriersum}
T_{\mu \nu }({{\mathbf x}},\tau )=\sum _{{\mathbf k}}\Theta_{\mu \nu
}({{\mathbf k}},\tau )e^{i{{\mathbf k}}{{\mathbf x}}}\, \, \, ,
\end{equation}
where ${{\mathbf k}}$ are discrete. If $V_{0}$ is sufficiently large
we can approximate the summation by the integral \begin{eqnarray}
\sum _{{\mathbf k}}\Theta_{\mu \nu }({{\mathbf k}},\tau )e^{i{{\mathbf
k}}{{\mathbf x}}}\approx \frac{V_{0}}{(2\pi )^{3}}\int d^{3}{{\mathbf
k}}\Theta_{\mu \nu }({{\mathbf k}},\tau )e^{i{{\mathbf k}}{{\mathbf
x}}}\, \,
\, ,\label{sumtoint}
\end{eqnarray}
and the corresponding inverse Fourier transform will be
\begin{equation}
\label{inversefourier}
\Theta_{\mu \nu }({{\mathbf k}},\tau )=\frac{1}{V_{0}}\int
_{V_{0}}d^{3}{{\mathbf x}}\,T_{\mu \nu }({{\mathbf x}},\tau
)e^{-i{{\mathbf
k}}{{\mathbf x}}}\, \, \, .
\end{equation}
Of course, the final results, such as the CMB power spectrum or
bispectrum,
do not depend on the choice of $V_{0}$. To ensure this independence,
we shall keep $V_{0}$ in all expressions where it appears below.                          

It is conventional to expand the temperature fluctuations over the
basis of
spherical harmonics, \begin{equation}
{\Delta T/T}({\hat{{\mathbf n}}})=\sum
_{lm}a_{lm}Y_{lm}({\hat{{\mathbf n}}}),\end{equation}
where $\hat{{\mathbf n}}$ is a
unit vector. The coefficients $a_{lm}$ can be decomposed into Fourier
modes, \begin{equation} \label{eq:alm-def}
a_{lm}=\frac{V_{0}}{(2\pi )^{3}}\left( -i\right) ^{l}4\pi \int
d^{3}{{\mathbf k}}\, \Delta _{l}\left( {{\mathbf k}}\right)
Y^{*}_{lm}({\hat{{\mathbf k}}}).
\end{equation}
Given the sources $\Theta_{\mu \nu }({{\mathbf k}},\tau )$, the
quantities
$\Delta _{l}({{\mathbf k}})$ are found by solving linearized
Einstein-Boltzmann equations and integrating along the line of sight,
using
a code similar to CMBFAST [Seljak \& Zaldarriaga, 1996]. 
This standard procedure
can be written
symbolically as the action of a linear operator ${\hat{B}}_{l}^{\mu
\nu
}(k)$ on the source energy-momentum tensor, $\Delta _{l}({{\mathbf
k}})={\hat{B}}_{l}^{\mu \nu }(k)\Theta_{\mu \nu }({{\mathbf k}},\tau
)$,
so the third moment of $\Delta _{l}({{\mathbf k}})$ is linearly
related to the
three-point correlator of $\Theta_{\mu \nu }({{\mathbf k}},\tau
)$. Below
we consider the quantities $\Delta _{l}({{\mathbf k}})$, corresponding
to a
set of realizations of active sources, as given. The numerical
procedure for computing $\Delta _{l}({{\mathbf k}})$ 
was developed in  [Albrecht \etal\, 1997] and in [Pogosian \& Vachaspati, 1999].    

The third moment of $a_{lm}$, namely $\left\langle
a_{l_{1}m_{1}}a_{l_{2}m_{2}}a_{l_{3}m_{3}}\right\rangle $, can be
expressed
as 
\begin{eqnarray} 
\left( -i\right) ^{l_{1}+l_{2}+l_{3}}\left( 4\pi
\right) ^{3}\! \! \frac{V_{0}^{3}}{(2\pi )^{9}}\! \! \int \! \!
d^{3}{{\mathbf k}}_{1}d^{3}{{\mathbf k}}_{2}d^{3}{{\mathbf
k}}_{3}Y^{*}_{l_{1}m_{1}}\! ({\hat{{\mathbf k}}}_{1})
Y^{*}_{l_{2}m_{2}}\! ({\hat{{\mathbf k}}}_{2})Y^{*}_{l_{3}m_{3}}\!
({\hat{{\mathbf k}}}_{3})\left\langle \Delta _{l_{1}}\! \! \left(
{{\mathbf
k}}_{1}\right) \Delta _{l_{2}}\! \! \left( {{\mathbf k}}_{2}\right)
\Delta
_{l_{3}}\! \! \left( {{\mathbf k}}_{3}\right) \right\rangle
.\label{eq:3alm-1}
\end{eqnarray}

A straightforward numerical evaluation of Eq.~(\ref{eq:3alm-1}) from given sources $\Delta _{l}\left(
{{\mathbf k}}\right) $ is prohibitively difficult, because it involves too many integrations of
oscillating functions.  However, we shall be able to reduce the computation to integrations over
scalars [a similar method was employed in Komatsu \& Spergel, 2001 and in Wang \& Kamionkowski, 2000].
Due to homogeneity, the 3-point function vanishes unless the triangle constraint is
satisfied,
\begin{equation} \label{eq:triangle} {{\mathbf k}}_{1}+{{\mathbf k}}_{2}+{{\mathbf
k}}_{3}=0.
\end{equation}
We may write \begin{eqnarray}
\left\langle \Delta _{l_{1}}\left( {{\mathbf k}}_{1}\right)
\Delta
_{l_{2}}\left( {{\mathbf k}}_{2}\right) \Delta _{l_{3}}\left(
{{\mathbf
k}}_{3}\right) \right\rangle  =  \delta ^{(3)}\left(
{{\mathbf k}}_{1}+{{\mathbf k}}_{2}+{{\mathbf k}}_{3}\right)
P_{l_{1}l_{2}l_{3}}\left( {{\mathbf k}}_{1},{{\mathbf
k}}_{2},{{\mathbf
k}}_{3}\right) ,\label{p3lvector}
\end{eqnarray}                                     
where the three-point function $P_{l_{1}l_{2}l_{3}}\left( {{\mathbf
k}}_{1},{{\mathbf k}}_{2},{{\mathbf k}}_{3}\right) $ is defined only
for values of ${\mathbf k}_{i}$ that satisfy
Eq.~(\ref{eq:triangle}). Given the scalar values $k_{1}$, $k_{2}$,
$k_{3}$, there is a unique (up to an overall rotation) triplet of
directions ${\hat{{\mathbf k}}}_{i}$ for which the RHS of
Eq.~(\ref{p3lvector}) does not vanish. The quantity
$P_{l_{1}l_{2}l_{3}}\left( {{\mathbf k}}_{1},{{\mathbf
k}}_{2},{{\mathbf k}}_{3}\right) $ is invariant under an overall
rotation of all three vectors ${{\mathbf k}}_{i}$ and therefore may be
equivalently represented by a function of \emph{scalar} values
$k_{1}$, $k_{2}$, $k_{3}$, while preserving all angular
information. Hence, we can rewrite Eq.~(\ref{p3lvector}) as
\begin{eqnarray} \left\langle \Delta _{l_{1}}\! \! \left(
{{\mathbf k}}_{1}\right) \Delta _{l_{2}}\! \! \left( {{\mathbf
k}}_{2}\right) \Delta _{l_{3}}\! \! \left( {{\mathbf k}}_{3}\right)
\right\rangle =  \delta ^{(3)}\left( {{\mathbf
k}}_{1}+{{\mathbf k}}_{2}+{{\mathbf k}}_{3}\right)
P_{l_{1}l_{2}l_{3}}(k_{1},k_{2},k_{3}).\label{p3lscalar}
\end{eqnarray}
Then, using the simulation volume $V_{0}$ explicitly, we have
\begin{equation}
\label{p3l0}
P_{l_{1}l_{2}l_{3}}\! \left( k_{1},k_{2},k_{3}\right) \! =\!
\frac{(2\pi
)^{3}}{V_{0}}\left\langle \Delta _{l_{1}}\! \! \left( {{\mathbf
k}}_{1}\right) \Delta _{l_{2}}\! \! \left( {{\mathbf k}}_{2}\right)
\Delta
_{l_{3}}\! \! \left( {{\mathbf k}}_{3}\right) \right\rangle .
\end{equation}
Given an arbitrary direction $\hat{{\mathbf k}}_{1}$ and the
magnitudes
$k_{1}$, $k_{2}$ and $k_{3}$, the directions $\hat{{\mathbf k}}_{2}$
and $\hat{{\mathbf k}}_{3}$ are specified up to overall rotations by
the
triangle constraint. Therefore, both sides of Eq.~(\ref{p3l0}) are
functions of scalar $k_{i}$ only. The expression on the RHS of Eq.
(\ref{p3l0}) is evaluated numerically by averaging over different
realizations of the sources \textit{and} over permissible directions
$\hat{{\mathbf k}}_{i}$; below we shall give more details of the
procedure.      

Substituting Eqs.~(\ref{p3lscalar}) and (\ref{p3l0}) into (\ref{eq:3alm-1}), Fourier transforming the
Dirac delta and using the Rayleigh identity, we can perform all angular integrations analytically and
obtain a compact form for the third moment,
\begin{equation}
\label{eq:3alm-res}
\left\langle a_{l_{1}m_{1}}a_{l_{2}m_{2}}a_{l_{3}m_{3}}\right\rangle
={\mathcal{H}}_{l_{1}l_{2}l_{3}}^{m_{1}m_{2}m_{3}}\int r^{2}dr\,
b_{l_{1}l_{2}l_{3}}(r),
\end{equation}
where, denoting the Wigner $3j$-symbol by
$\left( ^{\, \, l_{1}\, \; l_{2}\, \; l_{3}}_{m_{1}m_{2}m_{3}}\right)
$, we
have
\begin{eqnarray} {\mathcal{H}}_{l_{1}l_{2}l_{3}}^{m_{1}m_{2}m_{3}}
\equiv   \sqrt{\frac{\left( 2l_{1}+1\right) \left( 2l_{2}+1\right)
\left(
2l_{3}+1\right) }{4\pi }} \left(
\begin{array}{ccc}
l_{1} & l_{2} & l_{3}\\
0 & 0 & 0
\end{array}\right) \left( \begin{array}{ccc}
l_{1} & l_{2} & l_{3}\\
m_{1} & m_{2} & m_{3}
\end{array}\right) \, ,\label{eq:hlll}
\end{eqnarray}
and where we have defined the auxiliary quantities
$b_{l_{1}l_{2}l_{3}}$
using spherical Bessel functions $j_{l}$, \begin{eqnarray}
b_{l_{1}l_{2}l_{3}}(r) & \equiv  & \frac{8}{\pi
^{3}}\frac{V_{0}^{3}}{(2\pi
)^{3}}\int k_{1}^{2}dk_{1}\, k_{2}^{2}dk_{2}\, k_{3}^{2}dk_{3}\,
\nonumber
\\ & \times  &
j_{l_{1}}(k_{1}r)j_{l_{2}}(k_{2}r)j_{l_{3}}(k_{3}r)P_{l_{1}l_{2}l_{3}}(k_{1},k_{2},k_{3}).
\label{defineb}
\end{eqnarray}
The volume factor $V_{0}^{3}$ contained in this expression is correct:
as
shown in the next section, each term $\Delta _{l}$ includes a factor
$V_{0}^{-2/3}$, while the average quantity
$P_{l_{1}l_{2}l_{3}}(k_{1},k_{2},k_{3})\propto V_{0}^{-3}$
{[}cf.~Eq.~(\ref{p3l0}){]}, so that the arbitrary volume $V_{0}$ of
the
simulation cancels.
                                                                            
Our proposed numerical procedure therefore consists of computing the
RHS of
Eq.~(\ref{eq:3alm-res}) by evaluating the necessary integrals. For
fixed
$\left\{ l_{1}l_{2}l_{3}\right\} $, computation of the quantities
$b_{l_{1}l_{2}l_{3}}(r)$ is a triple integral over scalar $k_{i}$
defined
by Eq.~(\ref{defineb}); it is followed by a fourth scalar integral
over $r$
{[}Eq.~(\ref{eq:3alm-res}){]}. We also need to average over many
realizations of sources to obtain $P_{l_{1}l_{2}l_{3}}\! \left(
k_{1},k_{2},k_{3}\right) $. It was not feasible for us to precompute
the
values $P_{l_{1}l_{2}l_{3}}\! \left( k_{1},k_{2},k_{3}\right) $ on a
grid
before integration because of the large volume of data: for each set
$\left\{ l_{1}l_{2}l_{3}\right\} $ the grid must contain $\sim 10^{3}$
points for each $k_{i}$. Instead, we precompute $\Delta _{l}\! \!
\left(
{{\mathbf k}}\right) $ from one realization of sources and evaluate
the RHS
of Eq.~(\ref{p3l0}) on that data as an \emph{estimator} of
$P_{l_{1}l_{2}l_{3}}\! \left( k_{1},k_{2},k_{3}\right) $, averaging
over
allowed directions of $\hat{{\mathbf k}}_{i}$. The result is used for
integration in Eq.~(\ref{defineb}).

Because of isotropy and since the allowed sets of directions $\hat{{\mathbf k}}_{i}$ are planar, it is
enough to restrict the numerical calculation to directions $\hat{{\mathbf k}}_{i}$ within a fixed
two-dimensional plane. This significantly reduces the amount of computations and data storage, since
$\Delta _{l}\! \left( {{\mathbf k}}\right) $ only needs to be stored on a two-dimensional grid of
${\mathbf k}$.

In estimating $P_{l_{1}l_{2}l_{3}}\! \left( k_{1},k_{2},k_{3}\right) $ from Eq.~(\ref{p3l0}),
averaging over directions of $\hat{{\mathbf k}}_{i}$ plays a similar role to ensemble averaging over
source realizations.  Therefore if the number of directions is large enough (we used 720 for cosmic
strings), only a moderate number of different source realizations is needed. The main numerical
difficulty is the highly oscillating nature of the function $b_{l_{1}l_{2}l_{3}}(r)$. The calculation
of the bispectrum for cosmic strings presented in the next Section requires about 20 days of a
single-CPU workstation time per realization.

We note that this method is specific for the bispectrum and cannot be applied to compute higher-order
correlations. The reason is that higher-order correlations involve configurations of vectors ${\mathbf
k}_{i}$ that are not described by scalar values $k_{i}$ and not restricted to a plane. For instance, a
computation of a 4-point function would involve integration of highly oscillating functions over four
vectors ${\mathbf k}_{i}$ which is computationally infeasible.

{}From Eq.~(\ref{eq:3alm-res}) we derive the CMB angular bispectrum ${\mathcal{C}}_{l_{1}l_{2}l_{3}}$,
defined as [Gangui \& Martin, 2000b]
\begin{eqnarray}
\bigl \langle a_{l_{1}m_{1}}a_{l_{2}m_{2}}a_{l_{3}m_{3}}\bigr \rangle
=\left( \begin{array}{ccc} l_{1} & l_{2} & l_{3}\\
m_{1} & m_{2} & m_{3}
\end{array}\right) {\mathcal{C}}_{l_{1}l_{2}l_{3}}\, .
\end{eqnarray}
The presence of the 3$ j$-symbol guarantees that the third moment vanishes unless
$m_{1}+m_{2}+m_{3}=0$ and the $l_{i}$ indices satisfy the triangle rule $|l_{i}-l_{j}|\leq l_{k}\leq
l_{i}+l_{j}$. Invariance under spatial inversions of the three-point correlation function implies the
additional `selection rule' $l_{1}+l_{2}+l_{3}=\mbox {even}$, in order for the third moment not to
vanish. Finally, from this last relation and using standard properties of the 3$ j$-symbols, it
follows that the angular bispectrum ${\mathcal{C}}_{l_{1}l_{2}l_{3}}$ is left unchanged under any
arbitrary permutation of the indices $l_{i}$.
                                                                           
In what follows we will restrict our calculations to the angular bispectrum
$C_{l_{1}l_{2}l_{3}}$ in the `diagonal' case,
\textit{i.e.}~$l_{1}=l_{2}=l_{3}=l$.
This is a representative case and, in fact, the one most frequently
considered in the literature. Plots of the power spectrum are usually
done
in terms of $l(l+1)C_{l}$ which, apart from constant factors, is the
contribution to the mean squared anisotropy of temperature
fluctuations per
unit logarithmic interval of $l$. In full analogy with this, the
relevant
quantity to work with in the case of the bispectrum is
\begin{eqnarray} G_{lll} = l(2l+1)^{3/2}\left( \begin{array}{ccc}
l & l & l\\
0 & 0 & 0
\end{array}\right) C_{lll}\, .\label{eq:QtP}
\end{eqnarray}
For large values of the multipole index $l$, $G_{lll}\propto
l^{3/2}C_{lll}$.  Note also what happens with the 3$ j$-symbols
appearing in the definition of the coefficients
${\mathcal{H}}_{l_{1}l_{2}l_{3}}^{m_{1}m_{2}m_{3}}$: the symbol
$\left( ^{\, \, l_{1}\, \; l_{2}\, \; l_{3}}_{m_{1}m_{2}m_{3}}\right)
$ is absent from the definition of $C_{l_{1}l_{2}l_{3}}$, while in
Eq.~(\ref{eq:QtP}) the symbol $\left( ^{\, l\; l\; l}_{0\: 0\:
0}\right) $ is squared. Hence, there are no remnant oscillations due
to the alternating sign of $\left( ^{\, l\; l\; l}_{0\: 0\: 0}\right)$.

However, even more important than the value of $C_{lll}$ itself is the relation between the bispectrum
and the cosmic variance associated with it.  In fact, it is their comparison that tells us about the
observability `in principle' of the non-Gaussian signal. The cosmic variance constitutes a theoretical
uncertainty for all observable quantities and comes about due to the fact of having just one
realization of the stochastic process, in our case, the CMB sky [Scaramella \& Vittorio, 1991].

The way to proceed is to employ an estimator $\hat{C}_{l_{1}l_{2}l_{3}}$ for the bispectrum and
compute the variance from it. By choosing an unbiased estimator we ensure it satisfies
$C_{l_{1}l_{2}l_{3}}=\langle \hat{C}_{l_{1}l_{2}l_{3}}\rangle $. However, this condition does not
isolate a unique estimator. The proper way to select the {\it best unbiased} estimator is to compute
the variances of all candidates and choose the one with the smallest value.  The estimator with this
property was computed in [Gangui \& Martin, 2000b] and is
\begin{equation} \label{eq:clll-best}
\hat{C}_{l_{1}l_{2}l_{3}}=\! \! \! \sum _{m_{1},m_{2},m_{3}}\! \!
\left(
\begin{array}{ccc} l_{1} & l_{2} & l_{3}\\
m_{1} & m_{2} & m_{3}
\end{array}\right) a_{l_{1}m_{1}}a_{l_{2}m_{2}}a_{l_{3}m_{3}}.
\end{equation}
The variance of this estimator, assuming a mildly non-Gaussian
distribution, can be expressed in terms of the angular power spectrum
$C_{l}$ as follows
\begin{equation}
\label{eq:sigma}
 \sigma
^{2}_{\hat{C}_{l_{1}l_{2}l_{3}}}\! \! \! \!
=C_{l_{1}}C_{l_{2}}C_{l_{3}}\!
\left( 1\! +\! \delta _{l_{1}l_{2}}\! \! +\! \delta _{l_{2}l_{3}}\! \!
+\!
\delta _{l_{3}l_{1}}\! \! +\! 2\delta _{l_{1}l_{2}}\delta
_{l_{2}l_{3}}\right) .
\end{equation}
The theoretical signal-to-noise ratio for the bispectrum is then given
by
\begin{equation}
(S/N)_{l_{1}l_{2}l_{3}} =
|C_{l_{1}l_{2}l_{3}}/\sigma_{\hat{C}_{l_{1}l_{2}l_{3}}}|.
\end{equation}
In turn, for the diagonal case $l_{1}=l_{2}=l_{3}=l$ we have
\begin{equation}
(S/N)_{l} = |C_{lll}/\sigma _{\hat{C}_{lll}}|.
\end{equation}
                                                 
Incorporating all the specifics of the particular experiment, such as sky coverage, angular
resolution, etc., will allow us to give an estimate of the particular non-Gaussian signature
associated with a given active source and, if observable, indicate the appropriate range of multipole
$l$'s where it is best to look for it.
                                                             
\subsection{CMB bispectrum from strings}
\label{sec-bispectrum}      

To calculate the sources of perturbations we have used an updated version of the cosmic string model
first introduced by Albrecht \etal\ [1997] and further developed in [Pogosian \& Vachaspati, 1999],
where the wiggly nature of strings was taken into account. In these previous works the model was
tailored to the computation of the two-point statistics (matter and CMB power spectra). When dealing
with higher-order statistics, such as the bispectrum, a different strategy needs to be employed.

In the model, the string network is represented by a collection of uncorrelated straight string
segments produced at some early epoch and moving with random uncorrelated velocities. At every
subsequent epoch, a certain fraction of the number of segments decays in a way that maintains network
scaling. The length of each segment at any time is taken to be equal to the correlation length of the
network. This and the root mean square velocity of segments are computed from the velocity-dependent
one-scale model of Martins \& Shellard [1996].  The positions of segments are drawn from a uniform
distribution in space, and their orientations are chosen from a uniform distribution on a two-sphere.

The total energy of the string network in a volume $V$ at any time is $E=N\mu L$, where $N$ is the
total number of string segments at that time, $\mu $ is the mass per unit length, and $L$ is the
length of one segment. If $L$ is the correlation length of the string network then, according to the
one-scale model, the energy density is $\rho ={E/V}={\mu /L^{2}}$, where $V=V_{0}a^{3}$, the expansion
factor $a$ is normalized so that $a=1$ today, and $V_{0}$ is a constant simulation volume. It follows
that $N=V/L^{3}=V_{0}/\ell^{3}$, where $\ell=L/a$ is the comoving correlation length.  In the scaling
regime $\ell$ is approximately proportional to the conformal time $\tau $ and so the number of strings
$N(\tau )$ within the simulation volume $V_{0}$ falls as $\tau ^{-3}$.
                                                     
To calculate the CMB anisotropy one needs to evolve the string network over at least four orders of
magnitude in cosmic expansion. Hence, one would have to start with $N\gsim 10^{12}$ string segments in
order to have one segment left at the present time.  Keeping track of such a huge number of segments
is numerically infeasible.  A way around this difficulty was suggested in Ref.\cite{ABR97}, where the
idea was to consolidate all string segments that decay at the same epoch. The number of segments that
decay by the (discretized) conformal time $\tau _{i}$ is \begin{equation}
\label{eq:nd}
N_{d}(\tau _{i})=V_{0}\left( n(\tau _{i-1})-n(\tau _{i})\right) ,
\end{equation}
where $n(\tau )=[\ell(\tau )]^{-3}$ is the number density of strings
at time
$\tau $. The energy-momentum tensor in Fourier space, $\Theta^{i}_{\mu
\nu }$,
of these $N_{d}(\tau _{i})$ segments is a sum \begin{equation}
\label{emtsum}
\Theta^{i}_{\mu \nu }=\sum _{m=1}^{N_{d}(\tau _{i})}\Theta^{im}_{\mu
\nu }\,
\, \, , \end{equation}
where $\Theta^{im}_{\mu \nu }$ is the Fourier transform of the
energy-momentum
of the $m$-th segment. If segments are uncorrelated, then
\begin{equation}
\label{eq:theta2}
\langle\Theta^{im}_{\mu\nu}\Theta^{im'}_{\sigma\rho}\rangle =
\delta_{m m'}
\langle\Theta^{im}_{\mu\nu}\Theta^{im}_{\sigma\rho}\rangle
\end{equation}
and
\begin{equation}
\langle\Theta^{im}_{\mu\nu}\Theta^{im'}_{\sigma\rho}
\Theta^{im''}_{\gamma\delta}\rangle =
\delta_{m m'}\delta_{m m''}
\langle\Theta^{im}_{\mu\nu}\Theta^{im}_{\sigma\rho}
\Theta^{im}_{\gamma\delta}\rangle .
\end{equation}                      
Here the angular brackets $\langle \ldots \rangle $ denote the
ensemble average, which in our case means averaging over many
realizations
of the string network. If we are calculating power spectra, then the
relevant quantities are the two-point functions of $\Theta^{i}_{\mu
\nu }$,
namely
\begin{eqnarray} \langle \Theta^{i}_{\mu \nu }\Theta^{i}_{\sigma \rho
}\rangle =\langle \sum _{m=1}^{N_{d}(\tau _{i})}\sum
_{m'=1}^{N_{d}(\tau
_{i})}\Theta^{im}_{\mu \nu }\Theta^{im'}_{\sigma \rho }\rangle
.\label{thefix1} \end{eqnarray}
Eq.~(\ref{eq:theta2}) allows us to write \begin{eqnarray}
\langle \Theta^{i}_{\mu \nu }\Theta^{i}_{\sigma \rho }\rangle =\sum
_{m=1}^{N_{d}(\tau _{i})}\langle \Theta^{im}_{\mu \nu
}\Theta^{im}_{\sigma
\rho }\rangle =N_{d}(\tau _{i})\langle \Theta^{i1}_{\mu \nu
}\Theta^{i1}_{\sigma \rho }\rangle ,\label{thefix2}
\end{eqnarray}
where $\Theta^{i1}_{\mu \nu }$ is of the energy-momentum
of one of the segments that decay by the time $\tau _{i}$. The last
step in
Eq.~(\ref{thefix2}) is possible because the segments are statistically
equivalent. Thus, if we only want to reproduce the correct power
spectra in
the limit of a large number of realizations, we can replace the sum in
Eq.~(\ref{emtsum}) by \begin{equation} \label{thefix3}
\Theta^{i}_{\mu \nu }=\sqrt{N_{d}(\tau _{i})}\Theta^{i1}_{\mu \nu }.
\end{equation}
The total energy-momentum tensor of the network in Fourier space is a
sum over
the consolidated segments:
\begin{equation}
\label{emtsum1}
\Theta_{\mu \nu }=\sum _{i=1}^{K}\Theta^{i}_{\mu \nu }=\sum
_{i=1}^{K}\sqrt{N_{d}(\tau _{i})}\Theta^{i1}_{\mu \nu }\, .
\end{equation}
So, instead of summing over $\sum _{i=1}^{K}N_{d}(\tau _{i})\gsim
10^{12}$
segments we now sum over only $K$ segments, making $K$ a parameter.

For the three-point functions we extend the above procedure. Instead
of
Eqs.~(\ref{thefix1}) and (\ref{thefix2}) we now write
\begin{eqnarray}
\langle \Theta^{i}_{\mu \nu }\Theta^{i}_{\sigma \rho
}\Theta^{i}_{\gamma
\delta }\rangle \! = \! \langle \sum _{m=1}^{N_{d}(\tau _{i})}\sum
_{m'=1}^{N_{d}(\tau _{i})}\sum _{m''=1}^{N_{d}(\tau
_{i})}\Theta^{im}_{\mu \nu
}\Theta^{im'}_{\sigma \rho }\Theta^{im''}_{\gamma \delta }\rangle  
=\!\!\! \sum _{m=1}^{N_{d}(\tau _{i})}\langle \Theta^{im}_{\mu
\nu
}\Theta^{im}_{\sigma \rho }\Theta^{im}_{\gamma \delta }\rangle
= N_{d}(\tau
_{i})\langle \Theta^{i1}_{\mu \nu }\Theta^{i1}_{\sigma \rho
}\Theta^{i1}_{\gamma \delta }\rangle \, 
\label{newfix} \end{eqnarray}
Therefore, for the purpose of calculation of three-point functions,
the sum
in Eq.~(\ref{emtsum}) should now be replaced by \begin{equation}
\label{newfix1}
\Theta^{i}_{\mu \nu }=[N_{d}(\tau _{i})]^{1/3}\Theta^{i1}_{\mu \nu }\,
.
\end{equation}

Both expressions in Eqs.~(\ref{thefix3}) and (\ref{newfix1}), depend on the simulation volume,
$V_{0}$, contained in the definition of $N_{d}(\tau _{i})$ given in Eq.~(\ref{eq:nd}). This is to be
expected and is consistent with our calculations, since this volume cancels in expressions for
observable quantities.

Note also that the simulation model in its present form does not allow computation of CMB sky
maps. This is because the method of finding the two- and three-point functions as we described
involves {}``consolidated{}'' quantities $\Theta^{i}_{\mu \nu }$ which do not correspond to the
energy-momentum tensor of a real string network. These quantities are auxiliary and specially prepared
to give the correct two- or three-point functions after ensemble averaging.

\begin{figure}[htbp]
\includegraphics[width=9cm]{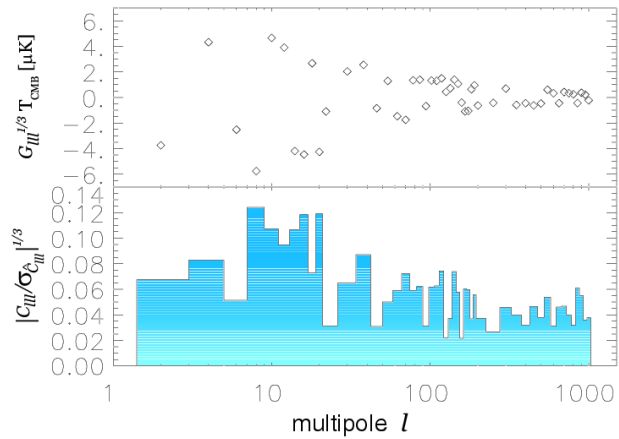}\\   
\caption{{\sl The CMB angular bispectrum in the `diagonal' case ($ G_{lll}^{1/3}$) from wiggly cosmic
strings in a spatially flat model with cosmological parameters $ \Omega _{\rm CDM}=0.3$, $ \Omega
_{\rm baryon}=0.05$, $ \Omega _{\Lambda }=0.65$, and Hubble constant $
H=0.65{\textrm{km}}{\textrm{s}}^{-1}{\textrm{Mpc}}^{-1}$ {[}upper panel{]}. In the lower panel we show
the ratio of the signal to theoretical noise $ |C_{lll}/\sigma _{\hat{C}_{lll}}|^{1/3}$ for different
multipole indices.  Normalization follows from fitting the power spectrum to the BOOMERANG and MAXIMA
data.}}
\label{fig:1}\end{figure}

In Fig.~\ref{fig:1} we show the results for $G_{lll}^{1/3}$ {[}cf.~Eq.~(\ref{eq:QtP}){]}. It was
calculated using the string model with $800$ consolidated segments in a flat universe with cold dark
matter and a cosmological constant. Only the scalar contribution to the anisotropy has been
included. Vector and tensor contributions are known to be relatively insignificant for local cosmic
strings and can safely be ignored in this model \cite{ABR97,PV99}\footnote{The contribution of vector
and tensor modes is large in the case of global strings [Turok, Pen \& Seljak, 1998; Durrer, Gangui \&
Sakellariadou, 1996].}. The plots are produced using a single realization of the string network by
averaging over $720$ directions of ${\mathbf k}_{i}$. The comparison of $G_{lll}^{1/3}$ (or
equivalently ${C}_{lll}^{1/3}$) with its cosmic variance {[}cf.~Eq.~(\ref{eq:sigma}){]} clearly shows
that the bispectrum (as computed from the present cosmic string model) lies hidden in the theoretical
noise and is therefore undetectable for any given value of $l$.

Let us note, however, that in its present stage the string code employed in these computations
describes Brownian, wiggly long strings in spite of the fact that long strings are very likely not
Brownian on the smallest scales, as recent field--theory simulations indicate.  In addition, the
presence of small string loops [Wu, \etal, 1998] and gravitational radiation into which they decay
were not yet included in this model. These are important effects that could, in principle, change the
above predictions for the string-generated CMB bispectrum on very small angular scales.

The imprint of cosmic strings on the CMB is a combination of different effects. Prior to the time of
recombination strings induce density and velocity fluctuations on the surrounding matter. During the
period of last scattering these fluctuations are imprinted on the CMB through the Sachs-Wolfe effect,
namely, temperature fluctuations arise because relic photons encounter a gravitational potential with
spatially dependent depth. In addition to the Sachs-Wolfe effect, moving long strings drag the
surrounding plasma and produce velocity fields that cause temperature anisotropies due to Doppler
shifts. While a string segment by itself is a highly non-Gaussian object, fluctuations induced by
string segments before recombination are a superposition of effects of many random strings stirring
the primordial plasma. These fluctuations are thus expected to be Gaussian as a result of the central
limit theorem.

As the universe becomes transparent, strings continue to leave their imprint on the CMB mainly due to
the Kaiser \& Stebbins [1984] effect.  As we mentioned in previous sections, this effect results in
line discontinuities in the temperature field of photons passing on opposite sides of a moving long
string.\footnote{The extension of the Kaiser-Stebbins effect to polarization will be treated below. In
fact, Benabed and Bernardeau [2000] have recently considered the generation of a B-type polarization
field out of E-type polarization, through gravitational lensing on a cosmic string.}  However, this
effect can result in non-Gaussian perturbations only on sufficiently small scales. This is because on
scales larger than the characteristic inter-string separation at the time of the radiation-matter
equality, the CMB temperature perturbations result from superposition of effects of many strings and
are likely to be Gaussian. Avelino \textit{et al.} [1998] applied several non-Gaussian tests to the
perturbations seeded by cosmic strings. They found the density field distribution to be close to
Gaussian on scales larger than $1.5 (\Omega _M h^2)^{-1}$ Mpc, where $\Omega _M$ is the fraction of
cosmological matter density in baryons and CDM combined. Scales this small correspond to the multipole
index of order $l \sim 10^4$.

\subsection{CMB polarization}
\label{sec-poladeff}    

The possibility that the CMB be polarized was first discussed by Martin~Rees in 1968, in the context
of anisotropic universe models. In spite of his optimism, and after many attempts during more than
thirty years, including some important upper limits [e.g., Keating, et al. 2001; Hedman, et al. 2001,
2002], there has been no positive detection of the polarization field until the DASI detection in
September 2002 [Leitch et al. 2002; Kovac et al. 2002].

Unlike previous experiments, DASI reached the required sensitivity to make a sounding discovery on
angular scales $\sim 0.\!\!^{\circ}5$. Along the same line, WMAP confirmed this detection with a
full-sky coverage and polarization data on five different frequencies on angular scales bigger than
$0.\!\!^{\circ}2$. Polarization is an important probe both for cosmological models and for the more
recent history of our nearby Universe. It arises from the interactions of CMB photons with free
electrons; hence, polarization can {\it only} be produced at the last scattering surface (its
amplitude depends on the duration of the decoupling process) and, unlike temperature fluctuations, it
is largely unaffected by variations of the gravitational potential after last scattering\footnote{With
the formation of the first stars and quasars, and the subsequent UV radiation emited by these
primitive sources, the hydrogen can re-ionize. As a consequence, the CMB will scatter again upon
ionized matter and will also modify its polarization, albeit on a different angular scale. Data from
first-year WMAP indicates that reionization did indeed take place somewhere around redshifts $z\sim
20$ (with big ``error'' bars), which, translated to the elapsed time since the big bang, represents
roughly a few hundred millon years.}. Future measurements of polarization will thus provide a clean
view of the inhomogeneities of the Universe at about 400,000 years after the Bang.

\begin{figure}[htbp]
\includegraphics[width=12cm]{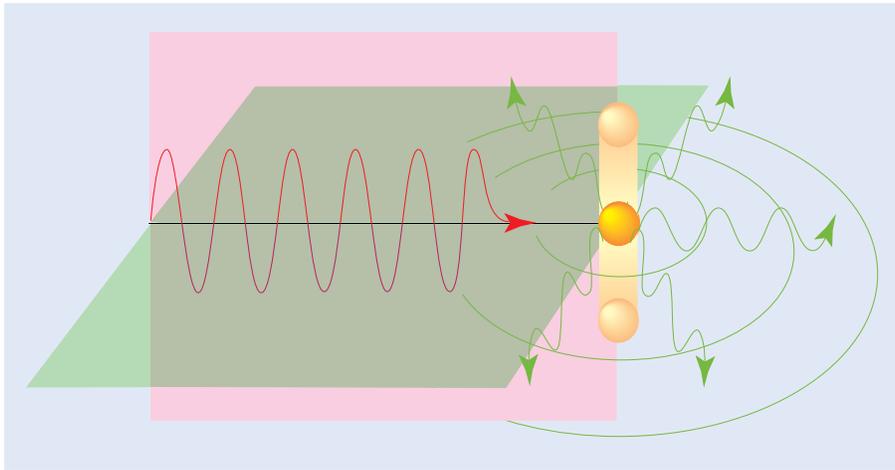}\\
\caption{{\sl 
An electromagnetic linearly polarized wave (in red) oscillates in a given plane (in pink). Reaching
an electron (orange ball) the wave induces the electron to also oscillate, making it emit radiation
(in green). This resulting electromagnetic wave is concentrated essentially in the (green) plane
orthogonal to the movement of the electron and it is polarized like the incident wave.}}
\label{figpol:1}\end{figure}          

For understanding polarization, a couple of things should be clear. First, the energy of the photons
is small compared to the mass of the electrons. Then, the CMB frequency does not change, since the
electron recoil is negligible. Second, the change in the CMB polarization (i.e., the orientation of
the oscillating electric field $\vec{\rm E}$ of the radiation) occurs due to a certain transition,
called {\it Thomson scattering}.  The transition probability per unit time is proportional to a
combination of the old ($\hat\epsilon^{\rm ~in}_\alpha$) and new ($\hat\epsilon^{\rm ~out}_\alpha$)
directions of polarization in the form $|\hat\epsilon^{\rm ~in}_\alpha\cdot\hat\epsilon^{\rm
~out}_\alpha|^2$. In other words, the initial direction of polarization will be favored.  Third, an
oscillating $\vec{\rm E}$ will push the electron to also oscillate; the latter can then be seen as a
dipole (not to be confused with the CMB dipole), and dipole radiation emits preferentially
perpendicularly to the direction of oscillation. These `rules' will help us understand why the CMB
should be linearly polarized \cite{agscien2}.

Previous to the recombination epoch, the radiation field is unpolarized.  In unpolarized light the
electric field can be decomposed into the two orthogonal directions (along, say, $\hat x$ and $\hat
z$) perpendicular to the line of propagation ($\hat y$). The electric field along $\hat\epsilon^{\rm
~in}_{\hat z}$ (suppose $\hat z$ is vertical) will make the electron oscillate also vertically. Hence,
the dipolar radiation will be maximal over the horizontal $xy$-plane.  Analogously, dipole radiation
due to the electric field along $\hat x$ will be on the $yz$-plane.  If we now look from the side
(e.g., from $\hat x$, on the horizontal plane and perpendicularly to the incident direction $\hat y$)
we will see a special kind of scattered radiation.  From our position we cannot perceive the radiation
that the electron oscillating along the $\hat x$ direction would emit, just because this radiation
goes to the $yz$-plane, orthogonal to us. Then, it is {\it as if} only the vertical component
($\hat\epsilon^{\rm ~in}_{\hat z}$) of the incoming electric field would cause the radiation we
perceive.  From the above rules we know that the highest probability for the polarization of the
outgoing radiation $\hat\epsilon^{\rm ~out}_\alpha$ will be to be aligned with the incoming one
$\hat\epsilon^{\rm ~in}_{\hat z}$, and therefore it follows that the outgoing radiation will be {\it
linearly} polarized.  Now, as both the chosen incoming direction and our position as observers were
arbitrary, the result will not be modified if we change them. Thomson scattering will convert
unpolarized radiation into linearly polarized one.

\begin{figure}[htbp]
\includegraphics[width=8.8cm]{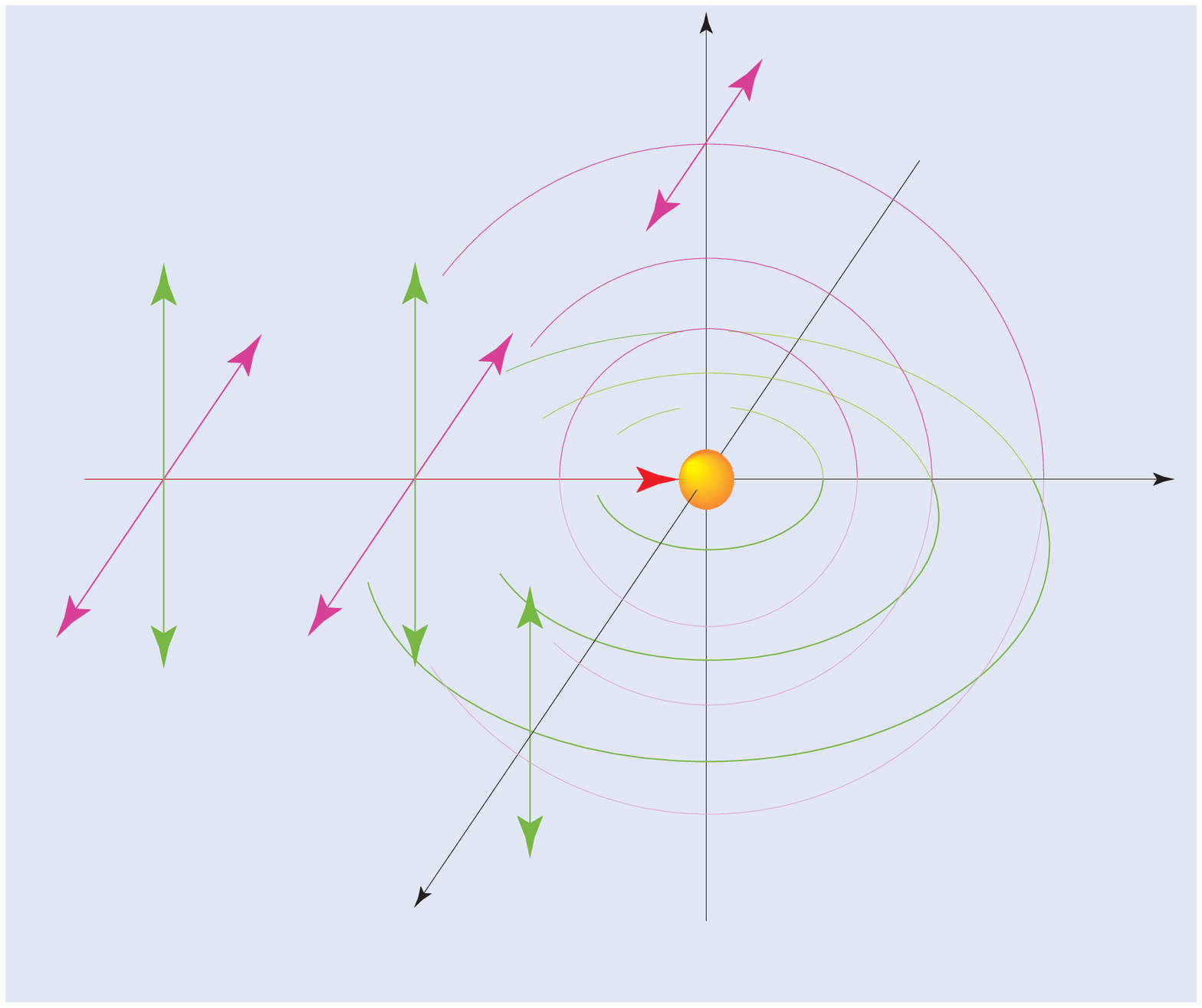}
\includegraphics[width=8.3cm]{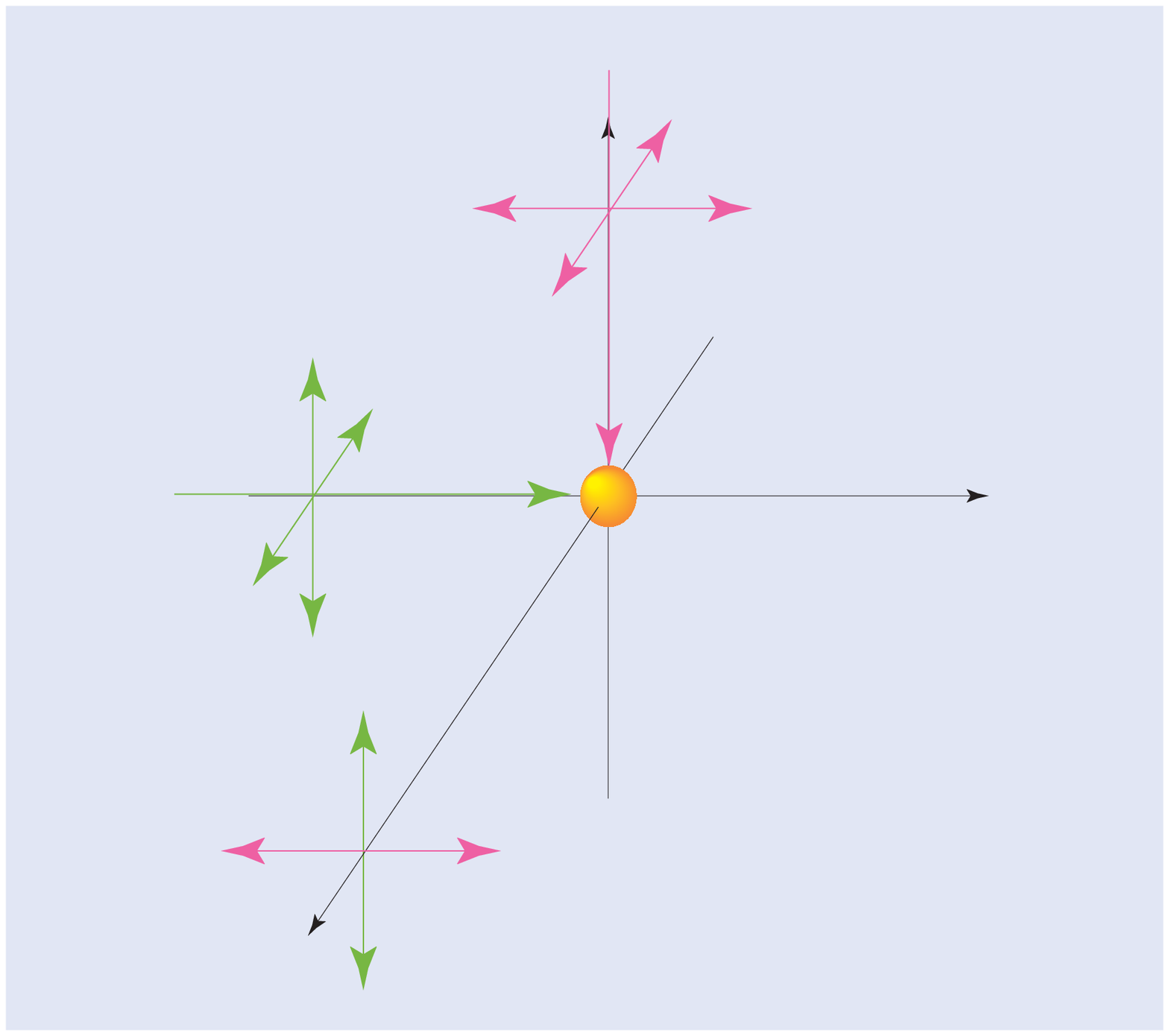}
\caption{{\sl 
Left panel: 
non-polarized electromagnetic wave can be decomposed into the sum of two linearly-polarized waves,
one along the line of sight (in pink), the other along a perpendicular direction (in
green). Scattered radiation due to the first wave is contained in the plane orthogonal to the line of
sight and cannot be detected. Only the second component (in green) reaches the observer and it is
polarized as the incident wave.
Right panel: 
when the charged particle receives non polarized waves from different directions (in green and pink),
it will re-emit the radiation, polarized also along different directions, to the observer.  If the
original radiation is not isotropic (say, the pink arrow is bigger than the green one), then one of
the resulting waves (in pink) will be slightly more intense than the other (in green), and the
observer will perceive a net excess of linear polarization.}}
\label{figpol:23}\end{figure}          

This however is not the end of the story. To get the total effect we need to consider all possible
directions from which photons will come to interact with the target electron, and sum them up. We see
easily that for an initial isotropic radiation distribution the individual contributions will cancel
out: just from symmetry arguments, in a spherically symmetric configuration no direction is
privileged, unlike the case of a net linear polarization which would select one particular direction.

Fortunately, we know the CMB is {\it not exactly} isotropic; to the millikelvin precision the dominant
mode is dipolar.  So, what about a CMB dipolar distribution~?  Although spatial symmetry does not help
us now, a dipole will not generate polarization either.  Take, for example, the radiation incident
onto the electron from the left to be more intense than the radiation incident from the right, with
average intensities above and below (that's a dipole); it then suffices to sum up all contributions to
see that no net polarization survives.  However, if the CMB has a {\it quadrupolar} variation in
temperature (that {\it it has}, first discovered by COBE, to tens of $\mu$K precision) then there will
be an excess of vertical polarization from left- and right-incident photons (assumed hotter than the
mean) with respect to the horizontal one from top and bottom light (cooler). From any point of view,
orthogonal contributions to the final polarization will be different, leaving a net linear
polarization in the scattered radiation.

There is one more point to emphasize.  Before recombination, ionized matter, electrons and radiation
formed a single fluid. In it, the inertia was provided by massive nucleons whilst the pressure was
that of radiation. And this fluid supported sound waves. In fact, the gravitational clumping tendency
of the effective mass in the perturbations was resisted by the restoring radiation pressure, and
therefore gravity-driven acoustic oscillations in both the fluid density and local velocity appeared.

Whereas the acoustic peaks in the temperature anisotropies correspond to the compression and
rarefaction maxima of the oscillating plasma, the polarization field responds to the local quadrupole
moment during the decoupling process. But this local quadrupole is mainly due to the Doppler shifts
induced by the velocity field of the plasma [Zaldarriaga \& Harari, 1995]. That is why we know with
certainty that polarization shows the uncontaminated dynamics of the primordial seeds at
recombination.

\begin{figure}[htbp]
\includegraphics[width=9.6cm]{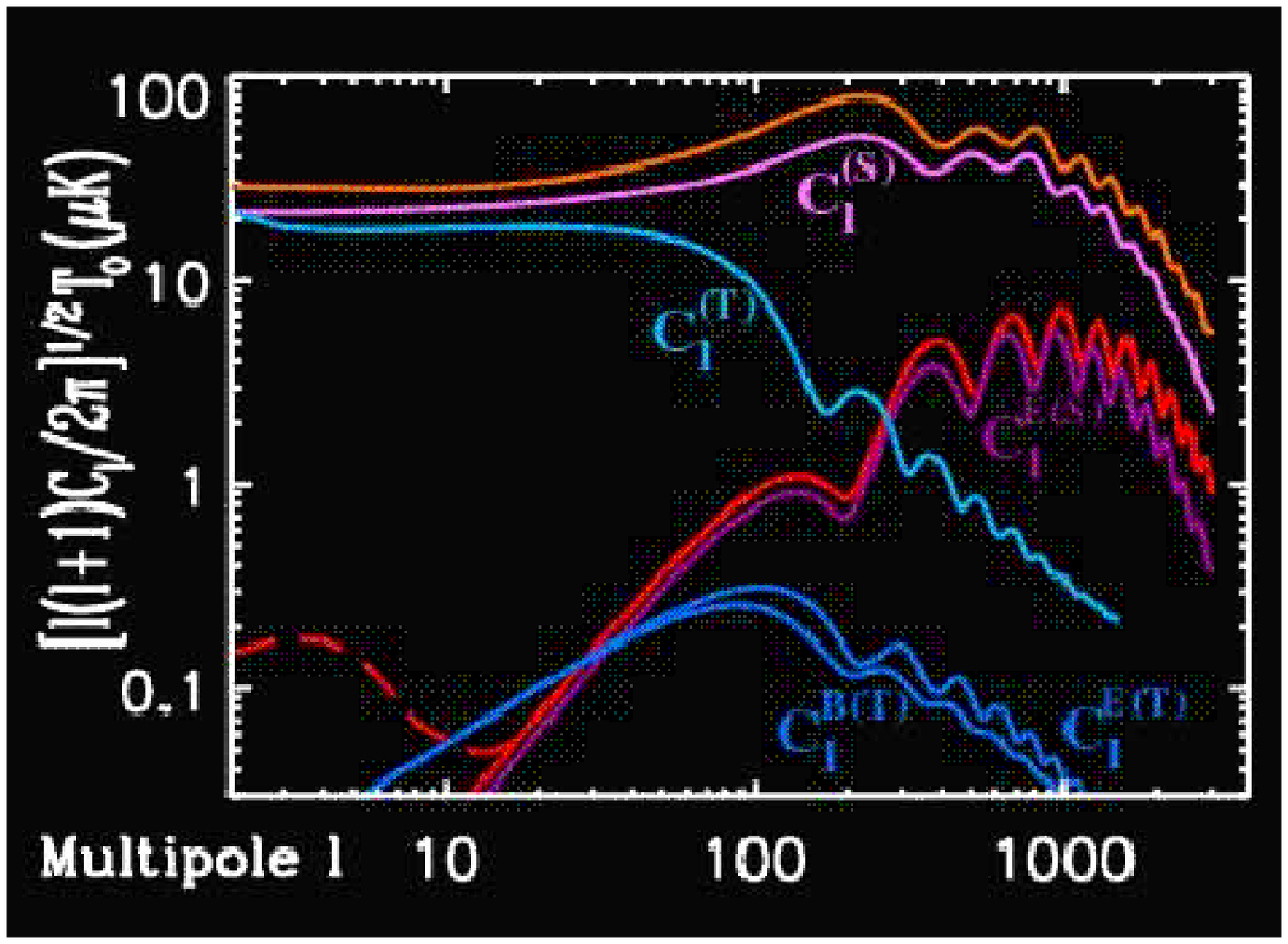}
\includegraphics[width=8cm]{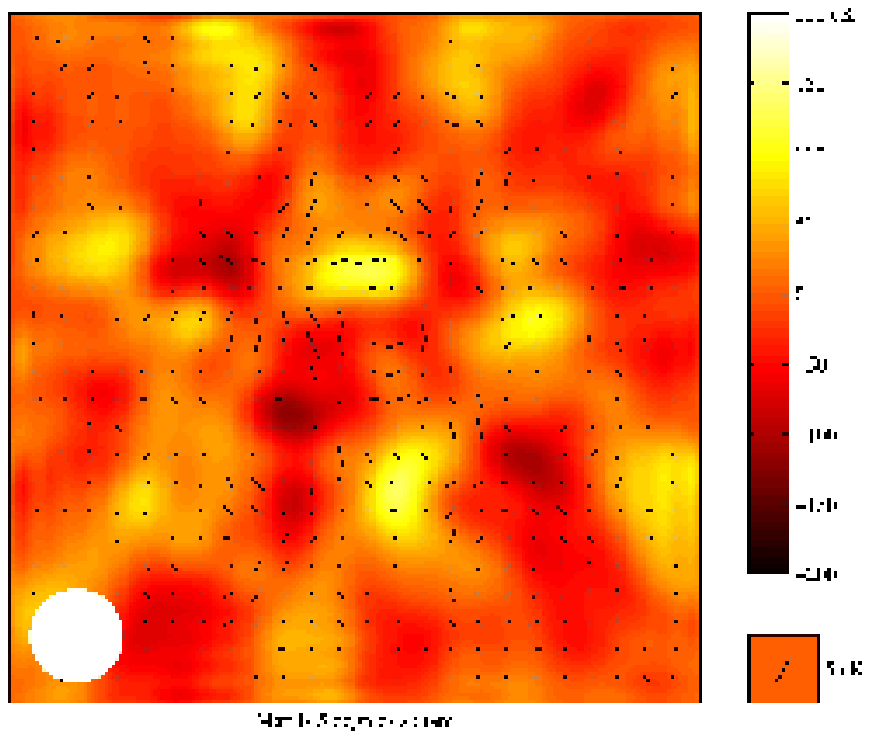}\\
\caption{{\sl Left panel: CMB Polarization for two different models.  Red and orange (unlabeled)
curves are the angular spectra derived for a $\Lambda$CHDM model, both with (red dashed line) and
without (red full line) reionization.  The temperature anisotropy spectrum from scalar perturbations
(proportional to $[ C_\ell ]^{1/2}$, orange curve) is virtually unchanged for both ionization
histories. The polarization spectrum ($\propto [ C^{\rm E(S)}_\ell ]^{1/2}$, red curves), although
indistinguishable for $\ell \gsim 20$, dramatically changes for small $\ell$'s; in this model the
Universe is reionized suddenly at low redshift with optical depth $\tau = 0.05$ [recall, however, that
recent first year data from the WMAP satellite indicates that $\tau = 0.17 \pm 0.04$].
Blue and violet curves represent a SCDM model {\it but} with a high tensor-mode amplitude, T/S=1 at
the quadrupole ($\ell=2$) level, with scale-invariant spectral indices $n_{\rm S}=1$ and $n_{\rm
T}=0$.  Separate scalar (noted $C^{\rm (S)}_\ell$) and tensor ($C^{\rm (T)}_\ell$) contributions to
temperature anisotropies are shown (top curves). Scalar modes only generate E-type polarization
($C^{\rm E(S)}_\ell$), which is smaller than the corresponding red curve of the $\Lambda$CHDM model
both due to differences in the models (notably $\Lambda\not=0$ for the red curves) and due to the
influence of tensors on the normalization at small $\ell$.  E- and B-type polarization from tensor
modes are also shown, respectively $C^{\rm E(T)}_\ell$ and $C^{\rm B(T)}_\ell$.  Model spectra were
computed with {\sc CMBFAST} and are normalized to $\delta {\rm T}_{\ell = 10} = 27.9\mu{\rm K}$.
Right panel: image of the intensity and polarization of the CMB made with the DASI telescope. The
small temperature variations of the CMB are shown in false color, with yellow hot and red cold. The
polarization at each spot in the image is shown by a black line. The length of the line shows the
strength of the polarization and the orientation of the line indicates the direction in which the
radiation is polarized. The size of the white spot in the lower left corner approximates the angular
resolution of the observations.  }}
\label{fig-mipola}
\end{figure}        

Within standard recombination models the predicted level of linear polarization on large scales is
tiny (see Figure \ref{fig-mipola}): the quadrupole generated in the radiation distribution as the
photons travel between successive scatterings is too small. Multiple scatterings make the plasma very
homogeneous and only wavelengths that are small enough (big $\ell$'s) to produce anisotropies over the
(rather short) mean free path of the photons will lead to a significant quadrupole, and thus also to
polarization.  Indeed, if the CMB photons last scattered at $z\sim 1100$, the SCDM model with $h=1$
predicts no more than 0.05 $\mu$K on scales greater than a few degrees. Hence, measuring polarization
at these scales represents an experimental challenge. 

However, CMB polarization increases remarkably around the degree-scale in standard models. In fact,
for $\theta < 1^\circ$ a bump with superimposed acoustic oscillations reaching $\sim 5 \mu$K is
generically forecasted. On these scales, like for the temperature anisotropies, the polarization field
shows acoustic oscillations. However, polarization spectra are sharper: temperature fluctuations
receive contributions from both density (dominant) and velocity perturbations and these, being out of
phase in their oscillation, partially cancel each other.  On the other hand, polarization is mainly
produced by velocity gradients in the baryon-photon fluid before last scattering, which also explains
why temperature and polarization peaks are located differently.  Moreover, acoustic oscillations
depend on the {\it nature} of the underlying perturbation; hence, we do not expect scalar acoustic
sound-waves in the baryon-photon plasma, propagating with characteristic adiabatic sound speed $c_{\rm
S}\sim c/\sqrt{3}$, close to that of an ideal radiative fluid, to produce the same peak-frequency as
that produced by gravitational waves, which propagate with the speed of light $c$ (see
Fig.\ref{fig-mipola}).

The main technical complication with polarization (characterized by a tensor field) is that it is not
invariant under rotations around a given direction on the sky, unlike the temperature fluctuation that
is described by a scalar quantity and invariant under such rotations.  The level of linear
polarization is conveniently expressed in terms of the {\it Stokes parameters} Q and U. It turns out
that there is a clever combination of these parameters that results in scalar quantities (in contrast
to the above noninvariant tensor description) but with different transformation properties under
spatial inversions ({\it parity} transformations).  Then, inspired by classical electromagnetism, any
polarization pattern on the sky can be separated into `electric' (scalar, unchanged under parity
transformation) and `magnetic' (pseudo-scalar, changes sign under parity) components (E- and B-type
polarization, respectively).

\subsubsection{CMB polarization from global defects}
\label{sec-polaglob}         

One then expands these different components in terms of spherical harmonics, very much like we did for
temperature anisotropies, getting coefficients $a^{m}_{\ell}$ for E and B polarizations and, from
these, the multipoles $C^{\rm E,B}_{\ell}$. The interesting thing is that (for symmetry reasons)
scalar-density perturbations will {\it not} produce any B polarization (a pseudo-scalar), that is
$C^{\rm B(S)}_\ell=0$.  We see then that an unambiguous detection of some level of B-type fluctuations
will be a signature of the existence (and of the amplitude) of a background of gravitational waves~!
[Seljak \& Zaldarriaga, 1997] (and, if present, also of rotational modes, like in models with
topological defects).

Linear polarization is a symmetric and traceless 2x2 tensor that requires 2 parameters to fully
describe it: $Q$, $U$ Stokes parameters.  These depend on the orientation of the coordinate system on
the sky. It is convenient to use $Q+iU$ and $Q-iU$ as the two independent combinations, which
transform under right-handed rotation by an angle $\phi$ as $(Q+iU)'=e^{-2i\phi}(Q+iU)$ and
$(Q-iU)'=e^{2i\phi}(Q-iU)$.  These two quantities have spin-weights $2$ and $-2$ respectively and can
be decomposed into spin $\pm 2$ spherical harmonics ${}_{\pm 2}Y_{lm}$
\begin{eqnarray}
(Q+iU)(\hat{\bbox{n}})&=&\sum_{lm}
a_{2,lm} \, {}_2Y_{lm}(\hat{\bbox{n}}) 
\\
(Q-iU)(\hat{\bbox{n}})&=&\sum_{lm}
a_{-2,lm} \, {}_{-2}Y_{lm}(\hat{\bbox{n}}). 
\end{eqnarray}
                       
Spin $s$ spherical harmonics form a complete orthonormal system for each value of $s$.  Important
property of spin-weighted basis: there exists spin raising and lowering operators $\edth$ and
$\baredth$.  By acting twice with a spin lowering and raising operator on $(Q+iU)$ and $(Q-iU)$
respectively one obtains quantities of spin 0, which are {\it rotationally invariant}. These
quantities can be treated like the temperature and no ambiguities connected with the orientation of
coordinate system on the sky will arise. Conversely, by acting with spin lowering and raising
operators on usual harmonics spin $s$ harmonics can be written explicitly in terms of derivatives of
the usual spherical harmonics.  Their action on ${}_{\pm 2}Y_{lm}$ leads to
\begin{eqnarray}
\baredth^2(Q+iU)(\hat{\bbox{n}})&=&
\sum_{lm}
\left({[l+2]! \over [l-2]!}\right)^{1/2}
a_{2,lm}Y_{lm}(\hat{\bbox{n}})
\\
\edth^2(Q-iU)(\hat{\bbox{n}})&=&\sum_{lm}
\left({[l+2]! \over [l-2]!}\right)^{1/2}
a_{-2,lm}Y_{lm}(\hat{\bbox{n}}). 
\end{eqnarray}                    
With these definitions the expressions for the expansion coefficients
of the two polarization variables become [Seljak \& Zaldarriaga, 1997]
\begin{eqnarray}
a_{2,lm}&=&\left({[l-2]! \over [l+2]!}\right)^{1/2}
\int d\Omega\; Y_{lm}^{*}(\hat{\bbox{n}})
\baredth^2(Q+iU)(\hat{\bbox{n}})
\\
a_{-2,lm}&=&\left({[l-2]! \over [l+2]!}\right)^{1/2}
\int d\Omega\;
Y_{lm}^{*}(\hat{\bbox{n}})\edth^2(Q-iU)(\hat{\bbox{n}}).
\label{alm}
\end{eqnarray}       
Instead of $a_{2,lm}$, $a_{-2,lm}$ it is convenient to introduce their
linear {\it electric} and {\it magnetic} combinations
\begin{eqnarray}
a_{E,lm}=-{1\over 2}(a_{2,lm}+a_{-2,lm}) \qquad
a_{B,lm}= {i\over 2}(a_{2,lm}-a_{-2,lm}). 
\label{aeb}
\end{eqnarray}
These two behave differently under {\it parity} transformation:
while $E$ remains unchanged $B$ changes the sign, in analogy
with electric and magnetic fields.

To characterize the statistics of the CMB perturbations only four power spectra are needed, those for
$X = T, E, B$ and the cross correlation between $T$ and $E$.  The cross correlation between $B$ and
$E$ or $B$ and $T$ vanishes because $B$ has the opposite parity of $T$ and $E$. As usual, the spectra
are defined as the rotationally invariant quantities
\begin{eqnarray}
C_{Xl}={1\over 2l+1}\sum_m \langle a_{X,lm}^{*} a_{X,lm}\rangle
\qquad
C_{Cl}={1\over 2l+1}\sum_m \langle a_{T,lm}^{*}a_{E,lm}\rangle
\label{Cls}
\end{eqnarray}
in terms of which on has
\begin{eqnarray}
\langle a_{X,l^\prime m^\prime}^{*} a_{X,lm}\rangle&=&
C_{Xl} \, \delta_{l^\prime l} \delta_{m^\prime m} 
\\
\langle a_{T,l^\prime m^\prime}^{*} a_{E,lm}\rangle&=&
C_{Cl} \, \delta_{l^\prime l} \delta_{m^\prime m} 
\\
\langle a_{B,l^\prime m^\prime}^{*} a_{E,lm}\rangle&=&
\langle a_{B,l^\prime m^\prime}^{*} a_{T,lm}\rangle=0. 
\label{stat}
\end{eqnarray}
                                        
According to what was said above, one expects some amount of polarization to be present in all
possible cosmological models. However, symmetry breaking models giving rise to topological defects
differ from inflationary models in several important aspects, two of which are the relative
contributions from scalar, vector and tensor modes and the coherence of the seeds sourcing the
perturbation equations. In the local cosmic string case one finds that in general scalar modes are
dominant, if one compares to vector and tensor modes in the usual decomposition of perturbations. The
situation with global topological defects is radically different and this leads to a very distinctive
signature in the polarization field.

\begin{figure}[htbp]
\includegraphics[width=7cm]{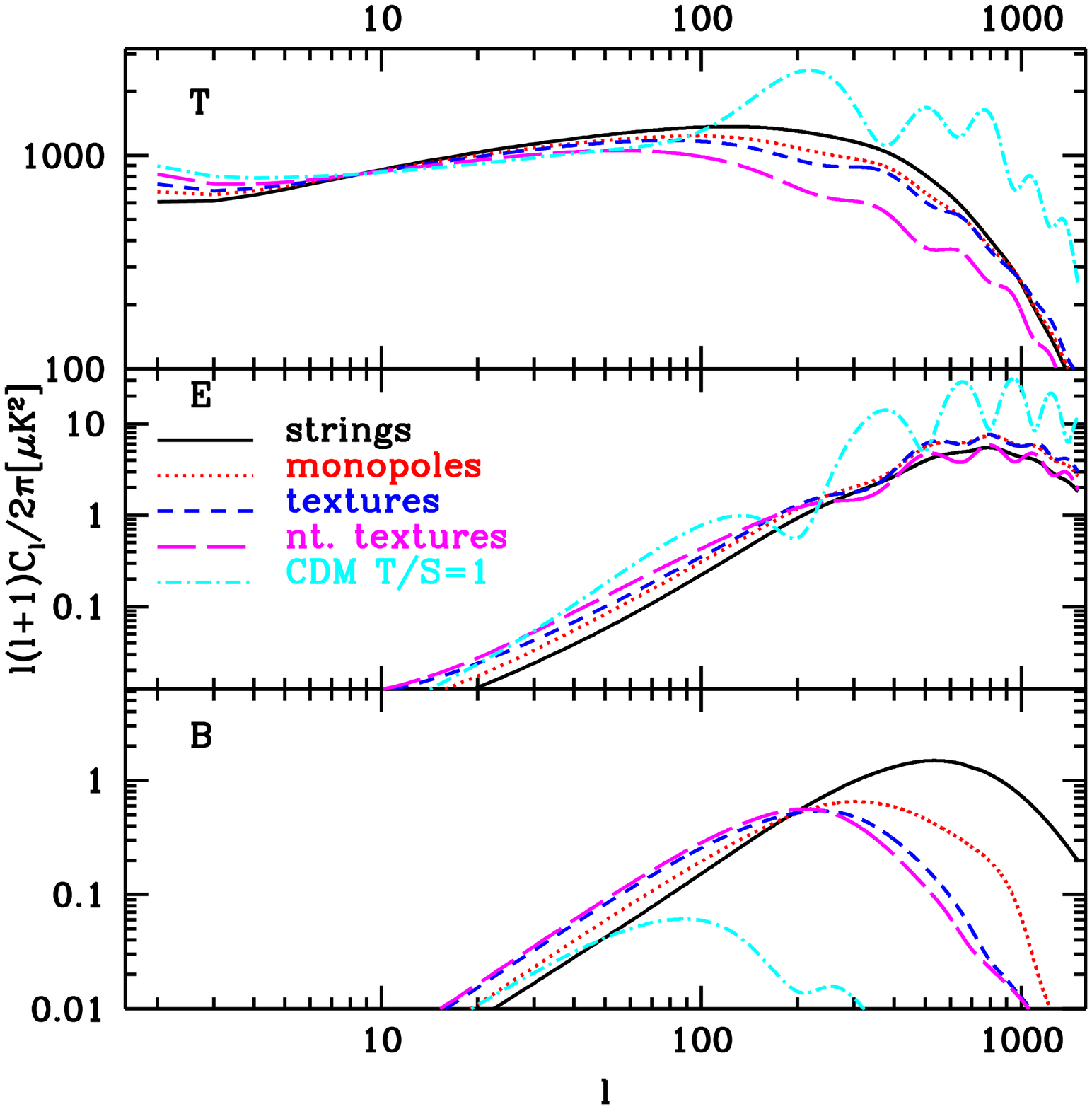}
\includegraphics[width=9.2cm]{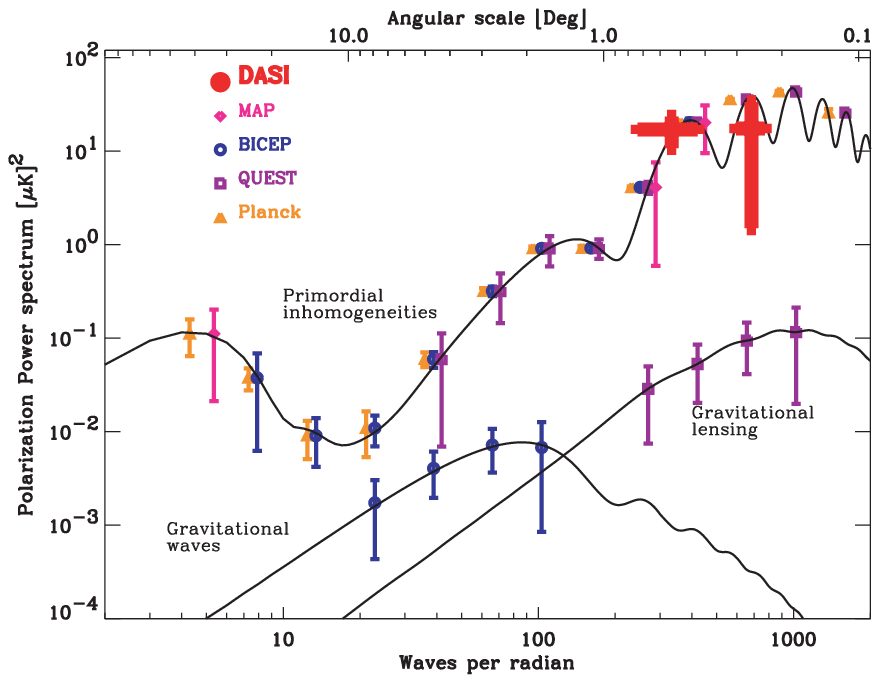}\\
\caption{{\sl Left panel: 
Power spectra of temperature (T), electric type polarization (E) and magnetic type
polarization (B) for global strings, monopoles, textures and nontopological textures [taken from
Seljak. \etal, 1997].  The corresponding spectra for a standard CDM model with $T/S=1$ is also shown
for comparison. B polarization turns out to be notably larger for all global defects considered if
compared to the corresponding predictions of inflationary models on small angular scales. 
Right panel: current and future polarization data by Hivon \& Kamionkowski [2002]. 
Top curve shows the prediction for the polarization from primordial inhomogeneities produced
by inflation. The large-angle bump in this curve is the enhancement from early
star formation (reionization). The lower curves are for inflationary gravitational-wave and
gravitational-lensing signals. Recently detected DASI data points are shown in red while the rest 
are expected data points for future experiments with more sensitivity. }}
\label{fig-polaglob}\end{figure}           

Temperature and polarization spectra for various symmetry breaking models were calculated by Seljak,
Pen \& Turok [1997] and are shown in figure \ref{fig-polaglob}. Both electric and magnetic components
of polarization are shown for a variety of global defects. They also plot for comparison the
corresponding spectra in a typical inflationary model, namely, the standard CDM model ($h=0.5$,
$\Omega=1$, $\Omega_{\rm baryon}=0.05$) but with equal amount of scalars and tensors perturbations
(noted $T/S=1$) which maximizes the amount of B component from inflationary models. In all the models
they assumed a standard reionization history. The most interesting feature they found is the large
magnetic mode polarization, with a typical amplitude of $\sim 1 \mu K$ on degree scales [exactly those
scales probed by Hedman, \etal, 2001]. For multipoles below $\ell \sim 100$ the contributions from $E$
and $B$ are roughly equal.  This differs strongly from the inflationary model predictions, where $B$
is much smaller than $E$ on these scales even for the extreme case of $T/S \sim 1$.  Inflationary
models only generate scalar and tensor modes, while global defects also have a significant
contribution from vector modes. As we mentioned above, scalar modes only generate $E$, vector modes
predominantly generate $B$, while for tensor modes $E$ and $B$ are comparable with $B$ being somewhat
smaller.  Together this implies that B can be significantly larger in symmetry breaking models than in
inflationary models.
In figure \ref{fig-polaglob} we also show the recent discovery of a tiny level of polarization by the
DASI collaboration together with predictions for future experiments, assuming an inflationary origin
for the temperature perturbation and polarization signals~\footnote{See {\tt
http://www.stanford.edu/$\sim$schurch} and {\tt
http://astro.caltech.edu/$\sim$lgg/bicep\_front.htm}}.

\subsubsection{String lensing and CMB polarization}
\label{sec-bb2000}      

Recent studies have shown that in realistic models of inflation cosmic string formation seems quite
natural in a post-inflationary preheating phase [Tkachev \etal , 1998, Kasuya \& Kawasaki, 1998].  So,
even if the gross features on CMB maps are produced by a standard (\eg, inflationary) mechanism, the
presence of defects, most particularly cosmic strings, could eventually leave a distinctive
signature. One such feature could be found resorting to CMB polarization: the lens effect of a string
on the small scale $E$-type polarization of the CMB induces a significant amount of $B$-type
polarization along the line-of-sight [Zaldarriaga \& Seljak, 1998; Benabed \& Bernardeau 2000]. This
is an effect analogous to the Kaiser-Stebbins effect for temperature maps.

In the inflationary scenario, scalar density perturbations generate a scalar polarization pattern,
given by \( E \)-type polarization, while tensor modes have the ability to induce both \( E \) and \(
B \) types of polarization. However, tensor modes contribute little on very small angular scales in
these models. So, if one considers, say, a standard \( \Lambda \mathrm{CDM} \) model, only scalar
primary perturbations will be present without defects. But if a few strings are left from a very early
epoch, by studying the patch of the sky where they are localized, a distinctive signature could come
to light.

In the small angular scale limit, in real space and 
in terms of the Stokes parameters  \( Q \) and \( U \) 
one can express the \( E \) and \( B \) fields as follows
\begin{eqnarray}
\label{EEBB}
E\equiv \Delta ^{-1}[({\partial x}^{2}-{\partial y}^{2})\,
Q+2{\partial x} {\partial y} \, U], \\
B\equiv \Delta ^{-1}[({\partial x}^{2}-{\partial y}^{2})\,
U-2{\partial x} {\partial y} \, Q]. 
\end{eqnarray}

The polarization vector is parallel transported along the geodesics.
The lens affects the polarization by displacing the apparent position
of the polarized light source.  Hence, the observed Stokes parameters
\( \hat{Q} \) and \( \hat{U} \) are given in terms of the
\emph{primary} (unlensed) ones by:
$\hat{Q} (\vec{\alpha} )=Q(\vec{\alpha} + \vec{\xi} )
 \, {\rm ~and~ } \,
 \hat{U} (\vec{\alpha} )=U(\vec{\alpha} + \vec{\xi} )$.
The displacement \( \vec{\xi} \) is given by the integration of the
gravitational potential along the line--of--sights.
Of course, here the `potential' acting as lens is the cosmic string
whose effect on the polarization field we want to study.

\begin{figure}[htbp]
\includegraphics[width=5cm]{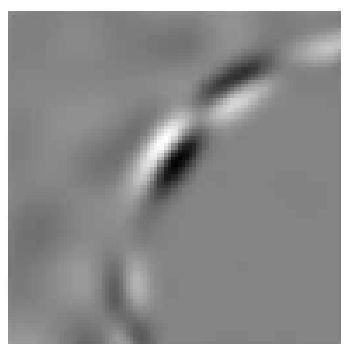}
%
\includegraphics[width=5cm]{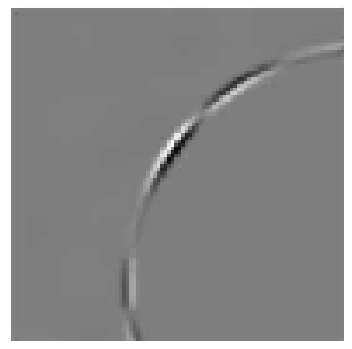}\\ 
\caption{{\sl Simulations for the \(B\) field in the case of a circular loop.  The angular size of the
figure is \(50'\times 50'\).  The resolution is 5' (left) and 1.2' (right).  The discontinuity in the
\(B\) field is sharper the better the resolution. Weak lensing of CMB photons passing relatively apart
from the position of the string core are apparent as faint patches outside of the string loop on the
left panel. [Benabed \& Bernardeau 2000].}}
\label{fig-bb2000}\end{figure}   

In the case of a straight string which is aligned along the $y$ axis,
the deflection angle (or half of the deficit angle) is \(4\pi G\mu \)
[Vilenkin \& Shellard, 2000] and this yields a displacement 
\(\xi_{x}=\pm \xi_{0}\) with
\be
\xi_{0}=4\pi G\mu
{\mathcal{D}}_{\textrm{lss,s}}/{\mathcal{D}}_{\textrm{s,us}}
\ee
with no displacement along the $y$ axis.
${\mathcal{D}}_{\textrm{lss,s}}$ and 
${\mathcal{D}}_{\textrm{s,us}}$  
are the cosmological angular distances
between the last scattering surface and the string, and
between the string and us, respectively. They can be computed, in an
Einstein-de Sitter universe (critical density, just dust and no
$\Lambda$), from
\be
{\mathcal{D}}_{\textrm{}}(z_1 , z_2)= 
{2c\over H_0}{1\over 1+z_2}[(1+z_1)^{-1/2}-(1+z_2)^{-1/2}]
\ee
by taking $z_1 = 0$ for us and $z_2\simeq 1000$ for the last scattering surface; see [Bartelmann \&
Schneider, 2001]. For the usual case in which the redshift of the string $z_{\rm s}$ is well below the
$z_{\textrm{lss}}$ one has ${\mathcal{D}}_{\textrm{lss,s}}/{\mathcal{D}}_{\textrm{lss,us}}\simeq
1/\sqrt{1+z_{\rm s}}$. Taking this ratio of order 1/2 (\ie, distance from us to the last scattering
surface equal to twice that from the string to the last scattering surface) yields $z_{\rm s}\simeq
3$. Plugging in some numbers, for typical GUT strings on has $G\mu \simeq 10^{-6}$ and so the typical
expected displacement is about less than 10 arc seconds.  Benabed \& Bernardeau [2000] compute the
resulting \( B \) component of the polarization and find that the effect is entirely due to the
discontinuity induced by the string, being nonzero just along the string itself. This clearly limits
the observability of the effect to extremely high resolution detectors, possibly post-Planck ones.

The situation for circular strings is different.  As shown by de Laix
\& Vachaspati [1996] the lens effect of such a string, when facing the
observer, is equivalent to the one of a static linear mass
distribution. Considering then a loop centered at the origin of the
coordinate system, the displacement field can be expressed very simply:
observing in a direction through the loop, $\vec{\xi}$ has to vanish, 
while outside of the loop the displacement decreases as \( {\alpha
_{l}}/{\alpha } \), \ie, inversely proportional to the angle.  
One then has [Benabed \& Bernardeau, 2000]
\be 
\vec{\xi}(\vec{\alpha})= -2\xi_{0}
{\alpha_{l} \over \alpha^{2}}
\vec{\alpha} \quad {\rm ~with~} \quad \alpha >\alpha_{l} ,
\ee
where $\alpha_{l}$ is the loop radius. 

This ansatz for the displacement, once plugged into the above equations, yields the \(B\) field shown
in both panels of Figure \ref{fig-bb2000}. A weak lensing effect is barely distinguishable outside the
string loop, while the strong lensing of those photons traveling close enough to the string is the
most clear signature, specially for the high resolution simulation.  One can check that the hot and
cold spots along the string profile have roughly the same size as for the polarization field in the
absence of the string loop.  The simulations performed show a clear feature in the maps, although
limited to low resolutions this can well be confused with other secondary polarization sources. It is
well known that point radio sources and synchrotron emission from our galaxy may contribute to the
foreground [de Zotti \etal\ 1999] and are polarized at a 10 \% level. Also lensing from large scale
structure and dust could add to the problem.


\section{Cosmic defects in perspective}
\label{sec-cdip}

Cosmic defects have proved very interesting and fruitful in high--energy physics and
astrophysics. Their generic production in grand unified theories has made defects an active field of
research for over two decades. Many of the interesting subjects now associated with defects were only
briefly mentioned in these notes, like the internal structure of defects --leading to persistent
currents in their cores-- and, as a consequence, the possible generation of primordial magnetic
fields. Also, primordial gravitational waves, extremely high--energy phenomena associated to cosmic
rays and uhecrons, electroweak baryogenesis and, finally, the very active condensed-matter--cosmology
interface, dubbed cosmos in the lab, equally --and unjustly-- received no attention [compensate for
this with references like Vilenkin \& Shellard, 2000; Hindmarsh \& Kibble, 1995; Kibble, 2002; Gangui,
2001b, for example -- yes, in that order.]. With regards to the most transparent test of current
cosmology, namely the CMB and matter power spectra, (not so) recent investigations have pointed out
severe problems in virtually all models where cosmic defects are the main source of the seeds of
structure in the universe\footnote{{\em ``There are no strings on me''} --Pinocchio}. In the case of
cosmic strings, however, these bad news were reached by the use of non--negligible, albeit
well--founded, approximations in order to cope with the limited range of realistic defect simulations
[bad or good news, depending on which side you are.] Although the whole method of unequal time
correlators employed by most of the groups can be regarded as a good approximation to reality during
both the matter and radiation eras, the important transition in between must be looked at more
carefully, as the above--mentioned correlators do not scale as expected. Recent, full Boltzmann
analyses aiming to solve this handicap are in progress [\eg, Landriau \& Shellard, 2002, 2003] and
already producing interesting results.




\section*{Acknowledgments}

\noindent 
I'd like to thank my collaborators in some of the topics covered in these lectures for their insights
and remarks. Thanks also to the other speakers and students for the many discussions during this very
instructive time we spent together, and to the members of the L.O.C., particularly Prof. M\'ario
Novello and Santiago Perez Bergliaffa from CBPF, for their superb job in organizing this charming
school. Finally, I'd like to acknowledge {\sc CONICET}, {\sc UBA} and {\sc Fundaci\'on Antorchas} for
their financial support.



\begin{thebibliography}{999}        


\bibitem{Abr57}
Abrikosov, A. A. [1957], 
 {\sl Sov. Phys. JETP}      {\bf 5},  1174 
[{\sl Zh. Eksp. Teor. Fiz.} {\bf 32}, 1442 (1957)]

\bibitem{AchucarroVachaspati2000}
Ach\'ucarro, A. \& Vachaspati, T. [2000],
{\sl Phys. Rept.} {\bf 327}, 347-426;  {\sl Phys. Rept.} {\bf 327}, 427.
[hep-ph/9904229]   
%
\bibitem{ABR97} Albrecht, A., Battye, R. \& Robinson, J. [1997],  
    {\sl Phys. Rev. Lett.} {\bf 79}, 4736.;
    {\sl Phys. Rev.} {\bf D59}, 023508 (1998).    

\bibitem{alb} 
Albrecht, A., Coulson, D., Ferreira, P. \& 
Magueijo, J. [1995],  Imperial Preprint /TP/94--95/30.



\bibitem{Alexanderetal99} 
Alexander, S., Brandenberger, R., Easther, R. \& Sornborger, A. [1999], 
hep-ph/9903254.




\bibitem{varyingc} 
Avelino, P.P. \& Martins, C.J.A.P. [2000],
\textit{Phys. Rev. Lett.} \textbf{85}, 1370. 

\bibitem{ASWA98} 
Avelino, P.P., Shellard, E.P.S., Wu, J.H.P. \& Allen, B. [1998], 
\textit{Astrophys. J.} {\bf 507}, L101. 




\bibitem{BartelmannSchneider2001}
Bartelmann, M. \& Schneider, P. [2001],
{\sl Phys. Rep.} {\bf 340}, 291.


\bibitem{Battyeetal99}
Battye, R. A., Bucher, M. \& Spergel, D. [1999], astro-ph/9908047 .

\bibitem{BattyeWeller2000}
Battye, R. \& Weller, J. [2000],
\textit{Phys. Rev.} \textbf{D61}, 043501.

\bibitem{coucou} 
Benabed, K. \& Bernardeau, F. [2000], 
\textit{Phys. Rev.} \textbf{D61}, 123510.  

\bibitem{wmabennett}
Bennett, C., \etal ~[2003], astro-ph/0302207.





\bibitem{BR90} 
Bennett, D.P. \&  Rhie, S.H. [1990], {\sl Phys. Rev. Lett.} {\bf 65}, 1709. 

\bibitem{BR} 
Bennett, D.P. \&  Rhie, S.H. [1993], {\sl Ap. J. Lett.} {\bf 406}, L7.

\bibitem{BRWein93}
Bennett, D.P., Rhie, S.H. \& Weinberg, D.H. [1993], preprint.







\bibitem{BorrillEtAl94}
Borrill, J., \etal ~[1994], {\sl Phys. Rev.} {\bf D50}, 2469.


\bibitem{Bouchet+Co2001}
Bouchet, F.R., Peter, P., Riazuelo, A. \& Sakellariadou, M. [2001], preprint astro-ph/0005022.



\bibitem{BrandenbergerRev}
Brandenberger, R. [1993], {\sl Topological Defects and Structure Formation}, 
EPFL lectures, Lausanne, Switzerland.  







\bibitem{BunkovGodfrin2000}
Bunkov, Y. \& Godfrin, H. [2000], (editors)
Proceedings of the NATO-ASI on topological defects and non-equilibrium
dynamics of symmetry breaking phase transitions (Kluwer, Dordrecht).



\bibitem{CallanColeman}
Callan, C. \& Coleman, S. [1977], {\sl Phys. Rev.} {\bf D16}, 1762.




\bibitem{Carter1990}
Carter, B, [1990], {\sl Phys. Rev.} {\bf D41}, 3869.

\bibitem{Carter1997}   
Carter, B. [1997], Tlaxcala lecture notes, hep-th/9705172.  




\bibitem{CPG}
Carter, B., Peter, P. \& Gangui, A. [1997],
{\sl Phys. Rev.} {\bf D55}, 4647. [hep-ph/9609401];




\bibitem{Contaldietal99}
Contaldi, C., Hindmarsh, M. \& Magueijo, J. [1999], 
\textit{Phys. Rev. Lett.} \textbf{82} 2034.

\bibitem{cope}
Copeland, E. [1993], in {\sl The physical universe: The interface 
between cosmology, astrophysics and particle physics}, eds. Barrow, J.D., \etal ~(Sringer--Verlag).


\bibitem{CowieHu1987}
Cowie, L. \& Hu, E. [1987], {\sl Ap. J.}{\bf 318} L33.





\bibitem{Davis87}
Davis, R. L. [1987], {\sl Phys. Rev. D} {\bf 35}, 3705. 

\bibitem{vortons}
Davis, R.L. \& Shellard, E.P.S. [1988],
{\sl Phys. Lett.} {\bf B207}, 404. 

\bibitem{boomerang} 
de Bernardis, P. {\it et al.} [2000], Nature 404, 995 [astro-ph/0004404] 


\bibitem{deLaixVacha} de Laix, A.A. \& Vachaspati, T. [1996], {\sl Phys. Rev.} {\bf D54}, 4780.

\bibitem{dezotti99} 
de Zotti, G. \etal [1999], astro--ph/9908058.


\bibitem{DolanJackiw74}
Dolan, L. \& Jackiw, R. [1974], {\sl Phys. Rev.} {\bf D9}, 3320. 



\bibitem{RuthReview2000}
Durrer, R. [2000], in Moriond meeting on Energy Densities in the Universe, astro-ph/0003363 .

\bibitem{DGS}
Durrer, R., Gangui, A. \& Sakellariadou, M. [1996], Phys. Rev. Lett. {\bf 76}, 579. [astro-ph/9507035].

\bibitem{DKM02}
Durrer, R., Kunz, M. \& Melchiorri, A. [2002], Phys.Rept. 364 (2002) 1-81. [astro-ph/0110348].

\bibitem{DZ} 
Durrer, R. \& Zhou, Z.H. [1995], Z\"urich University Preprint, ZH--TH19/95, astro--ph/9508016.


\bibitem{dvalietal98}
Dvali, G., Liu, H. \& Vachaspati, T. [1998], {\sl Phys. Rev. Lett.} {\bf 80}, 2281.
















\bibitem{agscien}
Gangui, A. [2001], \textit{Science} \textbf{291}, 837.

\bibitem{agscien2}
Gangui, A. [2003], \textit{Science} \textbf{299}, 1333.

\bibitem{boli01}
Gangui, A. [2001b], Topological Defects in Cosmology, 
Lecture Notes for the First Bolivian School on Cosmology, in press [astro-ph/0110285].

\bibitem{3point}
Gangui, A., Lucchin, F., Matarrese, S. \& Mollerach, S. [1994],
{\sl Ap. J.} {\bf 430}, 447.

\bibitem{GM99} Gangui, A. \& Martin, J. [2000a],
    {\sl Mon. Not. R. Astron. Soc.} \textbf{313}, 323.
    [astro-ph/9908009]

\bibitem{GM00} Gangui, A. \& Martin, J. [2000b],
    {\sl Phys. Rev.} {\bf D62}, 103004. [astro-ph/0001361]     

\bibitem{SilyYo}
Gangui, A. \& Mollerach, S. [1996], 
{\sl Phys. Rev.} {\bf D54}, 4750-4756. [astro-ph/9601069]     

\bibitem{GanguiPeter98}
Gangui, A. \& Peter, P. [1998],
Cosmological and Astrophysical Implications of Superconducting Cosmic Strings and Vortons, in 
Proceedings of the XXXIIIrd Rencontres de Moriond on `Fundamental Parameters in Cosmology',
pages 19-24, Editors: J.~Tran Thanh Van \etal, Editions Fronti\`eres, 1998.    

\bibitem{AgPpCb}
Gangui, A., Peter, P. \& Boehm, C. [1998],
{\sl Phys. Rev.} {\bf D57}, 2580. [hep-ph/9705204]


\bibitem{GPWna}
Gangui, A., Pogosian, L. \& Winitzki, S. [2001a], {\sl New Astronomy Reviews} 46, 681-691 (2002). 
[astro-ph/0112145]     

\bibitem{GPW}
Gangui, A., Pogosian, L. \& Winitzki, S. [2001b], {\sl Phys. Rev.} {\bf D64}, 043001.  







\bibitem{Goto1971}
Goto, T. [1971],
{\sl Prog. Theor. Phys.} {\bf 46}, 1560. 


\bibitem{Gott85}
Gott, R. [1985], {\sl Ap. J.}, {\bf 288}, 422.





\bibitem{Guth-Weinberg83}
Guth, A.H. \& Weinberg, E. [1983], {\sl Nucl. Phys.} {\bf B212}, 321.



\bibitem{maxima} Hanany, S., {\it et al},
{\sl Astrophys. J. Lett.} accepted (2000), astro-ph/0005123; 
Jaffe, A.H., {\it et al}, astro-ph/0007333.      


\bibitem{dasi} 
Halverson, N.W. \textit{et al}, preprint astro-ph/0104489.




\bibitem{Hedman_pola2001}
Hedman, M., Barkats, D., Gundersen, J. Staggs, S. \& Winstein,
B. [2001], 
{\sl Ap. J.}, {\bf 548}, L111-L114 [astro-ph/0010592];
Hedman, M. et al. [2002],  Astrophys. J. 573, L73 (2002).    


\bibitem{Higgs1964}
Higgs, P. [1964], {\sl Phys. Lett.} {\bf 12}, 132.


\bibitem{Hilton1953}
Hilton, P.J. [1953], {\sl Introduction to homotopy theory} (Cambridge: Cambridge University Press).


\bibitem{Hindmarsh+Kibble95}
Hindmarsh, M.B. \& Kibble, T.W.B. [1995], {\sl Rept. Prog. Phys.} {\bf 58}, 477-562 [hep-ph/9411342].

\bibitem{HivonKamion02}
Hivon, E. \& Kamionkowski, M. [2002], {\sl Science} 298, 1349.





\bibitem{HuSperWhite} 
Hu, W., Spergel, D. \& White, M.  [1997],
{\sl Phys. Rev.} {\bf D55}, 3288-3302.      







\bibitem{Kaiser+Stebbins84}
Kaiser, N. \&  Stebbins, A. [1984], {\sl Nature} {\bf 310}, 391.


\bibitem{KasuyaKawasaki98} 
Kasuya, S. \& Kawasaki, M, [1998], {\sl Phys. Rev.} {\bf D58}, 083516.

\bibitem{Keating01} 
Keating, B. et al. [2001], Astrophys. J. 560, L1 (2001).

\bibitem{Kibble76}
Kibble, T.W.B. [1976], {\sl J. Phys.} {\bf A9}, 1387.

\bibitem{Kibble}
Kibble, T.W.B. [1980], {\sl Phys. Rep.} {\bf 67}, 183.

\bibitem{Kib} 
Kibble, T.W.B. [1985], {\sl Nucl. Phys.} {\bf B252}, 227; {\bf B261}, 750 (1986).  

\bibitem{Kib02}
Kibble, T.W.B. [2002], Lectures at NATO ASI ``Patterns of symmetry breaking'', 
    Cracow [cond-mat/0211110].

\bibitem{KibbleEtAl82}
Kibble, T.W.B., Lazarides, G. \& Shafi, Q. [1982], 
{\sl Phys. Lett.} {\bf B113}, 237.

\bibitem{Kirzhnits-Linde74}
Kirzhnits, D.A. \& Linde, A.D. [1974], {\sl Sov. Phys. JETP} {\bf 40}, 628.

\bibitem{Kirzhnits-Linde76} 
Kirzhnits, D.A. \& Linde, A.D. [1976], {\sl Ann. Phys.} {\bf 101}, 195.






\bibitem{wmakogut}
Kogut, A., \etal ~[2003], astro-ph/0302213.   


\bibitem{KT90} 
Kolb, E.W. \&  Turner, M.S. [1990]
               {\sl The Early Universe} (New York: Addison--Wesley).


\bibitem{KS00} 
Komatsu, E. \& Spergel, D. [2000], preprint astro-ph/0005036;

\bibitem{Kovac02} 
Kovac, J. et al. [2002], Nature 420, 772-787.  [astro-ph/0209478] 



\bibitem{LemperiereShellard2002}
Lemperi\`ere, Y. \& Shellard, E.P.S. [2002], preprint hep-ph/0207199 . 

\bibitem{LandriauShellard2002}
Landriau, M. \&  Shellard, E.P.S. [2002], preprint astro-ph/0208540 .  

\bibitem{LandriauShellard2003}
Landriau, M. \&  Shellard, E.P.S. [2003], preprint astro-ph/0302166 .  

\bibitem{Langacker-Pi80}
Langacker, P. \& Pi, S.--Y. [1980], {\sl Phys. Rev. Lett.} {\bf 45}, 1. 

\bibitem{Langer92}
Langer, S. [1992], in {\sl Solids far from equilibrium}, Godr\`eche, C., 
ed. (Cambridge: Cambridge University Press).  

\bibitem{Leitch02} 
Leitch, E.M. et al. [2002], Nature 420, 763-771. astro-ph/0209476

\bibitem{liddle} 
Liddle, A. [1995], {\sl Phys. Rev.} {\bf D51}, 5347-5351.   









\bibitem{Linde83b}
Linde, A.D. [1983b], {\sl Nucl. Phys.} {\bf B216}, 421.




\bibitem{Linet86}
Linet, B. [1986], {\sl Phys. Rev.} {\bf D33} 1833. 











\bibitem{magalb} 
Magueijo, J. , Albrecht, A. , Ferreira, P. \& Coulson, D. [1996],
{\sl Phys. Rev.} {\bf D54}, 3727-3744.      

\bibitem{MagueijoBrandenberger2000}
Magueijo, J. \& Brandenberger, R. [2000],
Iran lectures on cosmic defects, astro-ph/0002030.  

\bibitem{MagueijoFerreira97}
Magueijo, J. \& Ferreira, P. [1997], {\sl Phys. Rev.} {\bf D55}, 3358.




\bibitem{MS96} 
Martins, C.J.A.P. \& Shellard,  E.P.S. [1996], {\sl Phys. Rev.} {\bf D54} 2535.



\bibitem{MS00} Martins, C.J.A.P. \& Shellard, E.P.S. [2000], hep-ph/0003298    



\bibitem{Mermin79}
Mermin, M. [1979] , {\sl Rev. Mod. Phys.} {\bf 51}, 591. 









\bibitem{Nambu1970}
Nambu, Y. [1970], in Proc. Int. Conf. on Symmetries and Quark Models, ed. Chand, R. 
(New York: Gordon and Breach). 



\bibitem{Notzold91}
Notzold, D. [1991], {\sl Phys. Rev.} {\bf D43}, R961.







\bibitem{wmappage}
Page, L., \etal ~[2003], astro-ph/0302220.



\bibitem{PeeblesRatra03}
Peebles, P.J.E. \& Ratra, B. [2002], Rev. Mod. Phys. (in press) astro-ph/0207347 . 


\bibitem{uetc} 
Pen, U.-L., Seljak, U. \& Turok, N. [1997], 
{\sl Phys. Rev. Lett.} {\bf 79}, 1611-1614.   

\bibitem{PST} 
Pen, U.--L., Spergel, D.N. \& Turok, N. [1994], {\sl Phys. Rev.} {\bf D49}, 692.



\bibitem{perlmutter99}
Perlmutter, S., \etal ~[1999], Astrophys. J. {\bf 517}, 565. 



\bibitem{Pogosian2001}
Pogosian, L. [2001], Int. J. Mod. Phys. A16S1C . astro-ph/0009307.

\bibitem{PV99} Pogosian, L. \& Vachaspati, T. [1999],
    {\sl Phys. Rev.} \textbf{D60}, 083504.          

\bibitem{PV00} Pogosian, L. \& Vachaspati, T. [2000],
    {\sl Phys. Rev.} \textbf{D62}, 105005.          

\bibitem{Preskill79}
Preskill, J. [1979], {\sl Phys. Rev. Lett.} {\bf 43}, 1365.






\bibitem{Rajaraman}
Rajaraman, R. [1982], {\sl Solitons and instantons} (Amsterdam: North--Holland).



\bibitem{Riessetal1998}
Riess, A.G, \etal ~[1998], Astron.J. 116, 1009-1038. astro-ph/9805201  



\bibitem{Sachs+Wolfe67}
Sachs, R. \& Wolfe, A. [1967], {\sl Ap. J.} {\bf 147}, 73.




\bibitem{Sazhinetal03} 
Sazhin, M., \etal ~[2003], to appear in MNRAS. [astro-ph/0302547]

\bibitem{Scaramella+Vittorio91}
Scaramella, R. \& Vittorio, N. [1991], {\sl Ap. J.} {\bf 375}, 439.

\bibitem{Scaramella+Vittorio93}
Scaramella, R. \& Vittorio, N. [1993], {\sl M.N.R.A.S.} {\bf 263}, L17.





\bibitem{SPT97}
Seljak, U., Pen, U.-L. \& Turok, N. [1997], {\sl Phys.Rev.Lett.} {\bf 79} 1615-1618.

\bibitem{cmbfast} Seljak, U. \& Zaldarriaga, M. [1996], {\sl Astrophys. J.} {\bf 469}, 437.   

\bibitem{UrMa}
Seljak, U. \& Zaldarriaga, M. [1997], {\sl Phys. Rev. Lett.} {\bf 78}, 2054.




\bibitem{SilkVilenkin84}
Silk, J. \& Vilenkin, A. [1984], {\sl Phys. Rev. Lett.} {\bf 53}, 1700.















\bibitem{Steenrod}
Steenrod, N. [1951], {\sl Topology of Fibre Bundles} (Princeton: Princeton University Press).









\bibitem{Tkachevetal98}
Tkachev, I., Khlebnikov, S., Kofman, L. \& Linde, A. [1998], {\sl Phys.Lett.} {\bf B440}, 262-268.







\bibitem{Tu} 
Turok, N. [1989], {\sl Phys. Rev. Lett.} {\bf 63}, 2625.

\bibitem{turok} 
Turok, N. [1996], 
{\sl Phys. Rev. Lett.} {\bf 77}, 4138-4141; 
{\sl Phys. Rev.} {\bf D54}, 3686-3689      

\bibitem{TPS98} 
Turok, N., Pen, U.-L. \& Seljak, U. [1998], 
{\it Phys. Rev.} {\bf D58}, 023506.  

\bibitem{TuSpe} 
Turok, N. \& Spergel, D.N. [1990], {\sl Phys. Rev. Lett.} {\bf 64}, 
2736.

\bibitem{TuZa} 
Turok, N. \& Zadrozny, J. [1990], {\sl Phys. Rev. Lett.} {\bf 65}, 
2331.







\bibitem{Vachaspati2001}
Vachaspati, T. [2001], ICTP summer school lectures, hep-ph/0101270.       

\bibitem{Vachaspati+Vilenkin91}
Vachaspati, T. \&  Vilenkin, A. [1991], 
{\sl Phys. Rev. Lett.} {\bf 67}, 1057.

\bibitem{Veeraraghavan+Stebbins90}
Veeraraghavan, S. \& Stebbins, A. [1990], {\sl Ap. J.} {\bf 365}, 37.

\bibitem{Vilenkin81}
Vilenkin, A. [1981], {\sl Phys. Rev.} {\bf D23}, 852.

\bibitem{Vilenkin83}
Vilenkin, A. [1983], {\sl Phys. Rev.} {\bf D27}, 2848.

\bibitem{Vilenkin84}  
Vilenkin, A. [1984], {\sl Phys. Rev. Lett.} {\bf 53}, 1016.    

\bibitem{Vilenkin85}
Vilenkin, A. [1985], {\sl Phys. Rep.} {\bf 121}, 263.  

\bibitem{Vilenkin90}
Vilenkin, A [1990],  {\sl Phys. Rev.} {\bf D41}, 3038.


\bibitem{Vilenkin+Shellard94}
Vilenkin, A. \& Shellard, E.P.S. [2000], {\sl Cosmic Strings and
other Topological Defects}, 2nd edition, (Cambridge: CUP). 


\bibitem{Vollick92}
Vollick, D.N. [1992], {\sl Phys. Rev.} {\bf D45}, 1884.


\bibitem{WalkerEtAl91}
Walker, P.N., \etal ~[1991], {\sl Ap. J.} {\bf 376}, 51.

Wang, L. \& Kamionkowski, M. [2000],
\textit{Phys. Rev.} \textbf{D61}, 063504. 


\bibitem{Weinberg74}
Weinberg, S. [1974], {\sl Phys. Rev.} {\bf D9}, 3357.








\bibitem{proty} 
Wu, J.-H.P. \textit{et al}, preprint astro-ph/9812156.








\bibitem{ZaHa95}
Zaldarriaga, M. \& Harari, D. [1995], Phys. Rev. D52, 3276.    

\bibitem{MaUr}
Zaldarriaga, M. \& Seljak, U. [1998],
{\sl Phys. Rev.} {\bf D58}, 023003.    


\bibitem{Zeldovichetal74}
Zel'dovich, Ya. B., I. Yu. Kobzarev \& L. B. Okun [1974], 
Zh. Eksp. Teor. Fiz. {\bf 67}, 3 [Sov. Phys. JETP {\bf 40}, 1 (1975)].         




\end{thebibliography}
\end{document}